\documentclass[leqno,draft,11pt,a4paper]{article}
\usepackage{amssymb,amsmath,latexsym,theorem}
\usepackage[includeheadfoot,margin=1in]{geometry}

\allowdisplaybreaks[2]

\newcommand\eq{\leftrightarrow}
\newcommand\EQ{\Leftrightarrow}
\newcommand\nequiv{\not\equiv}
\newcommand\LOR{\bigvee}
\newcommand\ET{\bigwedge}
\newcommand\model{\vDash}

\newcommand\fii{\varphi}
\newcommand\ep{\varepsilon}
\newcommand\roo{\varrho}

\newcommand\p[1]{\langle#1\rangle}
\newcommand\lh[1]{\lvert#1\rvert}
\newcommand\sset{\subseteq}
\newcommand\ssset{\subsetneq}
\newcommand\Sset{\supseteq}
\newcommand\nul{\varnothing}
\newcommand\res{\mathbin\restriction}

\newcommand\fl[1]{\lfloor#1\rfloor}
\newcommand\Fl[1]{\left\lfloor#1\right\rfloor}
\newcommand\cl[1]{\lceil#1\rceil}
\newcommand\fdiv{\genfrac\lfloor\rfloor{}{}}
\newcommand\cdiv{\genfrac\lceil\rceil{}{}}
\newcommand\rem{%
  \nonscript\mskip-\medmuskip\mkern5mu%
  \mathbin{\mathrm{rem}}\penalty900\mkern5mu%
  \nonscript\mskip-\medmuskip}
\newcommand\M{\mathit}
\newcommand\rsuv{\M{RSUV}}
\newcommand\PM[2]{\pm{}^{#1}_{#2}}

\DeclareMathOperator\im{im}
\DeclareMathOperator\bit{bit}
\DeclareMathOperator\card{card}
\DeclareMathOperator\Lh{lh}
\DeclareMathOperator\lcm{lcm}

\newcommand\cxt{\mathrm}
\newcommand\ptime{\cxt P}
\newcommand\nc{\cxt{NC}}
\newcommand\nci{\nc^1}

\newcommand\tc{\cxt{TC}^0}
\newcommand\Ac{\cxt{AC}^0}
\newcommand\LT{\cxt{DLOGTIME}}

\newcommand\sig{\Sigma^b_}
\newcommand\delt{\Delta^b_}
\newcommand\delz{\Delta_0}
\newcommand\Sig{\Sigma^B_}
\newcommand\io{\M{IOpen}}
\newcommand\idz{I\Delta_0}
\newcommand\pv{PV}
\newcommand\tpv{\pv_1}
\newcommand\sss{S^1_2}
\newcommand\vtc{\M{VTC}^0}
\newcommand\Div{\M{DIV}}
\newcommand\imul{\M{IMUL}}
\newcommand\vtcim{\vtc+\imul}
\newcommand\vl{\M{VL}}
\newcommand\vnl{\M{VNL}}
\newcommand\vnc{\M{\!VNC}^1}
\newcommand\php{\M{PHP}}
\newcommand\wphp{\M{WPHP}}
\newcommand\sch[1]{\ensuremath{\M{#1}}}
\newcommand\Min{\sch{MIN}}
\newcommand\ind{\sch{IND}}
\newcommand\comp{\sch{COMP}}
\newcommand\ac{\sch{AC}}
\newcommand\acr{\sch{AC^R}}
\newcommand\MF{\mathrm}
\newcommand\Divf{\MF{Div}}
\newcommand\powm{\MF{pow}}
\newcommand\imulm{\MF{imul}}
\newcommand\cyc{\M{Cyc}}
\newcommand\prm{\M{Prime}}
\newcommand\tot{\M{Tot}}
\newcommand\rec{\MF{Rec}}

\newcommand\Q{\mathbb Q}
\newcommand\Z{\mathbb Z}
\newcommand\zg[1]{(\Z/#1\Z)^\times}
\newcommand\plog{\mathrm{pl}}
\newcommand\cM{\mathcal M}

\newcommand\ob{\overline}
\newcommand\txto{${}\to{}$}

\newcommand\noproof{\leavevmode\unskip\bme\vadjust{}\nobreak\hfill$\Box$\par}
\newenvironment{Pf}[1][]
  {\par\noindent\textit{Proof:}\bme\ignorespaces}
  {\noproof\pagebreak[2]\vskip\medskipamount\ignorespacesafterend}
\newcommand\noqed{\let\noproof\relax}
\newcommand\bme{\hskip.75em\relax}

\theoremstyle{plain}
\newtheorem{Thm}{Theorem}[section]
\newtheorem{Prop}[Thm]{Proposition}
\newtheorem{Cor}[Thm]{Corollary}
\newtheorem{Lem}[Thm]{Lemma}
\theorembodyfont\upshape
\newtheorem{Def}[Thm]{Definition}
\newtheorem{Rem}[Thm]{Remark}
\newtheorem{Exm}[Thm]{Example}

\newcommand\numberthis{\addtocounter{equation}{1}\tag{\theequation}}

\DeclareRobustCommand*\magiclparen{\ifmmode(\else\textup(\nobreak\hskip0pt\relax\fi}
\DeclareRobustCommand*\magicrparen{\ifmmode)\else\textup)\fi}
\catcode`\(=\active \catcode`\)=\active
\let (=\magiclparen  \let )=\magicrparen

\author{Emil Je\v r\'abek\\[\medskipamount]
Institute of Mathematics, Czech Academy of Sciences\\
\small \v Zitn\'a 25,
115\:67 Praha 1,
Czech Republic,
email: \texttt{jerabek@math.cas.cz}
}

\title{Iterated multiplication in $\vtc$}

\begin{document}
\maketitle

\begin{abstract}
We show that $\vtc$, the basic theory of bounded arithmetic corresponding to the complexity class~$\tc$, proves the
$\imul$ axiom expressing the totality of iterated multiplication satisfying its recursive definition, by formalizing a
suitable version of the $\tc$ iterated multiplication algorithm by Hesse, Allender, and Barrington. As a
consequence, $\vtc$ can also prove the integer division axiom, and (by our previous results) the
$\rsuv$-translation of induction and minimization for sharply bounded formulas. Similar consequences hold for the
related theories $\delt1\text-\sch{CR}$ and~$C^0_2$.

As a side result, we also prove that there is a well-behaved $\delz$~definition of modular powering in
$\idz+\wphp(\delz)$.


\end{abstract}

\section{Introduction}\label{sec:introduction}

The underlying theme of this paper is \emph{feasible reasoning} about the elementary integer arithmetic operations $+$,
$\cdot$, $\le$: what properties of these operations can be proven using only concepts whose complexity does not exceed
that of $+$, $\cdot$, $\le$ themselves? There is a common construction in proof complexity that allows to make such
questions formal: given a (sufficiently well-behaved) complexity class~$C$, we can define a theory of arithmetic~$T$
that ``corresponds'' to~$C$. While the notion of correspondence is somewhat vague, what this typically means is that on
the one hand, the provably total computable (in a suitable sense) functions of~$T$ are exactly the $C$-functions, and
on the other hand, $T$ can reason with $C$-concepts: it proves induction, comprehension, minimization, or similar
schemata for formulas that express predicates computable in~$C$.

In our case, the right complexity class%
\footnote{Originally, $\tc$ was introduced as a nonuniform circuit class by Hajnal et al.~\cite{tc0}, but in this paper
we always mean the $\LT$-uniform version of the class, which gives a robust notion of ``fully uniform''
$\tc$ with several equivalent definitions across various computation models (cf.\ \cite{founif}). Likewise for~$\Ac$.}
is~$\tc$: the elementary arithmetic operations are all computable in~$\tc$, and while $+$ and~$\le$ are already in
$\Ac\ssset\tc$, multiplication is $\tc$-complete under $\Ac$ Turing-reductions. The arithmetical theory corresponding
to~$\tc$ that we will work with in this paper is $\vtc$, defined by Nguyen and Cook~\cite{ngu-cook} as a two-sorted
theory of bounded arithmetic in the style of Zambella~\cite{zamb:notes}. Earlier, Johannsen and
Pollett~\cite{joh-pol:c02,joh-pol:d1cr} introduced two theories corresponding to~$\tc$ in the framework of
single-sorted theories of Buss~\cite{buss}: $\delt1\text-\M{CR}$, which is equivalent to~$\vtc$ under the $\rsuv$
translation, and its extension~$C^0_2$. (Since $C^0_2$ is conservative over $\delt1\text-\M{CR}$ for a class of
formulas that encompasses the statements that we are interested in in this paper, there is no difference between these
theories for our purposes.)

While it is easy to show (and not particularly difficult to formalize in~$\vtc$) that $\tc$ includes $+$, $\cdot$, $-$,
and iterated addition $\sum_{i<n}X_i$, it is considerably harder to prove that it includes integer \emph{division} and
\emph{iterated multiplication} $\prod_{i<n}X_i$. The history of this result starts with Beame, Cook, and
Hoover~\cite{bch}, who proved (in present terminology) that division, iterated multiplication, and powering $X^n$
(with $n$ given in unary) are $\tc$ Turing-reducible to each other, and that they are all computable in
\emph{$\ptime$-uniform~$\tc$}. (In fact, \cite{bch} predates the definition of~$\tc$; they referred to~$\nci$ in the
paper. It is easy to observe though that their algorithms can be implemented using threshold circuits.) The basic idea
of~\cite{bch} is to compute iterated multiplication in the \emph{Chinese remainder representation (CRR)}, i.e., modulo a
sequence of small primes~$\vec m$, and then reconstruct the result in binary from CRR. The main source of nonuniformity
(or insufficient uniformity) in~\cite{bch} is the CRR reconstruction procedure: they require the CRR basis~$\vec m$ to
be fixed in advance (for a given input length), and supplied to the algorithm along with the product $\prod_im_i$.

The next breakthrough was achieved by Chiu, Davida, and Litow~\cite{cdl}, who devised a more efficient CRR
reconstruction procedure based on computation of the rank of CRR that did not rely on~$\prod_im_i$, and as a
consequence, proved that division and iterated multiplication are in $\cxt L$-uniform $\tc$, and in particular, in~$\cxt
L$ itself. (Their paper still refers to~$\nci$ rather than~$\tc$.) Subsequently, Hesse, Allender, and
Barrington~\cite{hab} proved the optimal result that division and iterated multiplication are in (fully uniform) $\tc$
by first reducing the remaining nonuniformity in CRR reconstruction to the modular powering function
$\powm(a,r,m)=a^r\bmod m$ (with all inputs in unary, and $m$~prime), and then showing that $\powm$ is in fact
computable in $\Ac\sset\tc$.

We mention that once we know that $\tc$ includes iterated multiplication, it follows easily that it can do many other
arithmetic functions: in particular, the basic operations $+$, $\cdot$, \dots\ (including iterated $\sum$ and $\prod$)
are $\tc$-computable not just in the integers, but also in~$\mathbb Q$ and more general number fields, and in rings of
polynomials; and we can compute rational approximations of analytic functions given by sufficiently nice power series,
such as trigonometric and inverse trigonometric functions, $\log$ and $\exp$ (for inputs of small magnitude). On the
arithmetical side, it was shown in Je\v r\'abek~\cite{ej:vtc0iopen} that the theory $\vtc$ augmented with
an iterated multiplication axiom is fairly powerful: by formalizing $\tc$ root approximation algorithms for
constant-degree univariate polynomials, it proves binary-number induction for quantifier-free formulas in the language
of ordered rings ($\io$), and even binary-number induction and minimization for $\rsuv$ translations of
$\sig0$~formulas in Buss's language.

In view of these developments, it is natural to ask whether $\tc$ integer division and iterated multiplication
algorithms can be formalized in the corresponding theory~$\vtc$. This problem was posed in the concluding section of
Nguyen and Cook~\cite{ngu-cook}, where it was attributed to A.~Atserias; it was then restated in Cook and
Nguyen~\cite[IX.7.6]{cook-ngu} and Je\v r\'abek~\cite[Q.~8.2]{ej:vtc0iopen}. Earlier,
Atserias~\cite{ats:wphp-conf,ats:wphp} asked whether $\idz$ can formalize a $\delz$~definition of modular
exponentiation (whose existence is another consequence of~\cite{hab}). Johannsen~\cite{joh:c02div}
(predating~\cite{cdl,hab}) devised a theory $C^0_2[\M{div}]$ extending~$C^0_2$ that corresponds to the $\tc$-closure of
division; the problem of formalizing division and iterated multiplication in~$\vtc$ is equivalent to the question if
$C^0_2\equiv C^0_2[\M{div}]$ (more precisely, if $C^0_2[\M{div}]$ is an extension of $C^0_2$ by a definition), but this
was not explicitly posed as a problem in~\cite{joh:c02div}.

To clarify, since all $\tc$~functions are provably total in~$\vtc$, it trivially follows that the theory can define
provably total functions that express the division and iterated multiplication algorithms of~\cite{hab}. However, the
theory does not necessarily \emph{prove} anything about such functions, besides the fact that they compute the correct
specific outputs for inputs given by standard constants. When we ask for formalization of division in~$\vtc$, what we
actually mean is whether the theory can prove an axiom $\Div$ postulating the existence of $\fl{Y/X}$ that satisfies
the defining property
\[X\ne 0\to\fl{Y/X}X\le Y<\bigl(\fl{Y/X}+1\bigr)X,\]
and likewise, formalization of iterated multiplication refers to an axiom $\imul$ stating the existence of iterated
products $\prod_{i<n}X_i$ satisfying the defining recurrence
\begin{align*}
\prod_{i<0}X_i&=1,\\
\prod_{i<n+1}X_i&=X_n\prod_{i<n}X_i.
\end{align*}
(The exact definitions of $\imul$ and $\Div$ are given in Section~\ref{sec:preliminaries}.) This requires much
more than just totality of the two functions. Note that whether we ask about the provability of $\imul$ or $\Div$ is
just a matter of convenience: it follows from the results of \cite{joh:c02div,ej:vtc0iopen} (formalizing the
reductions from~\cite{bch}) that $\imul$ implies $\Div$ over $\vtc$, and that $\vtc$ proves $\Div$ if and only if it
proves $\imul$. For the purposes of this paper, it will be more natural to work with $\imul$.

The reader may wonder what makes the formalization of the iterated multiplication algorithm from~\cite{hab} so
challenging. After all, the algorithm and its analysis are rather elementary, they do not rely on any sophisticated
number theory. It is true that the argument in~\cite{hab} does not really just consist of a single algorithm---it has a
complex structure with several interdependent parts:
\begin{enumerate}
\item\label{item:13} Show that iterated multiplication is in $\tc(\powm)$, using CRR reconstruction.
\item\label{item:14} Show that iterated multiplication with polylogarithmically small input is in $\Ac$, by scaling
down part~\ref{item:13}. 
\item\label{item:15} Show that $\powm$ is in~$\Ac$ using~\ref{item:14}, and plug it into~\ref{item:13}.
\end{enumerate}
However, this is not by itself a fundamental obstacle. What truly makes the formalization difficult is that the
analysis of the algorithms suffers from several problems of a ``chicken or egg'' type: which came first, the chicken or
the egg? Specifically:
\begin{itemize}
\item The analysis (proof of soundness) of the CRR reconstruction procedure in part~\ref{item:13} heavily relies on
iterated products and divisions: e.g., it refers to the product $\prod_im_i$ of primes from the CRR basis. However,
when working in $\vtc$, we need the soundness of the CRR reconstruction procedure to define such iterated products in
the first place.
\item Similarly, the analysis of the modular exponentiation algorithm in part~\ref{item:15} refers to results of
modular exponentiation such as $a^{\fl{n/d_i}}$, and in particular, it relies on Fermat's little theorem $a^n=1$.
However, the latter cannot be stated, let alone proved, without having a means to define modular exponentiation in the
first place.
\item A more subtle, but all the more important, issue is that in part~\ref{item:13}, the reduction of iterated modular
multiplication $\imulm(\vec a,m)=\prod_ia_i\bmod m$ ($m$ prime) to $\powm$ relies on cyclicity of the multiplicative
groups $\zg m$, which is notoriously difficult to prove in bounded arithmetic (cf.\ \cite[Q.~4.8]{ej:flt}). While this
may look more like an instance of ``sophisticated number theory'' at first sight, what makes it a chicken-or-egg
problem as well is that the cyclicity of $\zg m$ is in fact provable in $\vtcim$.
\end{itemize}

The main result of this paper is that $\imul$ is, after all, provable in~$\vtc$, and specifically, $\vtc$ can formalize
the soundness of a version of the Hesse, Allender, and Barrington~\cite{hab} algorithm. Our formalization follows the
basic outline of the original argument, adjusted to overcome the above-mentioned difficulties:
\begin{itemize}
\item Since we do not know how to prove directly the cyclicity of $\zg m$ in~$\vtc$, we formalize part~\ref{item:13}
using $\imulm$ as a primitive instead of $\powm$: that is, we prove $\imul$ in $\vtc(\imulm)$. We get around the
chicken-or-egg problems by developing many low-level properties of CRR in $\vtc(\imulm)$, in particular the effects of
simple CRR operations such as those used in the definition of the CRR reconstruction procedure. This is the most
technical part of the paper.
\item Part~\ref{item:14} is easy to formalize in the basic theory~$V^0$ (corresponding to~$\Ac$) by observing that
polylogarithmic cuts of models of~$V^0$ are models of $\vnl$, which improves a result of
M\"uller~\cite{muller:polylog}.
\item We avoid the chicken-or-egg problems in part~\ref{item:15} by modifying the modular powering algorithm so that it
does not need the auxiliary values $a^{\fl{n/d_i}}$ at all, using more directly the underlying idea from~\cite{hab} of
applying CRR to exponents. Since we need the weak pigeonhole principle to ensure there are enough ``good'' primes for
the CRR, the formalization proceeds in $V^0+\wphp$ rather than plain~$V^0$. By exploiting the conservativity of $V^0$
over $\idz$, we obtain the stand-alone result that there is a $\delz$~definition of $\powm$ (even for nonprime moduli)
whose defining recurrence is provable in~$\idz+\wphp(\delz)$, which may be of independent interest.
\item The results so far suffice to establish that over $\vtc$, $\imul$ is equivalent to the totality of $\imulm$, and
to the cyclicity of $\zg m$ for prime~$m$, which reduces to the statement that for any prime~$p$, all elements of
order~$p$ modulo~$m$ are powers of each other. Paying attention to how large products are needed to prove the last
statement for a given $m$ and~$p$, and vice versa, we show how to make progress on each turn around this circle of
implications, using a partial formalization of the structure theorem for finite abelian groups. This allows us to set
up a proof by induction to finish the derivation of $\imul$ in $\vtc$.
\end{itemize}

As a consequence of our main theorem, the above-mentioned results of~\cite{ej:vtc0iopen} on $\vtcim$ apply to
$\vtc$: that is, $\vtc$ proves the binary-number induction and minimization for $\rsuv$ translations of
$\sig0$~formulas. In terms of Johannsen and Pollett's theories, iterated multiplication and $\sig0$-minimization (in
Buss's language) are provable in $\delt1\text-\M{CR}$ and in~$C^0_2$, and the theory $C^0_2[\M{div}]$ is an extension of
$C^0_2$ by a definition (and therefore a conservative extension).

The paper is organized as follows. Section~\ref{sec:preliminaries} consists of preliminaries on $\vtc$ and related
theories. In Section~\ref{sec:primes}, we prove a suitable lower bound on the number of primes (to be used
for CRR) in~$\vtc$. Section~\ref{sec:divis-small-prim} formalizes a proof of division by small primes in $\vtc(\powm)$.
The core Section~\ref{sec:chin-rema-repr} formalizes various properties of CRR in $\vtc(\imulm)$, leading to a proof of
soundness of the CRR reconstruction procedure, and of $\imul$. In Section~\ref{sec:polylogarithmic-cut}, we discuss
polylogarithmic cuts and the ensuing results about~$V^0$. In Section~\ref{sec:modular-powering}, we construct modular
exponentiation in $V^0+\wphp$. We finish the proof of $\imul$ in $\vtc$ in Section~\ref{sec:gener-mult-groups}. In
Section~\ref{sec:tying-loose-ends}, we improve some of our auxiliary results to a more useful stand-alone form.
Section~\ref{sec:conclusion} concludes the paper.

\section{Preliminaries}\label{sec:preliminaries}

We will work with two-sorted (second-order) theories of bounded arithmetic in the style of Zambella~\cite{zamb:notes}.
Our main reference for these theories is Cook and Nguyen~\cite{cook-ngu}.

The language $L_2=\p{0,S,+,\cdot,\le,\in,\lh\cdot}$ of two-sorted bounded arithmetic is a first-order language with
equality with two sorts of variables, one for natural numbers (called \emph{small} or \emph{unary} numbers), and one
for finite sets of small numbers, which can also be interpreted as \emph{large} or \emph{binary} numbers so that $X$
represents $\sum_{u\in X}2^u$. Usually, variables of the number sort are written with lowercase letters $x,y,z,\dots$,
and variables of the set sort with uppercase letters $X,Y,Z,\dots$. The symbols $0,S,+,\cdot,\le$ of~$L_2$ provide the
standard language of arithmetic on the unary sort; $x\in X$ is the elementhood predicate, also written as $X(x)$, and
the intended meaning of the $\lh X$ function is the least strict upper bound on elements of~$X$. We write $x<y$ as an
abbreviation for $x\le y\land x\ne y$, and $\bit(X,i)$ for the indicator function of $i\in X$. 

Bounded quantifiers are introduced by
\begin{align*}
\exists x\le t\,\fii&\EQ\exists x\,(x\le t\land\fii),\\
\exists X\le t\,\fii&\EQ\exists X\,\bigl(\lh X\le t\land\fii\bigr),
\end{align*}
where $t$ is a term of unary sort not containing $x$ or~$X$ (resp.), and similarly for universal bounded quantifiers.
For any $i\ge0$, the class~$\Sig i$ consists of formulas that can be written as $i$~alternating (possibly empty) blocks
of bounded quantifiers, the first being existential, followed by a formula with only bounded first-order quantifiers.
Purely number-sort $\Sig0$ formulas without set-sort parameters (i.e., bounded formulas in the usual single-sorted
language of arithmetic) are called $\delz$. A formula is $\Sigma^1_1$ if it consists of a block
of (unbounded) existential quantifiers followed by a $\Sig0$ formula.

The theory~$V^0$ can be axiomatized by the basic axioms
\begin{align*}
&x+0=x&&x+Sy=S(x+y)\\
&x\cdot0=0&&x\cdot Sy=x\cdot y+x\\
&Sy\le x\to y<x&&\lh X\ne0\to\exists x\,\bigl(x\in X\land\lh X=Sx\bigr)\\
&x\in X\to x<\lh X&&\forall x\,(x\in X\eq x\in Y)\to X=Y
\end{align*}
and the bounded comprehension schema
\[\tag{$\fii$-\comp} \exists X\le x\,\forall u<x\,\bigl(u\in X\eq\fii(u)\bigr)\]
for $\Sig0$ formulas~$\fii$, possibly with parameters not shown (but with no occurrence of~$X$). We denote the set~$X$
whose existence is postulated by $\fii$-\comp\ as $\{u<x:\fii(u)\}$. Using $\comp$, $V^0$ proves the $\Sig0$-induction
schema $\Sig0\text-\ind$ and the $\Sig0$-minimization schema $\Sig0\text-\Min$; in particular, $V^0$ includes the
theory $\idz$ (the single-sorted theory of arithmetic axiomatized by induction for $\delz$~formulas over a base theory
such as Robinson's arithmetic) on the small number sort. In fact, $V^0$ is a conservative extension of~$\idz$
\cite[Thm.~V.1.9]{cook-ngu}.

Following~\cite{cook-ngu}, a set~$X$ can code a sequence (indexed by small numbers) of sets whose $u$th element is
$X^{[u]}=\bigl\{x:\p{u,x}\in X\bigr\}$, where $\p{x,y}=(x+y)(x+y+1)/2+y$. Likewise, we can code sequences of small
numbers using $X^{(u)}=\lh{X^{[u]}}$. (See below for a more efficient sequence encoding scheme.) While we stick to the
official notation in formal contexts such as when stating axioms, elsewhere we will generally write
$\vec X=\p{X_i:i<n}$ to indicate that $\vec X$ codes a sequence of length~$n$ whose $i$th element is~$X_i$. We denote
the length of the sequence as $\Lh(\vec X)=n$. (The official sequence coding system does not directly indicate the
length, hence we need to supply it using a separate first-order variable.)

There is a $\delz$-definition of the graph of $2^n$ such that $\idz$ proves that it is a partial function whose domain
is an initial segment closed under~$+$, and that it satisfies the defining recurrence $2^0=1$, $2^{n+1}=2\cdot2^n$ (see
e.g.\ H\'ajek and Pudl\'ak~\cite[\S V.3(c)]{hp}).
Thus, there is also a well-behaved $\delz$-definition of the function $\bit(x,i)=\fl{x2^{-i}}\rem2$, and
$\lh x=\min\{n:x<2^n\}$. In particular, in~$V^0$ there is a $\Sig0$-definable bijection identifying any small
number~$x$ with the corresponding large number, represented by the set $\{i<\lh x:\bit(x,i)=1\}$. Numbers of the form
$n=\lh x$, or equivalently, such that $2^n$ exists as a small number, will be called \emph{logarithmically small}. The
axiom $\Omega_1$ is defined as $\forall x\,\exists y\,\bigl(y=2^{{\lh x}^2}\bigr)$, or equivalently,
$\forall x\,\exists y\,\bigl(y=x^{\lh x}\bigr)$.

$\vtc$ is extends $V^0$ by the axiom
\[\forall n,X\:\exists Y\:\bigl(Y^{(0)}=0\land\forall i<n\,\bigl((i\notin X\to Y^{(i+1)}=Y^{(i)})
     \land(i\in X\to Y^{(i+1)}=Y^{(i)}+1)\bigr)\bigr),\]
asserting that for every set~$X$, there is a sequence~$Y$ supplying the counting function
$Y^{(i)}=\card(X\cap\{0,\dots,i-1\})$. Thus, in $\vtc$, there is a well-behaved $\Sig1$~definition of cardinality of
sets $\card(X)$ that provably satisfies
\begin{align}
\label{eq:67}\card(\nul)&=0,\\
\label{eq:68}\card(X\cup\{u\})&=\card(X)+1,\quad u\notin X.
\end{align}

$V^0$ can $\Sig0$-define $X+Y$ and $X<Y$, and prove that they make large numbers into a non-negative part of a discrete
totally ordered abelian group. Moreover, $\vtc$ can $\Sig1$-define iterated addition $\sum_{i<n}X^{[i]}$ satisfying the
recurrence
\begin{align}
\label{eq:69}\sum_{i<0}X^{[i]}&=0,\\
\label{eq:70}\sum_{i<n+1}X^{[i]}&=X^{[n]}+\sum_{i<n}X^{[i]},
\end{align}
and as a special case, it can $\Sig1$-define multiplication $X\cdot Y$, satisfying the axioms of non-negative parts of
discretely ordered rings. The embedding of small numbers to large numbers respects the arithmetic operations.

While we normally use set variables $X$, \ldots\ to represent nonnegative integers, we can also make them represent
arbitrary integers by reserving one bit for sign. We can extend $<$, $+$, $\cdot$, and $\sum_{i<n}X_i$ to signed
integers with no difficulty. We can also use fractions to represent rational numbers, but we have to be careful with
their manipulation: in particular, converting a bunch of fractions to a common denominator (such as when summing them)
requires the product of the denominators, and taking integer parts requires division with remainder (see below); one
case easy to handle is when all denominators are powers of~$2$. (Note that $2^n=\{n\}$ is easily definable in~$V^0$.)
Also, reducing fractions to lowest terms is impossible in general, as integer $\gcd$ is not known to be computable in
the $\nc$~hierarchy. (However, $\gcd$ of small integers can be done already in~$\idz$.)

When $Y=Q\cdot X+R$, where $0\le R<X$ (including the case of negative $Y$ and~$Q$), we will write%
\footnote{Conventionally, our $Y\rem X$ is written as just $Y\bmod X$. Since we will frequently mix this
notation with the $Y\equiv Y'\pmod X$ congruence notation, we want to distinguish the two more clearly than by relying
on the typographical difference between $Z=Y\bmod X$ and $Z\equiv Y\pmod X$, considering also that many authors write
the latter as $Z\equiv Y\mod X$, or even $Z=Y\mod X$.}
$Q=\fl{Y/X}$ and $R=Y\rem X$. We will also use the divisibility predicate $X\mid Y$, meaning $Y\rem X=0$, and the
congruence predicate $Y\equiv Y'\pmod X$, meaning $X\mid(Y-Y')$. (If the modulus~$X$ is the same throughout an
argument, we may write just $Y\equiv Y'$.) Since the provability of the totality of division in~$\vtc$ is equivalent to
the main result of this paper, we will need to make sure that the relevant quotients and remainders exist whenever we
employ these notations; in particular, $\idz$ proves that we can divide small numbers, $V^0$ can divide large
numbers by powers of~$2$, and we will prove in Section~\ref{sec:divis-small-prim} that $\vtc(\powm)$ can divide large
numbers by small primes.

Both notations $Y\rem X$ and $Y\equiv Y'\pmod X$ will establish contexts where everything inside $Y$ and~$Y'$ is
evaluated modulo~$X$ (except for nested $\bmod/\rem$ expressions modulo a different~$X'$); in particular, since $\idz$
proves that $x$ has an inverse modulo~$m$ when $\gcd(x,m)=1$, we may use $x^{-1}$ inside contexts evaluated modulo~$m$.
We denote by~$\zg m$ the group of units modulo~$m$: that is, with domain $\{x<m:\gcd(x,m)=1\}$ (which is just the
interval $[1,m-1]$ for $m$ prime) and the operation of multiplication modulo~$m$.

Following~\cite{ej:vtc0iopen}, we define the iterated multiplication axiom
\[\tag{$\imul$}\forall n,X\:\exists Y\:\forall u\le v<n\:\bigl(Y^{[u,u]}=1\land Y^{[u,v+1]}=Y^{[u,v]}\cdot X^{[v]}\bigr),\]
the meaning being that for any sequence $\p{X_i:i<n}$, we can find a triangular matrix $\p{Y_{u,v}:{u\le v\le n}}$ with
entries $Y_{u,v}=\prod_{i=u}^{v-1}X_i$.

Let us briefly recall the definitions of $\Ac$ and $\tc$ for context, even though we will not actually need to work
with complexity classes in this paper. A language~$L$ belongs to~$\Ac$ if it is computable by a
$\LT$-uniform family of constant-depth polynomial-size circuits using $\neg$ and unbounded-fan-in $\ET$ and
$\LOR$ gates. Equivalently, $L\in\Ac$ iff it is computable on an alternating Turing machine (with random-access input)
in time $O(\log n)$ using $O(1)$ alternations, iff $L$ (represented as a class of finite structures) is
$\cxt{FO}[+,\cdot]$-definable. A function $F(X)$ is in $\cxt{FAC}^0$ (and is called an $\Ac$~function) if
$\lh{F(X)}\le p\bigl(\lh X\bigr)$ for some polynomial~$p$, and the \emph{bit-graph} $\{\p{X,i}:\bit(F(X),i)=1\}$ is an
$\Ac$~language. A language $L$ is in $\tc$ iff it is computable by a $\LT$-uniform family of constant-depth
polynomial-size circuits using $\neg$ and unbounded-fan-in $\ET$, $\LOR$, and Majority gates (or more generally,
threshold gates), iff $L$ is computable in $O(\log n)$ time on a threshold Turing machine (see~\cite{par-sch}) using
$O(1)$ thresholds, iff it is definable in $\cxt{FOM}$ (first-order logic with majority quantifiers).

A predicate is $\Sig0$-definable in the standard model of~$L_2$ iff it is in~$\Ac$. The provably total
$\Sigma^1_1$-definable (= ``computable'') functions of~$V^0$ are exactly the $\Ac$~functions, and the provably total
$\Sigma^1_1$-definable functions of~$\vtc$ (or of $\vtcim$) are exactly the $\tc$~functions. Here, objects of the
set sort are represented as bit-strings in the usual way, and objects from the number sort are represented in unary;
see \cite[\S IV,\S A]{cook-ngu} for details.

We will need to work with various theories postulating totality of certain functions. Cook and Nguyen~\cite[\S
IX.2]{cook-ngu} developed a general framework for such theories under the slogan of \emph{theories~$VC$ associated with
complexity classes~$C$}. We refrain from this terminology as most of our theories will correspond to the \emph{same}
complexity class ($\tc$, sometimes $\Ac$), but we will adopt the machinery as such, using the notation
of~\cite{ej:vtc0iopen}.

For notational simplicity, we will formulate the setup for a single function of one variable $F(X)$ whose input and
output are binary numbers, but it applies just the same when we have several functions in several variables whose
inputs and outputs are a mix of binary and unary numbers. Thus, let $F(X)$ be a function with a $\Sig0$-definable graph
$\delta_F(X;Y)$ which is polynomially bounded, i.e., $\lh{F(X)}\le t(X)$ for some term~$t$. We assume that $V^0$
proves
\begin{gather}
\label{eq:65}\delta_F(X;Y)\land\delta_F(X;Y')\to Y=Y',\\
\label{eq:66}\delta_F(X;Y)\to\lh Y\le t(X).
\end{gather}
The totality of~$F$ is expressed by the sentence
\[\tag{$\tot_F$}\forall X\:\exists Y\:\delta_F(X;Y).\]
The \emph{aggregate function of~$F$} is the function~$F^*$ that maps (the code of) a sequence $\p{X_i:i<n}$ to
$\p{F(X_i):i<n}$. The graph of $F^*$ is defined by
\[\delta_F^*(n,X;Y)\EQ\forall i<n\:\delta_F\bigl(X^{[i]};Y^{[i]}\bigr),\]
and its totality is expressed by
\[\tag{$\tot_F^*$}\forall n\:\forall X\:\exists Y\:\delta_F^*(n,X;Y).\]
(Strictly speaking, $\delta_F^*$ does not define the graph of a function, as sequence codes are not completely unique.
This is why we write $\delta_F^*$ and~$\tot_F^*$ rather than $\delta_{F^*}$ and~$\tot_{F^*}$.) The \emph{Cook--Nguyen
(CN) theory} associated with~$\delta_F$ is $V^0(F)=V^0+\tot_F^*$.

The \emph{choice schema} (also called replacement or bounded collection) $\Sig0\text-\ac$ consists of the axioms
\[\forall P\:\bigl[\forall x<n\:\exists Y\le m\:\fii(x,Y,P)
   \to\exists W\:\forall x<n\,\fii\bigl(x,W^{[x]},P\bigr)\bigr]\]
for $\fii\in\Sig0$; a theory $T$ is closed under the \emph{choice rule} $\Sig0\text-\acr$ if
\[T\vdash\forall X\:\exists Y\:\fii(X,Y)
   \implies T\vdash\forall n\:\forall X\:\exists Y\:\forall i<n\:\fii\bigl(X^{[i]},Y^{[i]}\bigr)\]
for all $\fii\in\Sig0$.

The main properties of CN theories were summarized in \cite[Thm.~3.2]{ej:vtc0iopen} (mostly based on \cite[\S
IX.2]{cook-ngu}), which we repeat here:
\begin{Thm}\label{thm:cn-th}
Let $V^0(F)$ be a CN theory.
\begin{enumerate}
\item\label{item:8}
The provably total $\Sigma^1_1$-definable (or $\Sig1$-definable) functions of~$V^0(F)$ are exactly the functions in
the $\cxt{AC^0}$-closure (see \cite[\S IX.1]{cook-ngu}) of~$F$.
\item\label{item:9}
$V^0(F)$ has a universal extension~$\ob{V^0(F)}$ by definitions (and therefore conservative) in a
language~$L_{\ob{V^0(F)}}$ consisting of $\Sig1$-definable functions of~$V^0(F)$. The theory $\ob{V^0(F)}$ has quantifier
elimination for $\Sig0(F)$-formulas, and it proves $\Sig0(F)$-\comp, $\Sig0(F)$-\ind, and $\Sig0(F)$-\Min, where
$\Sig0(F)$ denotes the class of bounded formulas without second-order quantifiers in $L_{\ob{V^0(F)}}$.
\item\label{item:10}
$V^0(F)$ is closed under $\Sig0\text-\acr$, and $V^0(F)+\Sig0\text-\ac$ is $\forall\Sigma^1_1$-conservative
over~$V^0(F)$.\noproof
\end{enumerate}
\end{Thm}

A consequence of~\ref{item:10} is that whenever a CN theory proves $\tot_G$ for some $\Sig0$-defined function~$G$, it
also proves~$\tot_G^*$.

As a special case of Theorem~\ref{thm:cn-th} for a trivial function~$F$, $V^0$ has a universal extension $\ob{V^0}$ by
definitions in a language $L_{\ob{V^0}}$ (called $\mathcal L_{\boldsymbol{\mathit{FAC}}^0}$ in~\cite{cook-ngu})
consisting of $\Sig1$-definable functions of~$V^0$. Unlike general CN theories, it has the property that
$\Sig0(L_{\ob{V^0}})=\Sig0$ (more precisely, every $\Sig0(L_{\ob{V^0}})$ formula is equivalent to a~$\Sig0$ formula
over $\ob{V^0}$) by \cite[L.~V.6.7]{cook-ngu}. In particular, we will use the consequence that if $V^0\vdash\tot_F$,
then $\Sig0(F)=\Sig0$.

Note that any finite $\forall\Sig0$-axiomatized extension of~$V^0$ is trivially a CN theory: an axiom of the form
$\forall X\,\fii(X)$ with $\fii\in\Sig0$ is equivalent over~$V^0$ to $\tot_{F_\fii}$ and to~$\tot_{F_\fii}^*$ where
$\delta_{F_\fii}(X;Y)$ is $\fii(X)\land Y=0$. We still have that if $T=V^0+\forall X\,\fii(X)\vdash\tot_F$, then
$\Sig0(F)=\Sig0$ over~$T$ (by quantifier elimination for $\ob{V^0}$, $\forall X\,\fii(X)$ is equivalent to a
universal formula in~$\ob{V^0}$, thus using Herbrand's theorem, $F$ is defined by an $L_{\ob{V^0}}$ function symbol in
$\ob{V^0}+T$).

It is easy to show that $\vtc\vdash\tot_{\card}^*$ (see \cite[L.~IX.3.3]{cook-ngu}), hence $\vtc=V^0(\card)$ is a CN
theory. The $\Sig0(\card)$-definable predicates in the standard model are exactly the $\tc$~predicates.

As we already mentioned above, the whole setup may be formulated for several functions $F_0,\dots,F_k$ in place
of~$F$, thus we may define $V^0(F_0,\dots,F_k)$; formally, we may easily combine $F_0,\dots,F_k$ to a single function,
hence $V^0(F_0,\dots,F_k)$ is a CN theory. In particular, we will consider various theories of the form
$\vtc(F)=V^0(\card,F)$. More generally, we could iterate the construction to define CN theories over a fixed CN
theory (such as~$\vtc$) as a base theory in place of~$V^0$; that is, we can introduce $\vtc(F)$ when $F$ is given by a
$\Sig0(\card)$ formula~$\delta_F$ such that \eqref{eq:65} and~\eqref{eq:66} are provable in~$\vtc$. One can show that
the resulting theories are CN theories according to the original definition. In particular,
as explained in~\cite{ej:vtc0iopen}, $\vtcim$ is a CN theory.

Apart from $V^0$, $\vtc$, and $\vtcim$, we will consider the following CN theories (often in conjunction
with~$\vtc$).
\begin{itemize}
\item $\vtc(\Divf)$: given $Y$ and $X>0$, there are $\fl{Y/X}$ and $Y\rem X$; i.e, $\delta_\Divf(X,Y;Q,R)$ is
\[X=Q=R=0\lor(R<X\land Y=QX+R).\]
The $\tot_\Divf$ axiom is also denoted~$\Div$. As shown in~\cite{ej:vtc0iopen} (using results of
Johannsen\cite{joh:c02div}), $\vtc(\Divf)=\vtcim$.
\item $V^0(\powm)$: given $a$, $r$, and prime~$m$, we can compute $a^r\rem m$, or rather, the witnessing sequence
$Y=\p{a^i\rem r:i\le r}$. Formally, $\delta_\powm(a,r,m;Y)$ is
\[\bigl(\neg\prm(m)\land Y=0\bigr)\lor
  \bigl(\prm(m)\land Y^{(0)}=1\rem m\land\forall i<r\:Y^{(i+1)}=aY^{(i)}\rem m\bigr),\]
where $\prm(m)$ stands for $m>1\land\forall x,y\,(xy=m\to x=1\lor y=1)$, and here and below, we ignore issues with
non-uniqueness of sequence codes.
\item $V^0(\imulm)$: given a sequence $\p{a_i:i<n}$ and a prime~$m$, we can find (a witnessing sequence for)
$\prod_{i<n}a_i\rem m$. Formally, $\delta_\imulm(A,n,m;Y)$ is
\[\bigl(\neg\prm(m)\land Y=0\bigr)\lor
  \bigl(\prm(m)\land Y^{(0)}=1\rem m\land\forall i<n\:Y^{(i+1)}=Y^{(i)}A^{(i)}\rem m\bigr).\]
\item $V^0+\wphp$: $\wphp$ is the $\forall\Sig0$ axiom $\forall n\,\forall X\,\php^{2n}_n(X)$, where $\php^m_n(X)$ is
\[\forall x<m\:\exists y<n\:X(x,y)\to\exists x<x'<m\:\exists y<n\:\bigl(X(x,y)\land X(x',y)\bigr).\]
By results of Paris, Wilkie, and Woods~\cite{pww}, $V^0+\wphp\sset V_0+\Omega_1$. (This was locally improved by
Atserias~\cite{ats:wphp-conf,ats:wphp}, who showed $V^0\vdash\forall n\,\forall X\,\bigl(\exists r\,r=n^{(\log
n)^{1/k}}\to\php^{2n}_n(X)\bigr)$ for any constant~$k$.) We mention that $\vtc$ even proves
$\forall n\,\forall X\,\php^{n+1}_n(X)$ by~\cite[Thm.~IX.3.23]{cook-ngu}.
\item $\vl=V^0(\MF{Iter})$ (see \cite[\S IX.6.3]{cook-ngu}): given a function $F$ from $[0,a]$ to itself, we can compute
its iterates $F^i(0)$. Formally, $\delta_{\MF{Iter}}(a,F,n;Y)$ is
\[\bigl(\neg\M{Func}(F,a)\land Y=0\bigr)\lor
  \bigl(\M{Func}(F,a)\land Y^{(0)}=0\land\forall i<n\:F\bigl(Y^{(i)},Y^{(i+1)}\bigr)\bigr),\]
where $\M{Func}(F,a)$ is $\forall x\le a\,\exists!y\le a\,F(x,y)$.
\item $\vnl=V^0(\MF{Reach})$ (see \cite[\S IX.6.1]{cook-ngu}): given a relation $E\sset[0,a]\times[0,a]$ and~$d$, we can
define $E$-reachability (from~$0$) in $\le n$ steps. Formally, $\delta_{\MF{Reach}}(a,E,n;Y)$ is
\begin{multline*}
Y\sset[0,d]\times[0,a]\land\forall x\le a\:\bigl[\bigl(Y(0,x)\eq x=0\bigr)\\
\land\forall d<n\:\bigl(Y(d+1,x)\eq\exists y\le a\:\bigl(Y(d,y)\land(x=y\lor E(y,x))\bigr)\bigr)\bigr]
\end{multline*}
We will use the fact that $\vnl=V^0+\tot_{\MF{Reach}}$ (see \cite[L.~IX.6.7]{cook-ngu}).
\end{itemize}

For some of our axioms, we will also need formulas expressing that they hold restricted to some bound:
\begin{itemize}
\item $\imul[w]$ states the totality of the aggregate function of iterated multiplication $\prod_{i<n}X_i$ restricted
so that $\sum_{i<n}\lh{X_i}\le w$. Using the formulation of~$\imul$ as above, this can be expressed as
\[\forall n,X\:\exists Y\:\Bigl(\forall u\le n\:Y^{[u,u]}=1\land
   \forall u\le v<n\:\Bigl(\sum_{i=u}^v\lh{X_i}\le w\to Y^{[u,v+1]}=Y^{[u,v]}\cdot X^{[v]}\Bigr)\Bigr).\]
\item $\tot^*_\Divf[w]$ states the totality of the aggregate function of division restricted to arguments of length~$w$:
\[\forall n,X,Y\:\exists Q,R\:\forall i<n\:\bigl(0<\lh{X^{[i]}}\le w\land\lh{Y^{[i]}}\le w
  \to Y^{[i]}=Q^{[i]}X^{[i]}+R^{[i]}\land R^{[i]}<X^{[i]}\bigr).\]
\item $\tot^*_\imulm[w,-]$ states the totality of $\imulm^*$ restricted to $\prod_{i<n}a_i\rem m$ where $n\le w$:
\[\forall t,N,A,M\:\exists Y\:\forall u<t\:\bigl(N^{(u)}\le w
  \to\delta_\imulm\bigl(A^{[u]},N^{(u)},M^{(u)},Y^{[u]}\bigr)\bigr).\]
\item $\tot^*_\imulm[-,w]$ states the totality of $\imulm^*$ restricted to $\prod_{i<n}a_i\rem m$ where $m\le w$:
\[\forall t,N,A,M\:\exists Y\:\forall u<t\:\bigl(M^{(u)}\le w
  \to\delta_\imulm\bigl(A^{[u]},N^{(u)},M^{(u)},Y^{[u]}\bigr)\bigr).\]
\end{itemize}

Berarducci and D'Aquino~\cite{ber-daqui} proved that for any $\delz$-definable function $f(i)$, there exist a $\delz$
definition of the graph of the iterated product $\prod_{i<x}f(i)=y$ such that $\idz$ proves the recurrence
$\prod_{i<0}f(i)=1$ and (if either side exists) $\prod_{i<x+1}f(i)=f(x)\prod_{i<x}f(i)$. The argument relativizes,
hence it applies in~$V^0$ to functions defined by second-order objects: that is, we can construct a well-behaved
product $\prod_{i<n}x_i$ of a sequence $X=\p{x_i:i<n}$ as long as $\sum_{i<n}\lh{x_i}\le\lh w$ for some~$w$ (which
guarantees that the resulting product, if any, is a small number, and then by induction on~$n$, that it exists). In our
notation, this becomes:
\begin{Thm}[Berarducci, D'Aquino~\cite{ber-daqui}]\label{thm:imul-log}
$V^0$ proves $\forall w\,\imul\bigl[\lh w\bigr]$.
\noproof\end{Thm}
We will improve this result in Corollary~\ref{cor:vl-imul-plog}.

Paris and Wilkie~\cite{par-wil:coun-delz} showed how to count polylogarithmic-size sets in~$\idz$, and Paris, Wilkie
and Woods~\cite{pww} extended this to polylogarithmic sums. We can reformulate their results in the two-sorted setup as
follows.
\begin{Thm}\label{thm:count-polylog}
For any constant~$c$, $V^0$ proves:
\begin{enumerate}
\item\label{item:11}
For every $X$ and~$w$, either there exists a (unique) $s<\lh w^c$ and a bijection $F\colon X\to[0,s)$, or there exists
an injection $F\colon\bigl[0,\lh w^c\bigr)\to X$.
\item\label{item:12} For every $X$ and~$w$, there exists a sequence $\left<\sum_{i<n}X^{[i]}:n\le\lh w^c\right>$ that
satisfies \eqref{eq:69} and \eqref{eq:70} for $n<\lh w^c$.
\end{enumerate}
\end{Thm}
\begin{Pf}
In \cite[Thm.~5$'$]{par-wil:coun-delz}, \ref{item:11} is proved for $\delz$-definable sets in models
of~$\idz$; the argument is uniform in~$X$, hence it also applies to arbitrary sets~$X$ in models of~$V^0$. Likewise,
\ref{item:12} is proved for $\delz$-definable sequences of small numbers in \cite[Thm.~10]{pww}, and the argument
applies to arbitrary sequences of small numbers in~$V^0$.

In order to generalize it to sums of sequences of large numbers, we split each $X^{[i]}$ into $\lh w$-bit blocks:
$X^{[i]}=\sum_{j<2m}x_{i,j}2^{j\lh w}$, where $x_{i,j}<2^{\lh w}\le2w$, and $m\le\max_i\lh{X^{[i]}}/\lh w$. Notice that
for each $j<2m$, we have $\sum_{i<\lh w^c}x_{i,j}<\lh w^c2^{\lh w}\le2^{2\lh w}$ if $w$ is sufficiently large, hence we
may construct $Y_{\mathrm{even\vphantom{odd}}}$ and $Y_{\mathrm{odd}}$ such that
\begin{align*}
Y_{\mathrm{even\vphantom{odd}}}^{[n]}&=\sum_{j<m}2^{2j\lh w}\sum_{i<n}x_{i,2j},\\
Y_{\mathrm{odd}}^{[n]}&=\sum_{j<m}2^{(2j+1)\lh w}\sum_{i<n}x_{i,2j+1}
\end{align*}
for each $n\le \lh w^c$ by just concatenating suitably shifted copies of the small-number sums $\sum_{i<n}x_{i,j}$. If
we then define $Y$ such that $Y^{[n]}=Y_{\mathrm{even\vphantom{odd}}}^{[n]}+Y_{\mathrm{odd}}^{[n]}$, it satisfies the
required recurrence $Y^{[0]}=0$, $Y^{[n+1]}=Y^{[n]}+X^{[n]}$ for $n<\lh w^c$.
\end{Pf}

Some of our arguments will require rather tight bounds on the sizes of the objects involved, and in particular, on
sequence codes. Clearly, we need at least $\approx\sum_{i<n}\lh{X_i}$ bits to encode a sequence $\p{X_i:i<n}$, but the
encoding scheme from~\cite{cook-ngu} as defined above does not meet this lower bound: it uses
$\approx\bigl(n+\max_i\lh{X_i}\bigr)^2$ bits, which may be quadratically larger than the ideal size in unfavourable
conditions. We will now introduce a more efficient encoding scheme in~$\vtc$; it is based on the idea of
Nelson~\cite[\S10]{nels:pred}, but we repurpose it to directly encode sequences rather than just sets.

The encoding works as follows: the code of $\p{X_i:i<n}$ is a set
$X$ representing a pair of sets $R,B$ by $X=\{2x:x\in R\}\cup\{2x+1:x\in B\}$, where $B$ consists of the concatenation
of bits of all the $X_i$'s (in order), and $R$ is a ``ruler'' indicating where each $X_i$ starts in~$B$; that is,
$R=\{r_i:i<n\}$ with $0=r_0<r_1<\dots<r_{n-1}$, and $X_i$ is given by the bits $r_i,\dots,r_{i+1}-1$ of~$B$ (taking
$\max\bigl\{\lh B,r_{n-1}+1\bigr\}$ for $r_n$).

Formally, the sequence coded by~$X$ has length $\Lh(X)=\card\{x:X(2x)\}$, and for $i<\Lh(X)$, the $i$th element
of $X$, denoted~$X_i$, is
\[\bigl\{x:
  \exists r\:\bigl(X(2r)\land\card\{u<r:X(2u)\}=i\land X(2(x+r)+1)\land\forall i<x\:\neg X(2(r+i+1))\bigr)\bigr\}.\]
Here, all the quantifiers and comprehension variables can be bounded by~$\lh X$, hence $\Lh(X)$ and $X_i$ are
$\Sig0(\card)$-definable in~$\vtc$, and $\vtc$ proves that we can convert $X$ to a set
$Y=\bigl\{\p{i,x}:i<\Lh(X),x\in X_i\bigr\}$ that represents the same sequence using the sequence encoding
from~\cite{cook-ngu} (that is, $Y^{[i]}=X_i$ for all $i<\Lh(X)$). Conversely, if $Y$ represents a sequence of
length~$n$ using the encoding from~\cite{cook-ngu}, we can $\Sig0(\card)$-define
$r_i=\sum_{j<i}\max\bigl\{1,\lh{Y^{[i]}}\bigr\}$ and $X=\{2r_i:i<n\}\cup\bigl\{2(r_i+x)+1:i<n,Y(i,x)\bigr\}$
in~$\vtc$. Then $X$ represents under our new scheme the same sequence as~$Y$ and~$n$ (i.e., $\Lh(X)=n$ and
$X_i=Y^{[i]}$ for all $i<n$), and moreover,
\begin{equation}\label{eq:74}
\lh X\le2\sum_{i<n}\max\bigl\{1,\lh{X_i}\bigr\},
\end{equation}
thus the new encoding scheme realizes the optimal size bound up to a multiplicative constant. We can also encode
sequences of small numbers $\p{x_i:i<n}$ by sequences of the corresponding sets, i.e., $\p{X_i:i<n}$ where
$X_i=\{j<\lh{x_i}:\bit(x_i,j)=1\}$.

For general sequences, the efficient coding scheme requires%
\footnote{We could make $\Lh(X)$ and $X_i$ $\Sig0$-definable using a more elaborate definition of~$R$: e.g., indicate
the start of $X_i$ in $R$ not just by a single $1$-bit, but by $1+v_2(i)$ $1$-bits (followed by at least one $0$-bit).
We leave it to the reader's amusement to verify that this encoding is $\Sig0$-decodable, and that it can encode
$\p{X_i:i<n}$ using $O\bigl(n+\sum_i\lh{X_i}\bigr)$ bits. But crucially, proving the latter \emph{still}
requires~$\vtc$, or at least some form of approximate counting that allows close enough estimation of
$\sum_{j<i}\lh{X_j}$. Thus, we do not really accomplish much with this more complicated scheme.}%
~$\vtc$. However, for sequences of \emph{polylogarithmic} length (i.e., $\p{X_i:i<n}$ where $n\le\lh w^c$ for some~$w$
and a standard constant~$c$), it works already in~$V^0$: using Theorem~\ref{thm:count-polylog}, $\Lh(X)$ and $X_i$ are
well-defined in~$V^0$ (in fact, $\Sig0$-definable), and $V^0$ proves that a given sequence has a code
obeying~\eqref{eq:74}.

In the special case $c=1$, a sequence of small numbers $\p{x_i:i<n}$ such that $n$ and $\sum_i\lh{x_i}$ are bounded by
$\lh w$ has a code of length $O(\lh w)$, and as such, it can be represented by a \emph{small number}. Then the encoding
scheme does not involve any second-order objects at all, and it is $\delz$-definable in~$\idz$. When passing to~$\idz$,
the statement that a given sequence can be encoded to satisfy~\eqref{eq:74} becomes the theorem that for any
$\delz$-definable function $f(i)$ (possibly with parameters), if $n\le\lh w$ and $\sum_{i<n}\lh{f(i)}\le\lh w$, there
exists $x\le w^{O(1)}$ that encodes the sequence $\p{f(i):i<n}$.

\section{Prime supply}\label{sec:primes}

Since we will work extensively with the Chinese remainder representation, we will need lots of primes. To begin with,
if we want to represent a number~$X$ in CRR modulo a sequence of primes $\p{m_i:i<k}$, we must have $\prod_im_i>X$, thus
we need to get hold of sequences of primes such that $\sum_i\lh{m_i}$ exceeds any given small number.

Already in mid 19th century, Chebyshev proved using elementary methods that the number of primes below~$x$ is
$\Theta(x/\log x)$, or equivalently,
\begin{equation}\label{eq:72}
\sum_{p\le x}\log p=\Theta(x).
\end{equation}
(Here and below in this section, sums indexed by~$p$ are supposed to run over primes.) See e.g.\
Apostol~\cite[Thm.~4.6]{apos} for a nowadays-standard simple result of this type, based on considering the contribution
of various primes to the prime factorization of binomial coefficients (this form of the proof is due to Erd\H os and
Kalm\'ar). As we will see, it is fairly straightforward to formalize a version of Chebyshev's theorem in~$\vtc$. Similar
to~\cite{apos}, we will compute with sums of logarithms rather than with products of primes, factorials, and binomial
coefficients. For our purposes, the simple approximation of~$\log n$ by $\lh n$ is sufficient.

We mention that Woods~\cite{woods:phd} proved Sylvester's theorem in $\idz+\wphp(\delz)$ by formalizing similar
elementary arguments; our job is much easier as we can directly use bounded sums in~$\vtc$, which Woods avoided by
applying $\wphp$ to ingeniously constructed functions (he also needed much more elaborate approximations of
logarithms).

In fact, Nguyen~\cite{ngu:thesis} already proved a version of~\eqref{eq:72} with fairly good bounds in~$\vtc$,
also using elaborate approximate logarithms. We keep our argument below (which gives much worse bounds) as it is
simpler and more elementary than the proof in~\cite{ngu:thesis}, while making this paper more self-contained.

First, note that using
\begin{equation}\label{eq:73}
\lh x+\lh y-1\le\lh{xy}\le\lh x+\lh y,
\end{equation}
$\idz$ proves
\begin{equation}\label{eq:1}
y=\prod_{i<k}x_i\to\lh y\le\sum_{i<k}\lh{x_i}\land\lh y-1\ge\sum_{i<k}\bigl(\lh{x_i}-1\bigr).
\end{equation}
Considering a sequence of maximal length whose product is~$x$ (where we use the efficient sequence encoding), it is
easy to prove that every positive number is a product of a sequence of primes:
\[\idz\vdash\forall x>0\:\exists s=\p{p_i:i<k}\:\Bigl(x=\prod_{i<k}p_i\land\forall i<k\:\prm(p_i)\Bigr).\]
Moreover, the sequence code~$s$ is bounded by a polynomial in~$x$.

By double counting, $\vtc$ proves
\begin{align}
\label{eq:2}\sum_{\substack{n\le x\\n=\prod_jp_j}}\sum_j\lh{p_j}
&=\sum_{p\le x}\lh p\sum_{i\colon p^i\le x}\fdiv x{p^i},\\
\label{eq:3}\sum_{\substack{n\le x\\n=\prod_jp_j}}\sum_j\bigl(\lh{p_j}-1\bigr)
&=\sum_{p\le x}\bigl(\lh p-1\bigr)\sum_{i\colon p^i\le x}\fdiv x{p^i}.
\end{align}
Here and below in this section, sum indices such as $n$ and~$i$ are supposed to start at~$1$.

Our goal is to prove a lower bound on the number of primes (Theorem~\ref{thm:primes}), but we first need the following upper
bound, which is a formalization of a weak form of Mertens's theorem: $\sum_{p\le x}p^{-1}=O(\log\log x)$. The
reason is that when proving our lower bound, the crude approximation to $\log p$ provided by the $\lh p$ function will
introduce a copious amount of error into the calculations, and the lemma below is needed to bound the error.
\begin{Lem}\label{lem:sumx:p}
$\vtc$ proves
\[\sum_{p\le x}\sum_{i\colon p^i\le x}\fdiv x{p^i}\le16x\lh{\lh x}.\]
\end{Lem}
\begin{Pf}
Let $k=\lh x$. For any $l<\lh k$, we have
\begin{align*}
\cl{k2^{-(l+1)}}\sum_{p=2^{\cl{k2^{-(l+1)}}}}^{2^{\cl{k2^{-l}}}-1}\cdiv{2^k}p
&\le\sum_{p=2^{\cl{k2^{-(l+1)}}}}^{2^{\cl{k2^{-l}}}-1}\bigl(\lh p-1\bigr)\cdiv{2^k}p\\
&\le2^{k-\cl{k2^{-l}}}\sum_{p=2^{\cl{k2^{-(l+1)}}}}^{2^{\cl{k2^{-l}}}-1}\bigl(\lh p-1\bigr)\cdiv{2^{\cl{k2^{-l}}}}p\\
&\le2^{k+1-\cl{k2^{-l}}}\sum_{p=2^{\cl{k2^{-(l+1)}}}}^{2^{\cl{k2^{-l}}}-1}\bigl(\lh p-1\bigr)\fdiv{2^{\cl{k2^{-l}}}}p\\
&\le2^{k+1-\cl{k2^{-l}}}\sum_{\substack{n<2^{\cl{k2^{-l}}}\\n=\prod_{j<t}p_j}}\sum_{j<t}\bigl(\lh{p_j}-1\bigr)\\
&\le2^{k+1-\cl{k2^{-l}}}\sum_{n<2^{\cl{k2^{-l}}}}\bigl(\lh n-1\bigr)\\
&\le2^{k+1}\bigl(\cl{k2^{-l}}-1\bigr)
\end{align*}
using \eqref{eq:1} and~\eqref{eq:3}, thus
\[\sum_{p=2^{\cl{k2^{-(l+1)}}}}^{2^{\cl{k2^{-l}}}-1}\cdiv{2^k}p\le\frac{\cl{k2^{-l}}-1}{\cl{k2^{-(l+1)}}}2^{k+1}\le2^{k+2}.\]
Summing over all $l<\lh k$ gives
\[\sum_{p<2^k}\cdiv{2^k}p\le2^{k+2}\lh k,\]
thus
\[\sum_{p\le x}\cdiv xp\le2^{k+2}\lh k\le8x\lh{\lh x}\]
as $x<2^k\le2x$. Then estimating the geometric series
\[\sum_{i\colon p^i\le x}\fdiv x{p^i}\le2\cdiv xp\]
gives the result.
\end{Pf}
\begin{Thm}\label{thm:primes}
There is a standard constant~$c$ such that $\vtc$ proves
\[x\ge c\to\sum_{p\le x\lh x^{17}}\bigl(\lh p-1)\ge x.\]
\end{Thm}
\begin{Pf}
For any $0<x<y$, we have
\begin{align*}
x(\lh y-\lh x+1)&\le\sum_{y<n\le x+y}\lh n-\sum_{n\le x}\bigl(\lh n-1\bigr)\\
&\le\sum_{\substack{y<n\le x+y\\n=\prod_jp_j}}\sum_j\lh{p_j}
    -\sum_{\substack{n\le x\\n=\prod_jp_j}}\sum_j\bigl(\lh{p_j}-1\bigr)\\
&\le\sum_{p\le x+y}\lh p\sum_{i\colon p^i\le x+y}\left(\fdiv{x+y}{p^i}-\fdiv y{p^i}\right)
    -\sum_{p\le x}\bigl(\lh p-1\bigr)\sum_{i\colon p^i\le x}\fdiv x{p^i}\\
&\le\sum_{p\le x+y}\lh p\sum_{i\colon p^i\le x+y}\left(\fdiv{x+y}{p^i}-\fdiv x{p^i}-\fdiv y{p^i}\right)
    +\sum_{p\le x}\sum_{i\colon p^i\le x}\fdiv x{p^i}\\
&\le\sum_{p\le x+y}\lh p\sum_{i\colon p^i\le x+y}1+16x\lh{\lh x}\\
&\le\sum_{p\le x+y}\lh p+\sum_{p\le\sqrt{x+y}}\lh p\fdiv{\lh{x+y}-1}{\lh p-1}+16x\lh{\lh x}\\
&\le\sum_{p\le x+y}\lh p+2\lh{x+y}\fl{\sqrt{x+y}}+16x\lh{\lh x}
\end{align*}
using \eqref{eq:1}--\eqref{eq:3} and Lemma~\ref{lem:sumx:p}. Taking $y=x\lh x^{17}-x$, we have $\lh y\ge\lh{x+y}-1\ge\lh
x+17\lh{\lh x}-18$ and $\lh{x+y}\le\lh x+17\lh{\lh x}$ by~\eqref{eq:1}, thus
\[\sum_{p\le x\lh x^{17}}\lh p
  \ge\bigl(17\lh{\lh x}-17-16\lh{\lh x}\bigr)x-2\bigl(\lh x+17\lh{\lh x}\bigr)\fl{\sqrt{x\lh x^{17}}}
  \ge\bigl(\lh{\lh x}-18\bigr)x\]
and
\[\sum_{p\le x\lh x^{17}}\bigl(\lh p-1\bigr)\ge\frac12\sum_{p\le x\lh x^{17}}\lh p\ge\frac{\lh{\lh x}-18}2\,x\ge x\]
for large enough~$x$.
\end{Pf}

\section{Division by small primes}\label{sec:divis-small-prim}

We need a one more simple but important preparatory result: $\vtc(\powm)$, and a fortiori $\vtc(\imulm)$, can perform
division with remainder by \emph{small primes}. This is indispensable when working with the Chinese remainder
representation: it is required to define the CRR in the first place, but we will also extensively use it when studying
its properties.

Notice that $\powm$ directly provides $2^n\rem m$, and as it turns out, the bits of $\fl{2^n/m}$ can be explicitly
expressed in terms of $2^i\rem m$ as well. We then obtain $\fl{X/m}$ and $X\rem m$ for general~$X$ by summing over its
bits.
\begin{Lem}\label{lem:div-small-prm}
$\vtc(\powm)$ proves that we can divide by small primes:
\[\forall X\,\forall m\,\bigl(\prm(m)\to\exists Q\,\exists r<m\:X=mQ+r\bigr).\]
\end{Lem}
\begin{Pf}
We may assume $m$ is odd. Let us first consider $X=2^n$. Using $\powm$, define
\[Q_n=\sum_{i<n}2^i\bigl((2^{n-i}\rem m)\rem2\bigr).\]
We will prove
\begin{equation}\label{eq:4}
2^n=mQ_n+(2^n\rem m)
\end{equation}
by induction on~$n$. The statement holds for $n=0$. For the induction step, we have
\begin{align*}
Q_{n+1}&=\sum_{i<n+1}2^i\bigl((2^{n+1-i}\rem m)\rem2\bigr)\\
&=\bigl((2^{n+1}\rem m)\rem 2\bigr)+\sum_{i<n}2^{i+1}\bigl((2^{n-i}\rem m)\rem2\bigr)\\
&=2Q_n+\bigl((2^{n+1}\rem m)\rem 2\bigr),
\end{align*}
thus using the induction hypothesis,
\begin{align*}
mQ_{n+1}&=2mQ_n+\bigl((2^{n+1}\rem m)\rem2\bigr)m\\
&=2^{n+1}-2(2^n\rem m)+\bigl((2^{n+1}\rem m)\rem2\bigr)m.
\end{align*}
Now, either $2^n\rem m<m/2$, in which case
\[2^{n+1}\rem m=2(2^n\rem m)\qquad\text{and}\qquad(2^{n+1}\rem m)\rem2=0,\]
or $2^n\rem m>m/2$, in which case
\[2^{n+1}\rem m=2(2^n\rem m)-m\qquad\text{and}\qquad(2^{n+1}\rem m)\rem2=1.\]
Either way,
\[(2^{n+1}\rem m)+\bigl((2^{n+1}\rem m)\rem2\bigr)m=2(2^n\rem m),\]
hence $mQ_{n+1}=2^{n+1}-(2^{n+1}\rem m)$ as required.

Now, for general $X$, we have
\[X=\sum_{n\in X}2^n=m\sum_{n\in X}Q_n+x,\]
where
\[x=\sum_{n\in X}(2^n\rem m)\le\lh X\,m\]
is small, thus already $\idz$ can divide $x$ by~$m$, yielding
$X=m\bigl(Q_n+\fl{x/m}\bigr)+(x\rem m)$.
\end{Pf}

\section{Chinese remainder representation}\label{sec:chin-rema-repr}

We are coming to the core technical part of the paper. First, the basic definition:
\begin{Def}\label{def:crr}
(In $\vtc(\powm)$.)
If $\vec m=\p{m_i:i<k}$ is a sequence of distinct primes, the \emph{Chinese remainder representation (CRR)} of~$X$
modulo~$\vec m$ is the sequence $X\rem\vec m=\p{X\rem m_i:i<k}$, which is well-defined by Lemma~\ref{lem:div-small-prm}.
The sequence $\vec m$ is called the \emph{basis} of the CRR.
\end{Def}

Our goal in this section is to define in $\vtc(\imulm)$ a CRR reconstruction procedure, that is, a function that
recovers $X$ from $X\rem\vec m$ (under suitable conditions); this will in turn easily imply that $\vtc(\imulm)$ proves
$\imul$.

The principal problem we face when trying to formalize the CRR reconstruction procedure from~\cite{hab} is that the
argument involves various numbers constructed by iterated multiplication (and division), which we do not a priori know
to exist when working inside $\vtc(\imulm)$. Besides many references to the product $\prod_im_i$, the reconstruction
procedure for instance involves computing a CRR representation of a product of the form
$X\prod_{u<t}\frac12\bigl(1+\prod_ja_{u,j}\bigr)$ for a certain sequence of primes~$a_{u,j}$. We sidestep these
problems by developing in $\vtc(\imulm)$ low-level operations on CRR. We will systematically exploit the fact that even
though we cannot a priori convert a CRR representation to the number $X$ it represents, we \emph{can} compute certain
``shadows'' of~$X$: approximations to the ratio $X/\prod_im_i$, and $X\rem a$ for small primes~$a$; we will formally
define these quantities shortly in Definition~\ref{def:shadows}, but let us first introduce a few notational conventions in
order to save repetitive typing.
\begin{Def}\label{def:crr-notation}
(In $\vtc(\imulm)$.) In this section, $\vec m$ stands for a sequence of distinct primes, whose length is denoted~$k$:
$\vec m=\p{m_i:i<k}$. When we need another sequence of primes, we use $\vec a$ of length~$l$. We write
$\vec x<\vec m$ for $\vec x$ being a sequence of residues modulo~$\vec m$, i.e., $\vec x=\p{x_i:i<k}$ such that
$0\le x_i<m_i$ for each~$i<k$.

We put $[\vec m]=\prod_{i<k}m_i$ (evaluated using $\imulm$ modulo some prime specified
in the context), and likewise $[\vec m]_{\ne i}=\prod_{j\ne i}m_j$. If $\vec m$ and~$\vec a$ are sequences of primes,
$\vec m\perp\vec a$ denotes that each $m_i$ is coprime to (i.e., distinct from) each $a_j$. We interpret $\bmod/\rem$
notations modulo~$\vec m$ elementwise, so that, e.g., $X\rem\vec m$ means $\p{X\rem m_i:i<k}$ (as already indicated in
Definition~\ref{def:crr}), $\vec y=\vec x\rem\vec m$ means $y_i=x_i\rem m_i$ for each~$i<k$, and
$\vec x\equiv\vec y\pmod{\vec m}$ means $x_i\equiv y_i\pmod{m_i}$ for each $i<k$.

We will write $y=x\pm a$ for $x-a\le y\le x+a$; more generally, $y=x\PM ab$ abbreviates $x-b\le y\le x+a$.
\end{Def}

In the real world, if $\vec x$ is the CRR of $X$ modulo a basis~$\vec m$, we can reconstruct $X$ by
\[X\equiv\sum_{i<k}x_ih_i[\vec m]_{\ne i}\pmod{[\vec m]},\]
where $h_i=[\vec m]_{\ne i}^{-1}\rem m_i$ (recall that this denotes the multiplicative inverse of $[\vec m]_{\ne i}$
modulo~$m_i$). Thus, the right-hand side equals $X+r[\vec m]$ for some natural number~$r$, easily seen to satisfy
$r<\sum_im_i$. This integer is called the \emph{rank} of~$\vec x$, and it obeys
\begin{equation}\label{eq:5}
\sum_{i<k}\frac{x_ih_i}{m_i}=r+\frac X{[\vec m]}.
\end{equation}
This equation holds (with the same~$r$) in any field where the $\vec m$ are invertible: in particular,
evaluating~\eqref{eq:5} in~$\Q$ can provide $\tc$~approximations to $X/[\vec m]$, and evaluating it modulo a prime~$a$
coprime to~$\vec m$ yields the value of~$X$ modulo~$a$, that is, an extension of $\vec x$ to CRR modulo the basis
$\p{\vec m,a}$ (``basis extension'').

We need $\imul$ to make sense of \eqref{eq:5} in~$\Q$, hence we cannot use it directly in $\vtc(\imulm)$. However, we
will consider an approximation of rank and related quantities, and we will prove their various properties from first
principles, which will ultimately allow us to make CRR reconstruction work.
\begin{Def}\label{def:shadows}
(In $\vtc(\imulm)$.) Given $\vec x<\vec m$ and $n$, let $h_i=[\vec m]_{\ne i}^{-1}\rem m_i$ for $i<k$, and define
\begin{align*}
S_n(\vec m;\vec x)&=\sum_{i<k}\cdiv{2^nx_ih_i}{m_i},\\
r_n(\vec m;\vec x)&=\fl{2^{-n}S_n(\vec m;\vec x)},\\
\xi_n(\vec m;\vec x)&=2^{-n}\bigl(S_n(\vec m;\vec x)\rem2^n\bigr),\\
e_n(\vec m;\vec x;a)&=\Bigl(\sum_{i<k}x_ih_i[\vec m]_{\ne i}-[\vec m]r_n(\vec m;\vec x)\Bigr)\rem a
\end{align*}
for any prime~$a$, using Lemma~\ref{lem:div-small-prm} and $\imulm$. That is, $r_n\le\sum_im_i$ is an estimate of the rank
of~$\vec x$, $\xi_n\in[0,1]$ is a dyadic rational approximation of $X/[\vec m]$ per~\eqref{eq:5}, and $e_n<a$ is an
estimate of $X$ modulo~$a$. In order to make the notation less heavy, we may omit $\vec m$ if it is understood from the
context.

Observe that
\begin{equation}\label{eq:13}
e_n(\vec m;\vec x;m_i)=x_i.
\end{equation}
If $\vec a$ is a sequence of primes (which may include~$\vec m$), we let
$e_n(\vec m;\vec x;\vec a)=\p{e_n(\vec m;\vec x;a_j):j<l}$. This should be thought of as extension of $\vec x$ to CRR
modulo~$\vec a$.
\end{Def}
\begin{Exm}
Let $\vec x=\vec1$ (which is the CRR of $X=1$). Then for $n$ large enough, $e_n(\vec m;\vec x;a)=1$ and $\xi_n(\vec
m;\vec x)\approx1/[\vec m]$. See Lemma~\ref{lem:xi1-ub} for a formalization of this.
\end{Exm}

Note that the rank is a discrete quantity; while $2^{-n}S_n$ is an approximation of $\sum_ix_ih_i/m_i$ that can be
expected to converge in a reasonable way to the true value as $n$ gets larger, $r_n$ will make abrupt jumps. If $r_n$
happens to be the true rank, then $\xi_n$ should be a close approximation of $X/[\vec m]$, and $e_n$ has the correct
value, but if $r_n$ is off by~$1$, then $\xi_n$ is very far from the right value, and $e_n$ (another discrete
quantity) is also off. Thus, one of the annoying problems we need to deal with in $\vtc(\imulm)$ is that it is a priori
difficult to guess how large $n$ we need so that $r_n$ is ``correct''.

The remainder of this section is organized into two subsections. In Section~\ref{sec:auxil-prop-crr}, we will develop
computation with CRR in $\vtc(\imulm)$, in particular, we will show how various manipulations of CRR affect the related
$r_n$, $\xi_n$, and $e_n$ values. In Section~\ref{sec:chin-rema-reconstr}, we define and analyze the CRR reconstruction
procedure and derive $\imul$ in $\vtc(\imulm)$.

\subsection{Auxiliary properties of CRR}\label{sec:auxil-prop-crr}

Results in this section are nominally proved in the theory $\vtc(\imulm)$. In fact, the proofs will only use instances
of $\imulm$ modulo primes listed in the statements ($\vec m$, sometimes $\vec a$ or $\vec b$), which fact will become
relevant in Section~\ref{sec:gener-mult-groups}. However, we do not indicate this explicitly in an effort not to make
the notation more cluttered than it already is.

We start with two lemmas on basis extension. The first one is a formalization of the observation that if $\vec x<\vec m$
is the CRR of $X<[\vec m]$, and $\vec a\perp\vec m$, then the CRR of $[\vec a]X<[\vec m][\vec a]$ modulo the extended
basis $\p{\vec m,\vec a}$ is $\p{[\vec a]\vec x,\vec0}$. Here and below, operations on residue sequences
$\vec x<\vec m$ (such as multiplication by $[\vec a]$) are assumed to be evaluated modulo~$\vec m$.
\begin{Lem}\label{lem:extbymul}
$\vtc(\imulm)$ proves that for any $\vec x<\vec m$ and $\vec a,a\perp\vec m$,
\begin{align}
\label{eq:14}r_n(\vec m,a;a\,\vec x,0)&=r_n(\vec m;\vec x)+\sum_{i<k}x_i\fdiv{a\tilde h_i}{m_i},\\
\label{eq:16}e_n(\vec m,\vec a;[\vec a]\vec x,\vec0;\vec b)&=[\vec a]e_n(\vec m;\vec x;\vec b)\rem\vec b,\\
\label{eq:15}\xi_n(\vec m,\vec a;[\vec a]\vec x,\vec0)&=\xi_n(\vec m;\vec x),
\end{align}
where $\tilde h_i=(a[\vec m]_{\ne i})^{-1}\rem m_i$.
\end{Lem}
\begin{Pf}
Let $h_i=[\vec m]_{\ne i}^{-1}\rem m_i$. We have $a\tilde h_i\equiv h_i\pmod{m_i}$, i.e.,
\begin{equation}\label{eq:25}
a\tilde h_i=h_i+m_i\fdiv{a\tilde h_i}{m_i},
\end{equation}
thus
\begin{align*}
S_n(\vec m,a;a\,\vec x,0)
&=\sum_{i<k}\cdiv{2^nx_ia\tilde h_i}{m_i}=\sum_{i<k}\cdiv{2^nx_ih_i}{m_i}+2^n\sum_{i<k}x_i\fdiv{a\tilde h_i}{m_i}\\
&=S_n(\vec m;\vec x)+2^n\sum_{i<k}x_i\fdiv{a\tilde h_i}{m_i}.
\end{align*}
This gives \eqref{eq:14}, and~\eqref{eq:15} for $l=1$; the general case of~\eqref{eq:15} follows by induction%
\footnote{More precisely: for fixed $\vec a=\p{a_i:i<l}$, we prove by induction on $l'\le l$ that \eqref{eq:15} holds
for $\p{a_i:i<l'}$, which is a $\Sig0(\imulm)$ property. Most proofs by induction in this section should be
interpreted similarly.}
on~$l$.

We can again prove~\eqref{eq:16} by induction on~$l$, hence it is enough to show it for $l=1$. Obviously, we may also
assume $\Lh(\vec b)=1$. Computing modulo~$b$, we have
\begin{align*}
e_n(\vec m,a;a\vec x,0;b)
&\equiv \sum_{i<k}ax_i\tilde h_ia[\vec m]_{\ne i}-a[\vec m]r_n(\vec m,a;a\,\vec x,0)\\
&\equiv a\Bigl(\sum_{i<k}x_ia\tilde h_i[\vec m]_{\ne i}-[\vec m]r_n(\vec m,a;a\,\vec x,0)\Bigr)\\
&\equiv a\left(\sum_{i<k}x_ih_i[\vec m]_{\ne i}+[\vec m]\sum_{i<k}x_i\fdiv{a\tilde h_i}{m_i}-[\vec m]r_n(\vec m,a;a\,\vec x,0)\right)\\
&\equiv a\Bigl(\sum_{i<k}x_ih_i[\vec m]_{\ne i}-[\vec m]r_n(\vec m;\vec x)\Bigr)\\
&\equiv a\,e_n(\vec m;\vec x;b)
\end{align*}
using \eqref{eq:25} and~\eqref{eq:14}.
\end{Pf}

The second lemma formalizes the idea that $e_n(\vec m;\vec x;\vec m,\vec a)=\p{\vec x,e_n(\vec m;\vec x;\vec a)}$ is
the extension of $\vec x$ to the basis $\p{\vec m,\vec a}$ (representing the same number). Since the effect of basis
extension on the $\xi_n$ approximation is essentially division by $[\vec a]$, which we cannot do directly, we first
formulate the result for a single prime~$a$, and then we obtain a version for arbitrary~$\vec a$ using a crude
approximation of~$[\vec a]$.
\begin{Lem}\label{lem:ext}
$\vtc(\imulm)$ proves that for any $\vec x<\vec m$ and $a\perp\vec m$, if $n\ge\lh k$, then
\begin{align}
\label{eq:19}
a\,r_n\bigl(\vec m,a;e_n(\vec m;\vec x;\vec m,a)\bigr)&=r_n(\vec m;\vec x)+e_n(\vec m;\vec x;a)\tilde h+\sum_{i<k}x_i\fdiv{a\tilde h_i}{m_i},\\
\label{eq:20}
e_n\bigl(\vec m,a;e_n(\vec m;\vec x;\vec m,a);\vec b\bigr)&=e_n(\vec m;\vec x;\vec b),\\
\label{eq:17}
\xi_n\bigl(\vec m,a;e_n(\vec m;\vec x;\vec m,a)\bigr)&=\tfrac1a\xi_n(\vec m;\vec x)\PM{2^{-n}(k+1)(1-a^{-1})}0,
\end{align}
where $\tilde h=[\vec m]^{-1}\rem a$, $\tilde h_i=(a[\vec m]_{\ne i})^{-1}\rem m_i$.
\end{Lem}
\begin{Pf}
Put $h_i=[\vec m]_{\ne i}^{-1}\rem m_i$ and $y=e_n(\vec m;\vec x;a)$ so that
$e_n(\vec m;\vec x;\vec m,a)=\p{\vec x,y}$, and let $\roo$ denote the right-hand side of~\eqref{eq:19}. First,
using~\eqref{eq:25}, we have
\begin{align*}
[\vec m]\roo
&\equiv[\vec m]r_n(\vec x)+y+\sum_{i<k}x_im_i\fdiv{a\tilde h_i}{m_i}[\vec m]_{\ne i}\\
&\equiv\sum_{i<k}x_ih_i[\vec m]_{\ne i}+\sum_{i<k}x_im_i\fdiv{a\tilde h_i}{m_i}[\vec m]_{\ne i}
\equiv\sum_{i<k}x_ia\tilde h_i[\vec m]_{\ne i}\equiv0\pmod a,
\end{align*}
that is, $\roo/a$ is an integer. Observe that for any rational~$\omega$,
$a\cl\omega<a(\omega+1)=a\omega+a\le\cl{a\omega}+a$, hence
\[\cl{a\omega}\le a\cl\omega\le\cl{a\omega}+(a-1).\]
Using this, we obtain
\begin{align*}
2^n\bigl(\roo+\xi_n(\vec m;\vec x)\bigr)&=
S_n(\vec m;\vec x)+2^n\left(y\tilde h+\sum_{i<k}x_i\fdiv{a\tilde h_i}{m_i}\right)\\
&=\sum_{i<k}\cdiv{2^nx_ih_i}{m_i}+\sum_{i<k}\frac{2^nx_im_i\fl{a\tilde h_i/m_i}}{m_i}+2^ny\tilde h\\
&=\sum_{i<k}\cdiv{2^nax_i\tilde h_i}{m_i}+2^ny\tilde h\\
&=a\sum_{i<k}\cdiv{2^nx_i\tilde h_i}{m_i}+a\cdiv{2^ny\tilde h}a\PM0{(k+1)(a-1)}\\
&=aS_n(\vec m,a;\vec x,y)\PM0{(k+1)(a-1)}.
\end{align*}
On the one hand, this gives $\roo/a\le 2^{-n}S_n(\vec m,a;\vec x,y)<\bigl(r_n(\vec m,a;\vec x,y)+1\bigr)$,
thus $\roo/a\le r_n(\vec m,a;\vec x,y)$. On the other hand,
\[a\,r_n(\vec m,a;\vec x,y)\le2^{-n}a\,S_n(\vec m,a;\vec x,y)<\roo+1+2^{-n}(k+1)(a-1)\le\roo+a\]
as long as $2^n\ge k+1$, thus $r_n(\vec m,a;\vec x,y)<\roo/a+1$, i.e., $r_n(\vec m,a;\vec x,y)\le\roo/a$. This
proves~\eqref{eq:19}, whence also~\eqref{eq:17}:
\[a\,\xi_n(\vec m,a;\vec x,y)=2^{-n}a\,S_n(\vec m,a;\vec x,y)-\roo=\xi_n(\vec m;\vec x)\PM{2^{-n}(k+1)(a-1)}0.\]
To prove~\eqref{eq:20}, we may assume $\Lh(\vec b)=1$; working modulo~$b$,
\begin{align*}
e_n(\vec m,a;\vec x,y;b)
&\equiv\sum_{i<k}x_i\tilde h_ia[\vec m]_{\ne i}+y\tilde h[\vec m]-[\vec m]a\,r_n(\vec m,a;\vec x,y)\\
&\equiv\sum_{i<k}x_ih_i[\vec m]_{\ne i}
   +[\vec m]\left(\sum_{i<k}x_i\fdiv{a\tilde h_i}{m_i}+y\tilde h-a\,r_n(\vec m,a;\vec x,y)\right)\\
&\equiv\sum_{i<k}x_ih_i[\vec m]_{\ne i}-r_n(\vec m;\vec x)\\
&\equiv e_n(\vec m;\vec x;b)
\end{align*}
using \eqref{eq:25} and~\eqref{eq:19}.
\end{Pf}
\begin{Cor}\label{cor:extmore}
$\vtc(\imulm)$ proves that for any $\vec x<\vec m$ and $\vec a\perp\vec m$, if $n\ge\lh{k+l}$, then
\begin{gather}
\label{eq:26}
e_n\bigl(\vec m,\vec a;e_n(\vec m;\vec x;\vec m,\vec a);\vec b\bigr)=e_n(\vec m;\vec x;\vec b),\\
\label{eq:18}
2^{-\sum_j\lh{a_j}}\xi_n(\vec m;\vec x)\le\xi_n\bigl(\vec m,\vec a;e_n(\vec m;\vec x;\vec m,\vec a)\bigr)
  \le2^{-\sum_j(\lh{a_j}-1)}\xi_n(\vec m;\vec x)+2^{-n}(k+l).
\end{gather}
\end{Cor}
\begin{Pf}
By induction in~$l$, using \eqref{eq:20}, \eqref{eq:17}, and $2^{-\lh a}<\frac1a\le2^{-(\lh a-1)}$.
\end{Pf}

The CRR of~$1$, which is just the sequence~$\vec1$, will feature prominently in many calculations, as
$\xi_n(\vec m;\vec1)$ is our proxy for $1/[\vec m]$. The next lemma summarizes its most basic properties.
\begin{Lem}\label{lem:xi1-ub}
$\vtc(\imulm)$ proves: if $n\ge\lh k\ge1$, then
\begin{gather}
\label{eq:21}2^{-\sum_i\lh{m_i}}<\xi_n(\vec m;\vec1)<2^{-\sum_i(\lh{m_i}-1)}+2^{-n}(k+1),\\
\label{eq:27}e_n(\vec m;\vec1;\vec a)=\vec1.
\end{gather}
\end{Lem}
\begin{Pf}
Since $m_0\ge2$ and $2^n\ge2$, we have
$r_n(m_0;1)=\Fl{2^{-n}\cl{2^n/m_0}}=0$, thus
\[e_n(m_0;1;a)=1\cdot1\cdot1-0=1\]
for any $a$, i.e., $e_n(m_0;1;\vec a)=\vec1$. In particular, $e_n(m_0;1;\vec m)=\vec1$, hence
\[e_n(\vec m;\vec1;\vec a)=e_n\bigl(\vec m;e_n(m_0;1;\vec m);\vec a\bigr)=e_n(m_0;1;\vec a)=\vec1\]
by~\eqref{eq:26}. Moreover,
\[2^{-\lh{m_0}}<\frac1{m_0}\le2^{-n}\cdiv{2^n}{m_0}=\xi_n(m_0;1)<\frac1{m_0}+2^{-n}\le2^{-(\lh{m_0}-1)}+2^{-n},\]
thus
\begin{align*}
\xi_n(\vec m;\vec1)=\xi_n\bigl(\vec m;e_n(m_0;1;\vec m)\bigr)
&\le2^{-\sum_{i>0}(\lh{m_i}-1)}\xi_n(m_0;1)+2^{-n}k\\
&<2^{-\sum_i(\lh{m_i}-1)}+2^{-n}(k+1)
\end{align*}
using~\eqref{eq:18}. The other inequality is similar.
\end{Pf}

The next lemma expresses the fact that if $\vec x$ and~$\vec y$ are respectively the CRR of $X,Y<[\vec m]$,
then $\vec x+\vec y$ (modulo~$\vec m$) is the CRR of $(X+Y)\bmod[\vec m]$, which is $X+Y-c[\vec m]$ for $c\in\{0,1\}$.
The first version we prove here also allows $c=-1$ (which is impossible in the real world); we will fix this
discrepancy in Corollary~\ref{cor:sum} below, under a stronger requirement on~$n$.
\begin{Lem}\label{lem:sum}
$\vtc(\imulm)$ proves: if $\vec x,\vec y<\vec m$, $\vec z=(\vec x+\vec y)\rem\vec m$, and $n\ge\lh k$, then there exists
$c\in\{-1,0,1\}$ such that
\begin{align}
\label{eq:22}
r_n(\vec m;\vec z)&=r_n(\vec m;\vec x)+r_n(\vec m;\vec y)+c-\sum_{x_i+y_i\ge m_i}h_i,\\
\label{eq:23}
e_n(\vec m;\vec z;a)&\equiv e_n(\vec m;\vec x;a)+e_n(\vec m;\vec y;a)-c[\vec m]\pmod a,\\
\label{eq:24}
\xi_n(\vec m;\vec z)&=\xi_n(\vec m;\vec x)+\xi_n(\vec m;\vec y)-c\PM0{2^{-n}k},
\end{align}
where $h_i=[\vec m]_{\ne i}^{-1}\rem m_i$.
\end{Lem}
\begin{Pf}
Let $I=\{i<k:x_i+y_i\ge m_i\}$, so that $z_i=x_i+y_i$ for $i\notin I$, and $z_i=x_i+y_i-m_i$ for $i\in I$. Then
\begin{align*}
2^{-n}S_n(\vec m;\vec z)
&=2^{-n}\sum_{i<k}\cdiv{2^n(x_i+y_i)h_i}{m_i}-\sum_{i\in I}h_i\\
&=2^{-n}S_n(\vec m;\vec x)+2^{-n}S_n(\vec y;\vec m)-\sum_{i\in I}h_i\PM0{2^{-n}k}\\
&=r_n(\vec m;\vec x)+r_n(\vec y;\vec m)-\sum_{i\in I}h_i+\xi_n(\vec m;\vec x)+\xi_n(\vec y;\vec m)\PM0{2^{-n}k}.
\end{align*}
Since $k\le2^n$ and $0\le\xi_n(\vec m;\vec x)+\xi_n(\vec y;\vec m)<2$, this readily implies \eqref{eq:22}
and~\eqref{eq:24}. Moreover,
\begin{align*}
e_n(\vec m;\vec z;a)
&\equiv\sum_{i<k}x_ih_i[\vec m]_{\ne i}+\sum_{i<k}y_ih_i[\vec m]_{\ne i}-\sum_{i\in I}h_i[\vec m]\\
&\qquad-\Bigl(r_n(\vec m;\vec x)+r_n(\vec m;\vec y)+c-\sum_{i\in I}h_i\Bigr)[\vec m]\\
&\equiv e_n(\vec m;\vec x;a)+e_n(\vec m;\vec y;a)-c[\vec m]
\end{align*}
modulo~$a$.
\end{Pf}

The following lemma can be read as stating that $0<X<[\vec m]\implies1\le X\le[\vec m]-1$. While this sounds like a
triviality, it is in fact an important result implying that (for large enough~$n$) $\xi_n$ cannot take arbitrary
values in $[0,1]$, but it is a discrete quantity coming in steps of $1/[\vec m]$ (i.e., $\xi_n(\vec m;\vec1)$). Among
other consequences, this will eventually allows us to prove a bound on~$n$ above which $r_n$ and $e_n$ stabilize.
\begin{Lem}\label{lem:xi-ub-lb}
$\vtc(\imulm)$ proves that for any $\vec0\ne\vec x<\vec m$,
\begin{equation}
\label{eq:11}\min\bigl\{\xi_n(\vec m;\vec x),1-\xi_n(\vec m;\vec x)\bigr\}\ge\xi_n(\vec m;\vec1)-2^{-n}(3k).
\end{equation}
\end{Lem}
\begin{Pf}
The statement is vacuous for~$k=0$, and it also holds trivially unless $2^n>3k$ and $\xi_n(\vec m;\vec1)>2^{-n}(3k)$.
We claim that this condition implies
\begin{equation}\label{eq:35}
2^n\ge\max_{i<k}m_i.
\end{equation}
If $k\ge2$, then $\xi_n(\vec m;\vec1)>2^{-n}(3k)$ gives
\[2^{-n}(3k)<2^{-\sum_i(\lh{m_i}-1)}+2^{-n}(k+1)\]
by~\eqref{eq:21}, hence
\[\max_{i<k}m_i\le2^{1+\sum_i(\lh{m_i}-1)}\le(2k-1)2^{\sum_i(\lh{m_i}-1)}<2^n.\]
If $k=1$, then $m_0\ge2$ and $n\ge1$ ensure $\cl{2^n/m_0}\le2^{n-1}<2^n$, thus $\xi_n(m_0;1)=2^{-n}\cl{2^n/m_0}$. If
$2^n\xi_n(m_0;1)>3k=3$, we obtain $2^n/m_0>3$, and a fortiori $2^n\ge m_0$.

Now, let us prove \eqref{eq:11} by induction on~$k$. For $k=1$, we have $\xi_n(m_0;1)=2^{-n}\cl{2^n/m_0}$, and
\eqref{eq:35} ensures $\cl{2^n(m_0-1)/m_0}<2^n$, thus $\xi_n(m_0;x)=2^{-n}\cl{2^nx/m_0}$, and we obtain
\[\xi_n(m_0;1)\le\xi_n(m_0;x)\le1-\xi_n(m_0;1)+2^{-n}.\]

Assume \eqref{eq:11} holds for~$k\ge1$, we will prove it for~$k+1$. Let $\p{\vec0,0}\ne\p{\vec x,y}<\p{\vec m,m_k}$. As
above, we assume $\xi_n(\vec m,m_k;\vec1,1)>3(k+1)2^{-n}$, thus $2^n\ge m_k$ by~\eqref{eq:35}, which ensures
$\cl{2^n(m_k-1)/m_k}<2^n$.

We have
\begin{equation}\label{eq:12}
\xi_n(\vec m,m_k;\vec1,1)\le\frac1{m_k}\xi_n(\vec m;\vec1)+2^{-n}(k+1)(1-m_k^{-1})
\end{equation}
by~\eqref{eq:17}. We distinguish two cases. If $\vec x=\vec0$, let $\tilde y=y[\vec m]^{-1}\rem m_k$; then $1\le\tilde y\le m_k-1$, and
\[\xi_n(\vec m,m_k;\vec x,y)=2^{-n}\cdiv{2^n\tilde y}{m_k}=\frac{\tilde y}{m_k}\PM{2^{-n}}0,\]
hence
\[\min\bigl\{\xi_n(\vec m,m_k;\vec x,y),1-\xi_n(\vec m,m_k;\vec x,y)\bigr\}\ge\frac1{m_k}-2^{-n}
   \ge\xi_n(\vec m,m_k;\vec1,1)-2^{-n}(k+2)\]
using~\eqref{eq:12}.

If $\vec x\ne\vec0$, let $y'=\bigl(y-e_n(\vec m;\vec x;m_k)\bigr)\rem m_k$, and $\tilde y=y'[\vec m]^{-1}\rem m_k$. Then
\begin{align*}
\xi_n(\vec m,m_k;\vec x,y)
&=\xi_n\bigl(\vec m,m_k;e_n(\vec m;\vec x;\vec m,m_k)\bigr)+\xi_n(\vec m,m_k;\vec 0,y')-c\PM0{2^{-n}(k+1)}\\
&=\frac1{m_k}\xi_n(\vec m;\vec x)+2^{-n}\cdiv{2^n\tilde y}{m_k}-c\pm2^{-n}(k+1)\\
&=\frac1{m_k}\bigl(\xi_n(\vec m;\vec x)+\tilde y\bigr)-c\pm2^{-n}(k+2)
\end{align*}
for some $c\in\{-1,0,1\}$ using \eqref{eq:24} and~\eqref{eq:17}. Since $0\le\tilde y\le m_k-1$, we have
\[\min\bigl\{\tfrac1{m_k}\bigl(\xi_n(\vec m;\vec x)+\tilde y\bigr),
    1-\tfrac1{m_k}\bigl(\xi_n(\vec m;\vec x)+\tilde y\bigr)\bigr\}
  \ge\frac1{m_k}\bigl(\xi_n(\vec m;\vec1)-2^{-n}(3k)\bigr)\]
by the induction hypothesis, thus
\begin{align*}
\min\bigl\{\xi_n(\vec m,m_k;\vec x&,y)+c,1-\bigl(\xi_n(\vec m,m_k;\vec x,y)+c\bigr)\bigr\}\\
&\ge\xi_n(\vec m,m_k;\vec1,1)-2^{-n}\bigl(3km_k^{-1}+(k+1)(1-m_k^{-1})+k+2\bigr)\\
&\ge\xi_n(\vec m,m_k;\vec1,1)-2^{-n}\bigl(km_k^{-1}+k+1+(k+1)(2-m_k^{-1})\bigr)\\
&\ge\xi_n(\vec m,m_k;\vec1,1)-2^{-n}\bigl(3(k+1)\bigr)>0
\end{align*}
using~\eqref{eq:12} and $m_k\ge2$, which implies $c=0$ and the result.
\end{Pf}

\begin{Cor}\label{cor:sum}
$\vtc(\imulm)$ proves that if $n\ge\lh k+2+\sum_{i<k}\lh{m_i}$, then Lemma~\ref{lem:sum} holds with $c\in\{0,1\}$.
\end{Cor}
\begin{Pf}
If, say, $\vec x=0$, then $\vec z=\vec y$, and the statement holds with $c=0$. Thus, we may assume
$\vec x\ne\vec0\ne\vec y$. If $c=-1$, then
\[1>\xi_n(\vec m;\vec z)=\xi_n(\vec m;\vec x)+\xi_n(\vec m;\vec y)+1\PM0{2^{-n}k}\]
implies
\[2^{-n}k>\xi_n(\vec m;\vec x)+\xi_n(\vec m;\vec y)\ge2\xi_n(\vec m;\vec1)-2^{-n}(6k)>2^{1-\sum_i\lh{m_i}}-2^{-n}(6k)\]
using Lemmas \ref{lem:xi-ub-lb} and~\ref{lem:xi1-ub}, thus
\[2^{1-\sum_i\lh{m_i}}<2^{-n}(7k)<2^{\lh k+3-n}.\]
This is a contradiction if $n\ge\lh k+2+\sum_{i<k}\lh{m_i}$.
\end{Pf}

The next crucial lemma states how large $n$ needs to be so that $r_n(\vec m;\vec x)$ is the true rank, and
$e_n(\vec m;\vec x;\vec a)$ the correct basis extension of~$\vec x$; it also gives the rate of convergence
of~$\xi_n(\vec m;\vec x)$. This will considerably simplify our subsequent arguments, as we can fix the rank and basis
extension functions independently of any extraneous parameters, and it will make calculations with $\xi_n$
self-correcting, preventing accumulation of errors (we may temporarily switch to $\xi_{n'}$ with $n'\ge n$ as large as
we want to make any given argument work with sufficient accuracy, and get back to $\xi_n$ using~\eqref{eq:7}).
\begin{Lem}\label{lem:r-indep}
$\vtc(\imulm)$ proves: if $n'\ge n\ge\lh k+2+\sum_{i<k}\lh{m_i}$, then for all $\vec x<\vec m$ and $\vec a$,
\begin{align}
\label{eq:6}r_n(\vec m;\vec x)&=r_{n'}(\vec m;\vec x),\\
\label{eq:8}e_n(\vec m;\vec x;\vec a)&=e_{n'}(\vec m;\vec x;\vec a),\\
\label{eq:7}\xi_n(\vec m;\vec x)&=\xi_{n'}(\vec m;\vec x)\PM{2^{-n}k}0.
\end{align}
\end{Lem}
\begin{Pf}
If $\vec x=\vec0$, all quantities in \eqref{eq:6}--\eqref{eq:7} are~$0$, thus we may assume $\vec x\ne\vec0$ (whence
$k\ge1$). Put $h_i=[\vec m]_{\ne i}^{-1}\rem m_i$. Since
\[2^{n'-n}\fdiv{2^nx_ih_i}{m_i}\le\fdiv{2^{n'}x_ih_i}{m_i},\qquad2^{n'-n}\cdiv{2^nx_ih_i}{m_i}\ge\cdiv{2^{n'}x_ih_i}{m_i},\]
we have
\begin{equation}\label{eq:9}
2^{-n'}S_{n'}(\vec x)=2^{-n'}\sum_{i<k}\cdiv{2^{n'}x_ih_i}{m_i}
  \le2^{-n'}\sum_{i<k}2^{n'-n}\cdiv{2^nx_ih_i}{m_i}=2^{-n}S_n(\vec x)
\end{equation}
and
\begin{align*}
2^{-n}S_n(\vec x)&
\le2^{-n}\sum_{i<k}\fdiv{2^nx_ih_i}{m_i}+2^{-n}k
=2^{-n'}\sum_{i<k}2^{n'-n}\fdiv{2^nx_ih_i}{m_i}+2^{-n}k\\
&\le2^{-n'}\sum_{i<k}\fdiv{2^{n'}x_ih_i}{m_i}+2^{-n}k\le2^{-n'}S_{n'}(\vec x)+2^{-n}k.\numberthis\label{eq:10}
\end{align*}
Thus, using \eqref{eq:11} and Lemma~\ref{lem:xi1-ub},
\[r_n(\vec x)+1>2^{-n'}S_{n'}(\vec x)
\ge r_n(\vec x)+\xi_n(\vec x)-2^{-n}k\ge r_n(\vec x)+2^{-\sum_i\lh{m_i}}-2^{-n}(4k)\ge r_n(\vec x)\]
as long as $n\ge\sum_i\lh{m_i}+\lh k+2$. Then $r_{n'}(\vec x)=r_n(\vec x)$; \eqref{eq:8} follows as the only
dependence of $e_n$ on~$n$ is through~$r_n$, and \eqref{eq:7} follows from \eqref{eq:9} and~\eqref{eq:10}.
\end{Pf}

\begin{Def}\label{def:indep-r}
For $\vec x<\vec m$, we define $r(\vec m;\vec x)=r_n(\vec m;\vec x)$ and $e(\vec m;\vec x;\vec a)=e_n(\vec m;\vec x;\vec
a)$, where $n=\lh k+2+\sum_{i<k}\lh{m_i}$.
\end{Def}

The meaning of the next lemma is that if $\vec m$ is odd, the CRR of $(1+[\vec m])/2$ is $2^{-1}\rem\vec m$ (i.e., the
sequence of inverses of~$2$ modulo each~$m_i$). The CRR reconstruction procedure will involve such factors.
\begin{Lem}\label{lem:half-odd}
$\vtc(\imulm)$ proves: if $\vec x<\vec m\perp2$, $\vec a\perp2$, $k>0$, and $n\ge\lh{k+1}+4+\sum_i\lh{m_i}$, then
\begin{align}
\label{eq:28}
e(\vec m;2^{-1}\rem\vec m;\vec a)&\equiv2^{-1}(1+[\vec m])\pmod{\vec a},\\
\label{eq:30}
\xi_n(\vec m;2^{-1}\rem\vec m)&=\frac12+\xi_n(\vec m,2;\vec1,1).
\end{align}
\end{Lem}
\begin{Pf}
We may assume $\Lh(\vec a)=1$. Working modulo~$a$, we have
\begin{align*}
2e(\vec m;2^{-1}\rem\vec m;a)&=e(\vec m,2;\vec1,0;a)\\
&\equiv e(\vec m,2;\vec1,1;a)-[\vec m]+\bigl(r(\vec m,2;\vec 1,1)-r(\vec m,2;\vec 1,0)\bigr)2[\vec m]\\
&\equiv1-[\vec m]+\bigl(r(\vec m,2;\vec 1,1)-r(\vec m,2;\vec 1,0)\bigr)2[\vec m]
\end{align*}
by~\eqref{eq:16}, the definition of~$e_n$, and~\eqref{eq:27}. Now, the definition of~$S_n$ gives
\begin{equation}\label{eq:31}
S_n(\vec m,2;\vec1,1)-S_n(\vec m,2;\vec1,0)=\cdiv{2^n}2=2^{n-1},
\end{equation}
thus
\[r(\vec m,2;\vec 1,1)-r(\vec m,2;\vec 1,0)
  =\begin{cases}1,&\text{if $\xi_n(\vec m,2;\vec1,1)<\frac12$,}\\0,&\text{otherwise}\end{cases}\]
for any $n\ge\lh{k+1}+4\sum_i\lh{m_i}$. However, \eqref{eq:21} ensures $\xi_n(\vec m,2;\vec1,1)<\frac12$ as long as
$k\ge1$ and $2^n\ge4(k+2)$, hence
\[2e(\vec m;2^{-1}\rem\vec m;a)\equiv1+[\vec m]\pmod a\]
as required. Also, \eqref{eq:31} ensures
\[\xi_n(\vec m;2^{-1}\rem\vec m)=\xi_n(\vec m,2;\vec 1,0)=\frac12+\xi_n(\vec m,2;\vec1,1)\]
using Lemma~\ref{lem:extbymul}.
\end{Pf}

The following lemma shows that if $X$ (which is not too big w.r.t.\ $\vec m$) has CRR $\vec x$, then $e(\vec x;\vec a)$
is $X\rem\vec a$ as expected, and $\xi_n(\vec x)\approx X/[\vec m]$ (formulated with $\xi_n(\vec1)$).

As a corollary, we obtain that $X$ (which is not too big) is uniquely determined by its CRR.
\begin{Lem}\label{lem:Xcrr}
$\vtc(\imulm)$ proves: if $\lh X\le\sum_{i<k}\bigl(\lh{m_i}-1\bigr)$, $\vec x=X\rem\vec m$, and $n\ge\lh
k+2+\sum_{i<k}\lh{m_i}$, then
\begin{align}
\label{eq:32}
e(\vec m;\vec x;\vec a)&=X\rem\vec a,\\
\label{eq:33}
X\bigl(\xi_n(\vec m;\vec1)-2^{1-n}k\bigr)&\le\xi_n(\vec m;\vec x)\le X\xi_n(\vec m;\vec1).
\end{align}
\end{Lem}
\begin{Pf}
We may assume $\Lh(\vec a)=1$. If we fix~$X$, we can prove the statement for $\fl{2^{-t}X}$, $t\le\lh X$, by reverse
induction on~$t$; that is, it suffices to show that it holds for $X=0$ (trivial) and $X=1$ (Lemma~\ref{lem:xi1-ub}), and
that it holds for $X\ge2$ assuming it holds for $\fl{X/2}$. To facilitate the induction argument, we strengthen the
lower bound for $X\ge1$ to
\begin{equation}\label{eq:34}
X\xi_n(\vec1)-(2X-1)2^{-n}k\le\xi_n(\vec x)\le X\xi_n(\vec1).
\end{equation}
Assume that \eqref{eq:32} and~\eqref{eq:34} hold for $Y=\fl{X/2}$, and put $\vec y=\fl{X/2}\rem\vec m$. Using
Corollary~\ref{cor:sum}, there is a constant $c\in\{0,1\}$ such that
\begin{align*}
e(2\vec y;a)&\equiv2e(\vec y;a)-c[\vec m]\pmod a,\\
\xi_n(2\vec y)&=2\xi_n(\vec y)-c\PM0{2^{-n}k}.
\end{align*}
However, since
\[2\xi_n(\vec y)\le2Y\xi_n(\vec1)\le X2^{-\sum_i(\lh{m_i}-1)}\le X2^{-\lh X}<1\]
by the induction hypothesis and Lemma~\ref{lem:xi1-ub}, we must have $c=0$, thus
\[e(2\vec y;a)\equiv2e(\vec y;a)\equiv2Y\pmod a,\]
and
\[2Y\xi_n(\vec1)-\bigl(4Y-2+1\bigr)2^{-n}k\le\xi_n(2\vec y)\le2Y\xi_n(\vec1)\]
using the induction hypothesis. If $X=2Y$, then $\vec x=2\vec y$ and we are done. If $X=2Y+1$ and $\vec x=2\vec
y+\vec1$, we apply
Corollary~\ref{cor:sum} once again: there is $c'\in\{0,1\}$ such that
\begin{align*}
e(\vec x;a)&\equiv e(2\vec y;a)+1-c'[\vec m]\pmod a,\\
\xi_n(\vec x)&=\xi_n(2\vec y)+\xi_n(\vec1)-c'\PM0{2^{-n}k}
\end{align*}
using \eqref{eq:27}. As above,
\[\xi_n(2\vec y)+\xi_n(\vec1)\le(2Y+1)\xi_n(\vec1)\le X2^{-\sum_i(\lh{m_i}-1)}\le X2^{-\lh X}<1,\]
thus $c'=0$, and
\begin{gather*}
e(\vec x;a)\equiv e(2\vec y;a)+1\equiv2Y+1\equiv X\pmod a,\\
X\xi_n(\vec1)-(2X-1)2^{-n}k\le X\xi_n(\vec1)-(4Y)2^{-n}k\le\xi_n(\vec x)\le X\xi_n(\vec1)
\end{gather*}
as required.
\end{Pf}
\begin{Cor}\label{cor:crr-inj}
$\vtc(\imulm)$ proves: if $\lh X,\lh Y\le\sum_{i<k}\bigl(\lh{m_i}-1\bigr)$ and $X\equiv Y\pmod{\vec m}$,
then $X=Y$.
\end{Cor}
\begin{Pf}
Put $\vec x=X\rem\vec m$ and $\vec y=Y\rem\vec m$. If, say $X<Y$, then
\[\xi_n(\vec x)\le(Y-1)\xi_n(\vec1)<Y\bigl(\xi_n(\vec1)-2^{1-n}k\bigr)\le\xi_n(\vec y)\]
by Lemma~\ref{lem:Xcrr} as long as $n\ge\lh k+2+\sum_{i<k}\lh{m_i}$ and $Y2^{1-n}k<\xi_n(\vec1)$. (Since
$Y<2^{\sum_i(\lh{m_i}-1)}$ and $\xi_n(\vec1)>2^{-\sum_i\lh{m_i}}$ by~\eqref{eq:21}, this holds if we take $n\ge\lh
k+2+2\sum_i\lh{m_i}$.) Then it follows that $\vec x\ne\vec y$.
\end{Pf}

The final, and most complicated, technical result in this subsection expresses that given the CRR of $X<[\vec m]$ in
basis~$\vec m$, and the CRR of $Y<[\vec a]$ in basis~$\vec a$ (where $\vec m\perp\vec a$), we obtain the CRR of
$XY<[\vec m][\vec a]$ in the basis $\p{\vec m,\vec a}$ by extending both original CRRs to the combined basis, and
multiplying them elementwise (modulo each prime).

We first need a simple ``reciprocity lemma'' relating inverses of two primes modulo each other.
\begin{Lem}\label{lem:inv-rec}
$\idz$ proves that if $m$ and $a$ are distinct primes, then
\begin{equation}\label{eq:38}
m\bigl(m^{-1}\rem a\bigr)-a\bigl((-a^{-1})\rem m\bigr)=1.
\end{equation}
\end{Lem}
\begin{Pf}
We have $m(m^{-1}\rem a)\equiv1\pmod a$, i.e., $m(m^{-1}\rem a)=1+au$ for some~$u$. Since $0<m(m^{-1}\rem a)<am$, we
have $0\le u<m$, and $-au\equiv1\pmod m$, thus $u=(-a^{-1})\rem m$.
\end{Pf}

\begin{Def}\label{def:prod}
If $\vec x,\vec y<\vec m$, then $\vec x\times\vec y$ denotes the elementwise product $\p{x_iy_i\rem m_i:i<k}$. More
generally, we will write $\prod_{u<t}\vec x_u$ for the elementwise product of terms $\p{x_{u,i}:i<k}=\vec x_u<\vec m$,
$u<t$.
\end{Def}
\begin{Lem}\label{lem:prod}
$\vtc(\imulm)$ proves: let $\vec m\perp\vec a$, $\vec x<\vec m$, $\vec y<\vec a$, and
$n\ge\lh{k+l}+2+\sum_{i<k}\lh{m_i}+\sum_{j<l}\lh{a_j}$. Then
\begin{align}
\label{eq:36}
e\bigl(\vec m,\vec a;e(\vec m;\vec x;\vec m,\vec a)\times e(\vec a;\vec y;\vec m,\vec a);\vec b\bigr)
&=e(\vec m;\vec x;\vec b)\times e(\vec a;\vec y;\vec b),\\
\label{eq:37}
\xi_n\bigl(\vec m,\vec a;e(\vec m;\vec x;\vec m,\vec a)\times e(\vec a;\vec y;\vec m,\vec a)\bigr)
&=\xi_n(\vec m;\vec x)\,\xi_n(\vec a;\vec y)\pm2^{-n}(k+l).
\end{align}
\end{Lem}
\begin{Pf}
If, say, $\vec x=\vec0$, then both sides of \eqref{eq:36} and~\eqref{eq:37} are~$0$, thus we may assume $\vec
x\ne\vec0\ne\vec y$. Put $\vec u=e(\vec a;\vec y;\vec m)$ and $\vec v=e(\vec m;\vec x;\vec a)$, so that
\[e(\vec m;\vec x;\vec m,\vec a)\times e(\vec a;\vec y;\vec m,\vec a)
   =\p{\vec x\times\vec u,\vec y\times\vec v}.\]
Let
\begin{align*}
h_i&=[\vec m]_{\ne i}^{-1}\rem m_i,&\tilde h_i&=[\vec a]^{-1}h_i\rem m_i,\\
h'_j&=[\vec a]_{\ne j}^{-1}\rem a_j,&\tilde h'_j&=[\vec m]^{-1}h'_j\rem a_j.
\end{align*}
For any $i<k$ and $j<l$, Lemma~\ref{lem:inv-rec} gives
\begin{align*}
\cdiv{2^nx_ih_i}{m_i}\cdiv{2^ny_jh'_j}{a_j}
&=2^{2n}\frac{x_ih_iy_jh'_j}{m_ia_j}\PM{2^n(m_i+a_j)}0\\
&=2^nx_ih_i\bigl(m_i^{-1}\rem a_j\bigr)\frac{2^ny_jh'_j}{a_j}\\
&\qquad-2^ny_jh'_j\bigl((-a_j^{-1})\rem m_i\bigr)\frac{2^nx_ih_i}{m_i}\PM{2^n(m_i+a_j)}0\\
&=2^nx_ih_i\bigl(m_i^{-1}\rem a_j\bigr)\cdiv{2^ny_jh'_j}{a_j}\\
&\qquad-2^ny_jh'_j\bigl((-a_j^{-1})\rem m_i\bigr)\cdiv{2^nx_ih_i}{m_i}
 \PM{2^n(m_i+a_j+m_ia_j^2)}{2^nm_i^2a_j}
\end{align*}
Then, expanding the definition,
\begin{align*}
\xi_n(\vec x)\,\xi_n(\vec y)
&=\bigl(2^{-n}S_n(\vec x)-r(\vec x)\bigr)\bigl(2^{-n}S_n(\vec y)-r(\vec y)\bigr)\\
&=2^{-2n}\sum_{\substack{i<k\\j<l}}\cdiv{2^nx_ih_i}{m_i}\cdiv{2^ny_jh'_j}{a_j}\\
&\qquad-2^{-n}r(\vec y)\sum_{i<k}\cdiv{2^nx_ih_i}{m_i}
 -2^{-n}r(\vec x)\sum_{j<l}\cdiv{2^ny_jh'_j}{a_j}
 +r(\vec x)\,r(\vec y)\\
&=2^{-n}\sum_{i<k}\cdiv{2^nx_ih_i}{m_i}\Bigl(-\sum_{j<l}y_jh'_j\bigl((-a_j^{-1})\rem m_i\bigr)-r(\vec y)\Bigr)\\
&\qquad+2^{-n}\sum_{j<l}\cdiv{2^ny_jh'_j}{a_j}\Bigl(\sum_{i<k}x_ih_i\bigl(m_i^{-1}\rem a_j\bigr)-r(\vec x)\Bigr)\\
&\qquad+r(\vec x)\,r(\vec y)\PM{2^{-n}\bigl(\sum_im_i+\sum_ja_j+\sum_im_i\sum_ja_j^2\bigr)}{2^{-n}\sum_im_i^2\sum_ja_j}.
\end{align*}
For any $i<k$,
\[h_i\Bigl(-\sum_{j<l}y_jh'_j\bigl((-a_j^{-1})\rem m_i\bigr)-r(\vec y)\Bigr)
\equiv h_i[\vec a]^{-1}e(\vec a;\vec y;m_i)\equiv\tilde h_iu_i\pmod{m_i},\]
thus
\[s_i=\frac{\tilde h_iu_i+h_i\Bigl(\sum_{j<l}y_jh'_j\bigl((-a_j^{-1})\rem m_i\bigr)+r(\vec y)\Bigr)}{m_i}\]
is a (small) integer, and we have
\begin{align*}
\cdiv{2^nx_ih_i}{m_i}\Bigl(-\sum_{j<l}y_j&h'_j\bigl((-a_j^{-1})\rem m_i\bigr)-r(\vec y)\Bigr)\\
&=\frac{2^nx_ih_i}{m_i}\Bigl(-\sum_{j<l}y_jh'_j\bigl((-a_j^{-1})\rem m_i\bigr)-r(\vec y)\Bigr)
  \PM0{m_i\sum_ja_j^2}\\
&=\frac{2^nx_i\tilde h_iu_i}{m_i}-2^nx_is_i\PM0{m_i\sum_ja_j^2}\\
&=\cdiv{2^nx_iu_i\tilde h_i}{m_i}-2^nx_is_i\PM0{m_i\sum_ja_j^2+1}.
\end{align*}
Likewise,
\[\cdiv{2^ny_jh'_j}{a_j}\Bigl(\sum_{i<k}x_ih_i\bigl(m_i^{-1}\rem a_j\bigr)-r(\vec x)\Bigr)
 =\cdiv{2^ny_jv_j\tilde h'_j}{a_j}-2^ny_jt_j\PM{a_j\sum_im_i^2}{\sum_im_i},\]
where
\[t_j=\frac{\tilde h'_jv_j-h'_j\Bigl(\sum_{i<k}x_ih_i\bigl(m_i^{-1}\rem a_j\bigr)-r(\vec x)\Bigr)}{a_j}\]
is an integer. Thus, continuing the computation above,
\begin{align*}
\xi_n(\vec x)\,\xi_n(\vec y)
&=2^{-n}\cdiv{2^nx_iu_i\tilde h_i}{m_i}+2^{-n}\cdiv{2^ny_jv_j\tilde h'_j}{a_j}\\
&\qquad-\sum_{i<k}x_is_i-\sum_{j<l}y_jt_j+r(\vec x)\,r(\vec y)
 \PM{2^{-n}\bigl(\sum_im_i+\sum_ja_j+\sum_im_i\sum_ja_j^2+\sum_im_i^2\sum_ja_j\bigr)}
    {2^{-n}\bigl(\sum_im_i^2\sum_ja_j+\sum_im_i\sum_ja_j^2+k+l\sum_im_i\bigr)}\\
&=S_n(\vec x\times\vec u,\vec y\times\vec v)
  -\sum_{i<k}x_is_i-\sum_{j<l}y_jt_j+r(\vec x)\,r(\vec y)\\
&\qquad\pm{\textstyle2^{-n}\Bigl(\sum_im_i\sum_ja_j^2+\sum_im_i^2\sum_ja_j+\sum_im_i\sum_ja_j\Bigr)}\\
&=S_n(\vec x\times\vec u,\vec y\times\vec v)
  -\sum_{i<k}x_is_i-\sum_{j<l}y_jt_j+r(\vec x)\,r(\vec y)\pm{\textstyle2^{-n}\sum_im_i^2\sum_ja_j^2}
\end{align*}
(using $(m_i-1)(a_j-1)\ge2$, which implies $m_ia_j^2+m_i^2a_j+m_ia_j\le m_i^2a_j^2$).
By Lemmas \ref{lem:xi1-ub} and~\ref{lem:xi-ub-lb}, there is $n_0$ such that
\[\min\bigl\{\xi_n(\vec x)\,\xi_n(\vec y),1-\xi_n(\vec x)\,\xi_n(\vec y)\bigr\}>\textstyle2^{-n}\sum_im_i^2\sum_ja_j^2\]
for all $n\ge n_0$. It follows that
\begin{equation}\label{eq:39}
r(\vec x\times\vec u,\vec y\times\vec v)=\sum_{i<k}x_is_i+\sum_{j<l}y_jt_j-r(\vec x)\,r(\vec y),
\end{equation}
and
\[n\ge n_0\implies\xi_n(\vec x\times\vec u,\vec y\times\vec v)
  =\xi_n(\vec x)\,\xi_n(\vec y)\pm{\textstyle2^{-n}\sum_im_i^2\sum_ja_j^2}.\]
In order to prove~\eqref{eq:37} for all $n\ge\lh{k+l}+2+\sum_i\lh{m_i}+\sum_j\lh{a_j}$, we pick $n'\ge\max\{n,n_0\}$
such that $2^{n'}>2^{2n+1}\sum_im_i^2\sum_ja_j^2$, and we apply Lemma~\ref{lem:r-indep}:
\begin{align*}
\xi_n(\vec x\times\vec u,\vec y\times\vec v)
&=\xi_{n'}(\vec x\times\vec u,\vec y\times\vec v)\PM{2^{-n}(k+l)}0\\
&=\xi_{n'}(\vec x)\,\xi_{n'}(\vec y)\PM{2^{-n}(k+l)}0\pm{\textstyle2^{-n'}\sum_im_i^2\sum_ja_j^2}\\
&=\bigl(\xi_n(\vec x)\PM0{2^{-n}k}\bigr)\bigl(\xi_n(\vec y)\PM0{2^{-n}l}\bigr)\PM{2^{-n}(k+l)}0
    \pm{\textstyle2^{-n'}\sum_im_i^2\sum_ja_j^2}\\
&=\xi_n(\vec x)\,\xi_n(\vec y)\pm\bigl(2^{-n}(k+l)+\textstyle2^{-n'}\sum_im_i^2\sum_ja_j^2\bigr)\\
&=\xi_n(\vec x)\,\xi_n(\vec y)\pm\bigl(2^{-n}(k+l)+2^{-2n-1}\bigr).
\end{align*}
Since the terms on both sides are integer multiples of $2^{-2n}$, this implies
\[\xi_n(\vec x\times\vec u,\vec y\times\vec v)=\xi_n(\vec x)\,\xi_n(\vec y)\pm2^{-n}(k+l).\]

It remains to prove~\eqref{eq:36}. We may assume $\Lh(\vec b)=1$, i.e., $\vec b=\p b$. The result is easy to check if
$b=m_i$ or $b=a_j$, hence we may assume $b\perp\vec m,\vec a$. Using Lemma~\ref{lem:inv-rec} again, we compute modulo~$b$:
\begin{align*}
[\vec m]^{-1}[\vec a]^{-1}e(\vec x;b)\,e(\vec y;b)
&\equiv\Bigl(\sum_{i<k}x_ih_im_i^{-1}-r(\vec x)\Bigr)\Bigl(\sum_{j<l}y_jh'_ja_j^{-1}-r(\vec y)\Bigr)\\
&\equiv\sum_{\substack{i<k\\j<l}}x_ih_iy_jh'_jm_i^{-1}a_j^{-1}\\
&\qquad-r(\vec y)\sum_{i<k}x_ih_im_i^{-1}-r(\vec x)\sum_{j<l}y_jh'_ja_j^{-1}+r(\vec x)\,r(\vec y)\\
&\equiv\sum_{\substack{i<k\\j<l}}x_ih_iy_jh'_j\bigl(a_j^{-1}(m_i^{-1}\rem a_j)-m_i^{-1}((-a_j^{-1})\rem m_i)\bigr)\\
&\qquad-r(\vec y)\sum_{i<k}x_ih_im_i^{-1}-r(\vec x)\sum_{j<l}y_jh'_ja_j^{-1}+r(\vec x)\,r(\vec y)\\
&\equiv\sum_{i<k}x_ih_im_i^{-1}\Bigl(-\sum_{j<l}y_jh'_j\bigl((-a_j^{-1})\rem m_i\bigr)-r(\vec y)\Bigr)\\
&\qquad+\sum_{j<l}y_jh'_ja_j^{-1}\Bigl(\sum_{i<k}x_ih_i\bigl(m_i^{-1}\rem a_j\bigr)-r(\vec x)\Bigr)+r(\vec x)\,r(\vec y)\\
&\equiv\sum_{i<k}x_im_i^{-1}(\tilde h_iu_i-m_is_i)+\sum_{j<l}y_ja_j^{-1}(\tilde h'_jv_j-a_jt_j)+r(\vec x)\,r(\vec y)\\
&\equiv\sum_{i<k}x_iu_i\tilde h_im_i^{-1}+\sum_{j<l}y_jv_j\tilde h'_ja_j^{-1}\\
&\qquad-\Bigl(\sum_{i<k}x_is_i+\sum_{j<l}y_jt_j-r(\vec x)\,r(\vec y)\Bigr)\\
&\equiv\sum_{i<k}x_iu_i\tilde h_im_i^{-1}+\sum_{j<l}y_jv_j\tilde h'_ja_j^{-1}-r(\vec x\times\vec u,\vec y\times\vec v)\\
&\equiv[\vec m]^{-1}[\vec a]^{-1}e(\vec x\times\vec u,\vec y\times\vec v;b)
\end{align*}
by~\eqref{eq:39}.
\end{Pf}

\subsection{Chinese remainder reconstruction and iterated products}\label{sec:chin-rema-reconstr}

We now introduce the CRR reconstruction procedure. The definition mostly follows the proof of \cite[Thm.~4.1]{hab},
inlining the construction from \cite[L.~4.5]{hab}. (The latter lemma shows how to compute the CRR of $\fl{X/[\vec a]}$
from the CRR of~$X$; since we cannot yet define what $[\vec a]$ is in the first place, we do not know how to formulate
the lemma in a stand-alone way.)

\begin{Def}\label{def:crr-rec}
(In $\vtc(\imulm)$.) If $\vec x<\vec m$ and $\vec a$ is a subsequence of~$\vec m$, let $\vec x\res\vec a$ denote the
corresponding subsequence of~$\vec x$. (Thus, in fact, $\vec x\res\vec a=e(\vec m;\vec x;\vec a)$.)

Let $\rec(\vec m;\vec x)$ denote the $\Sig0(\card,\imulm)$-definable function formalizing the following algorithm.
Given a nonempty $\vec m\perp2$ and $\vec x<\vec m$, let $s=2+\sum_{i<k}\lh{m_i}$,
and using Theorem~\ref{thm:primes}, let $\vec a=\p{a_{u,j}:u<s,j<l}$ be a sequence of distinct odd primes such that
$\vec a\perp\vec m$ and
\begin{equation}\label{eq:29}
\sum_{j<l}\bigl(\lh{a_{u,j}}-1\bigr)>2s
\end{equation}
for all $u<s$. We write $\vec a_u=\p{a_{u,j}:j<l}$ and $\vec a_{<t}=\p{a_{u,j}:u<t,j<l}$. For each $t\le s$, we define
residue sequences $\vec w_t<\p{\vec m,\vec a_{<t}}$ and $\vec y_t<\vec m$ by
\begin{align*}
\vec w_t&=\Bigl(2^{-t}\prod_{u<t}\bigl(1+[\vec a_u]\bigr)\Bigr)e(\vec m;\vec x;\vec m,\vec a_{<t})\rem\p{\vec m,\vec a_{<t}},\\
\vec y_t&=[\vec a_{<t}]^{-1}\bigl(\vec w_t\res\vec m-e(\vec a_{<t};\vec w_t\res\vec a_{<t};\vec m)\bigr)\rem\vec m,\\
\intertext{and for $t<s$, we define a residue sequence $\vec z_t<\vec m$ and a (possibly negative) number $b_t$ by}
\vec z_t&=(\vec y_t-2\vec y_{t+1})\rem\vec m,\\
b_t&=\begin{cases}-1,&\text{if }\vec z_t\equiv-\vec1\pmod{\vec m},\\z_{t,0}&\text{otherwise.}\end{cases}\\
\intertext{(Here, $z_{t,0}<m_0$ is the $0$th component of $\vec z_t=\p{z_{t,i}:i<k}$.) Finally, we define}
\rec(\vec m;\vec x)&=\sum_{t<s}2^tb_t.
\end{align*}
\end{Def}

To get the basic intuition: in the real world, if $\vec x$ is the CRR of~$X$ in basis~$\vec m$, then $\vec w_t$ is the
CRR of $X\prod_{u<t}\bigl(1+[\vec a_u]\bigr)/2$ in basis $\p{\vec m,\vec a_{<t}}$, and $\vec y_t$ is the CRR of
$\Fl{X\prod_{u<t}\bigl(1+[\vec a_u]\bigr)/\bigl(2[\vec a_u]\bigr)}=\fl{X2^{-t}}$ in basis~$\vec m$ (using the fact that
$[\vec a_u]$ is large enough so that $\bigl(1+[\vec a_u]\bigr)/\bigl(2[\vec a_u]\bigr)$ exceeds $1/2$ only by a
negligible amount). Thus, $\vec z_t$ is the CRR of $\bit(X,t)=b_t$, and $\rec(\vec m;\vec x)=X$.

In particular, in reality $b_t\in\{0,1\}$, whereas our argument in $\vtc(\imulm)$ will only establish that $\vec z_t$
is the CRR of one of $-1$, $0$, $1$, $2$, which is extracted as~$b_t$ (see Lemma~\ref{lem:rec-bits}); a priori,
$\rec(\vec m;\vec x)$ may be negative.

Since we cannot refer in $\vtc(\imulm)$ to the product $X\prod_{u<t}\bigl(1+[\vec a_u]\bigr)/2$ that we do not know to
exist, we base our analysis instead on $\xi_n$ estimation: in particular, we aim to show $\xi_n(\vec
y_t)\approx\xi_n(\vec w_t)\approx2^{-t}\xi_n(\vec x)$. To this end, we first need to rewrite the definition of~$\vec
w_t$ as a recurrence:
\begin{Lem}\label{lem:wt-ind}
$\vtc(\imulm)$ proves: using the notation from Definition~\ref{def:crr-rec},
\begin{align}
\label{eq:41}
e(\vec m,\vec a_{<t};\vec w_t;\vec m,\vec a)
&=e(\vec m;\vec x;\vec m,\vec a)\times\prod_{u<t}e(\vec a_u;2^{-1}\rem\vec a_u;\vec m,\vec a),\\
\label{eq:42}
\vec w_{t+1}
&=e(\vec m,\vec a_{<t};\vec w_t;\vec m,\vec a_{\le t})\times e(\vec a_t;2^{-1}\rem\vec a_t;\vec m,\vec a_{\le t}),
\end{align}
for all $t<s$.
\end{Lem}
\begin{Pf}
By Lemma~\ref{lem:half-odd}, the definition of $\vec w_t$ amounts to
\begin{equation}\label{eq:40}
\vec w_t=e(\vec m;\vec x;\vec m,\vec a_{<t})\times\prod_{u<t}e(\vec a_u;2^{-1}\rem\vec a_u;\vec m,\vec a_{<t}).
\end{equation}
In light of this, for any given~$t$, \eqref{eq:41} implies~\eqref{eq:42}: we have
\begin{align*}
e(\vec m,\vec a_{<t}&;\vec w_t;\vec m,\vec a_{\le t})\times e(\vec a_t;2^{-1}\rem\vec a_t;\vec m,\vec a_{\le t})\\
&=e(\vec m;\vec x;\vec m,\vec a_{\le t})\times\prod_{u<t}e(\vec a_u;2^{-1}\rem\vec a_u;\vec m,\vec a_{\le t})
  \times e(\vec a_t;2^{-1}\rem\vec a_t;\vec m,\vec a_{\le t})\\
&=e(\vec m;\vec x;\vec m,\vec a_{\le t})\times\prod_{u\le t}e(\vec a_u;2^{-1}\rem\vec a_u;\vec m,\vec a_{\le t})\\
&=\vec w_{t+1}.
\end{align*}

Thus, it suffices to prove \eqref{eq:41} by induction on~$t$. For $t=0$, the statement follows from $\vec w_0=\vec x$.
Assuming \eqref{eq:41} holds for $t$, we also have \eqref{eq:42}, therefore
\begin{align*}
e(\vec m,\vec a_{\le t};\vec w_{t+1};\vec m,\vec a)
&=e\bigl(\vec m,\vec a_{\le t};
      e(\vec m,\vec a_{<t};\vec w_t;\vec m,\vec a_{\le t})\times e(\vec a_t;2^{-1}\rem\vec a_t;\vec m,\vec a_{\le t});
      \vec m,\vec a\bigr)\\
&=e(\vec m,\vec a_{<t};\vec w_t;\vec m,\vec a)\times e(\vec a_t;2^{-1}\rem\vec a_t;\vec m,\vec a)\\
&=e(\vec m;\vec x;\vec m,\vec a)\times\prod_{u<t}e(\vec a_u;2^{-1}\rem\vec a_u;\vec m,\vec a)\times
   e(\vec a_t;2^{-1}\rem\vec a_t;\vec m,\vec a)\\
&=e(\vec m;\vec x;\vec m,\vec a)\times\prod_{u\le t}e(\vec a_u;2^{-1}\rem\vec a_u;\vec m,\vec a)
\end{align*}
by Lemma~\ref{lem:prod}.
\end{Pf}

Now we can estimate $\xi_n(\vec w_t)$ and $\xi_n(\vec y_t)$ using the properties developed in
Section~\ref{sec:auxil-prop-crr}.
\begin{Lem}\label{lem:xi-yt}
$\vtc(\imulm)$ proves: using the notation from Definition~\ref{def:crr-rec}, let $n\ge\lh k+2+\sum_i\lh{m_i}$. Then for all
$t\le s$,
\begin{equation}\label{eq:43}
\xi_n(\vec m;\vec y_t)=2^{-t}\xi_n(\vec m;\vec x)\PM{2^{-n}k+2^{-2s}}{2^{-n}k+\xi_n(\vec m;\vec1)}.
\end{equation}
\end{Lem}
\begin{Pf}
Let us first assume that $n$ is sufficiently large. We start with a bound on~$\xi_n(\vec w_t)$. We have
\[\xi_n(\vec m;\vec w_0)=\xi_n(\vec m;\vec x).\]
By Lemmas \ref{lem:wt-ind}, \ref{lem:prod}, \ref{lem:half-odd}, and~\ref{lem:xi1-ub}, we have
\begin{align*}
\xi_n(\vec m,\vec a_{\le t};\vec w_{t+1})
&=\xi_n(\vec m,\vec a_{<t};\vec w_t)\left(\tfrac12+\xi_n(\vec a_t,2;\vec1)\right)\pm2^{-n}\bigl(k+s(t+1)\bigr)\\
&=\xi_n(\vec m,\vec a_{<t};\vec w_t)\left(\tfrac12\PM{2^{-\sum_j(\lh{a_{t,j}}-1)}}0\right)\pm2^{-n}\bigl(k+s(t+1)\bigr)\\
&=\xi_n(\vec m,\vec a_{<t};\vec w_t)\left(\tfrac12\PM{2^{-2s-1}}0\right)\pm2^{-n}\bigl(k+s(t+1)\bigr),
\end{align*}
thus by induction on~$t\le s$, we obtain
\[\xi_n(\vec m,\vec a_{<t};\vec w_t)=2^{-t}\xi_n(\vec m;\vec x)\PM{2^{1-n}(k+ts)+2^{-2s}}{2^{1-n}(k+ts)}.\]
Notice that
\[\vec w_t-e(\vec a_{<t};\vec w_t\res\vec a_{<t};\vec m,\vec a_{<t})=\p{[\vec a_{<t}]\vec y_t,\vec0},\]
thus by Lemma~\ref{lem:extbymul} and Corollary~\ref{cor:sum}, there is $c_t\in\{0,1\}$ such that
\begin{align*}
\xi_n(\vec m;\vec y_t)&=\xi_n(\vec m,\vec a_{<t};[\vec a_{<t}],\vec0)\\
&=\xi_n(\vec m,\vec a_{<t};\vec w_t)
  -\xi_n\bigl(\vec m,\vec a_{<t};e(\vec a_{<t};\vec w_t\res\vec a_{<t};\vec m,\vec a_{<t})\bigr)
  +c_t\PM{2^{-n}(k+ts)}0
\end{align*}
(for $n$ large enough, $c_t$ is independent of~$n$ due to Lemma~\ref{lem:r-indep}). Now, since
\[e(\vec a_{<t};\vec w_t\res\vec a_{<t};\vec m,\vec a_{<t})
  =e(\vec m;\vec1;\vec m,\vec a_{<t})\times e(\vec a_{<t};\vec w_t\res\vec a_{<t};\vec m,\vec a_{<t}),\]
we have
\begin{align*}
e(\vec a_{<t};\vec w_t\res\vec a_{<t};\vec m,\vec a_{<t})
&\le\xi_n(\vec m;\vec1)\,\xi_n(\vec a_{<t};\vec w_t\res a_{<t})+2^{-n}(k+ts)\\
&\le\bigl(1-2^{-\sum_{u,j}\lh{a_{u,j}}}+2^{-n}(3ts)\bigr)\xi_n(\vec m;\vec1)+2^{-n}(k+ts)\\
&\le\xi_n(\vec m;\vec1)-2^{-\tilde s}+2^{-n}(k+4ts)
\end{align*}
by Lemmas \ref{lem:prod}, \ref{lem:xi-ub-lb}, and~\ref{lem:xi1-ub}, where $\tilde s=s+\sum_{u,j}\lh{a_{u,j}}$. It follows that
\[1-\xi_n(\vec m;\vec1)+2^{-n}(3k)\ge\xi_n(\vec m;\vec y_t)\ge
 c_t-\xi_n(\vec m;\vec1)+2^{-\tilde s}-2^{-n}(k+4ts),\]
which implies $c_t=0$ by considering $n$ large enough so that $2^{-\tilde s}>2^{-n}(4k+4ts)$. Thus,
\begin{align*}
\xi_n(\vec m;\vec y_t)
&=\xi_n(\vec m,\vec a_{<t};\vec w_t)\PM{2^{-n}(k+ts)}{\xi_n(\vec m;\vec1)+2^{-n}(k+4ts)}\\
&=2^{-t}\xi_n(\vec m;\vec x)\PM{2^{-2s}+2^{-n}(3k+3ts)}{\xi_n(\vec m;\vec1)+2^{-n}(3k+6ts)}.
\end{align*}

In order to obtain the bound as stated in the lemma, we use Lemma~\ref{lem:r-indep} as in the proof of
Lemma~\ref{lem:prod}: for sufficiently sufficiently large~$n'$,
\begin{align*}
\xi_n(\vec m;\vec y_t)&=\xi_{n'}(\vec m;\vec y_t)\PM{2^{-n}k}0\\
&=2^{-t}\xi_{n'}(\vec m;\vec x)\PM{2^{-2s}+2^{-n}k+2^{-n'}(3k+3ts)}{\xi_{n'}(\vec m;\vec1)+2^{-n'}(3k+6ts)}\\
&=2^{-t}\xi_n(\vec m;\vec x)\PM{2^{-2s}+2^{-n}k+2^{-n'}(3k+3ts)}{\xi_n(\vec m;\vec1)+2^{-n}k+2^{-n'}(3k+6ts)}.
\end{align*}
For large enough $n'$, we may drop the $2^{-n'}(3k+6ts)$ terms, as all the remaining terms are integer multiples of
$2^{-z}$ for $z=\max\{n+t,2s\}$.
\end{Pf}

The next task is to make sense of $\vec z_t$ and~$b_t$: the basic idea is to derive
$\xi_n(\vec z_t)=O\bigl(\xi_n(\vec1)\bigr)$ from the bounds on $\xi_n(\vec y_t)$, and then use discreteness of the
$\xi_n$ values (Lemma~\ref{lem:xi-ub-lb}) to infer that $\vec z_t$ is the CRR of an $O(1)$ integer, which is~$b_t$.
\begin{Lem}\label{lem:rec-bits}
$\vtc(\imulm)$ proves: using the notation from Definition~\ref{def:crr-rec}, $\vec y_0=\vec x$, $\vec y_s=\vec0$, and
for each $t<s$, we have $b_t\in\{-1,0,1,2\}$ and $\vec z_t=b_t\rem\vec m$. Moreover,
\begin{equation}\label{eq:45}
\xi_n(\vec m;\vec y_t)=2\xi_n(\vec m;\vec y_{t+1})+b_t\xi_n(\vec m;\vec1)\PM{2^{-n}k}{2^{-n}(3k)}
\end{equation}
for $n\ge\lh k+2+\sum_i\lh{m_i}$.
\end{Lem}
\begin{Pf}
The first identity follows immediately from the definition. By Lemmas \ref{lem:xi-yt} and~\ref{lem:xi1-ub},
\[\xi_n(\vec y_s)\le2^{-s}+2^{-2s}+2^{-n}k<2^{2-s}-2^{-n}(3k)<\xi_n(\vec1)-2^{-n}(3k)\]
for large enough~$n$, which implies $\vec y_s=\vec0$ by Lemma~\ref{lem:xi-ub-lb}.

Let $t<s$. By Corollary~\ref{cor:sum} and Lemma~\ref{lem:xi-yt}, we have
\[\xi_n(2\vec y_{t+1})=2\xi_n(\vec y_{t+1})\PM0{2^{-n}k}
=2^{-t}\xi_n(\vec x)\PM{2^{1-2s}+2^{1-n}k}{2\xi_n(\vec1)+2^{-n}(3k)}\]
(the right-hand side is $<1$ for $n$ large enough, hence the constant~$c$ from Lemma~\ref{lem:sum} cannot be~$1$). Using
Corollary~\ref{cor:sum} again, there is $c_t\in\{0,1\}$ (independent of~$n$ if $n$ is large enough) such that
\[\xi_n(\vec z_t)=\xi_n(\vec y_t)-\xi_n(2\vec y_{t+1})+c_t\PM{2^{-n}k}0
  =c_t\PM{2\xi_n(\vec1)+2^{-2s}+2^{-n}(5k)}{\xi_n(\vec1)+2^{1-2s}+2^{-n}(3k)},\]
thus for large enough~$n$, we have
\[\Bigl(c_t=0\quad\text{and}\quad\xi_n(\vec z_t)\le\frac52\xi_n(\vec1)\Bigr)
  \qquad\text{or}\qquad
  \Bigl(c_t=1\quad\text{and}\quad\xi_n(\vec z_t)\ge1-\frac32\xi_n(\vec1)\Bigr).\]
We claim that this implies
\begin{equation}\label{eq:44}
\vec z_t=-\vec1\rem\vec m\qquad\text{or}\qquad
   \vec z_t=\vec0\qquad\text{or}\qquad\vec z_t=\vec1\qquad\text{or}\qquad\vec z_t=\vec2.
\end{equation}
Assume first $\xi_n(\vec z_t)\le\frac52\xi_n(\vec1)$. Either $\vec z_t=\vec0$ and we are done, or
\[\xi_n(\vec z_t)\ge\xi_n(\vec 1)-2^{-n}(3k)\]
by Lemma~\ref{lem:xi-ub-lb}, and $\vec z'_t=\vec z_t-\vec1$ satisfies
\[\xi_n(\vec z'_t)=\xi_n(\vec z_t)-\xi_n(\vec1)+c\PM{2^{-n}k}0\ge c-2^{-n}(3k)\]
for some $c\in\{0,1\}$ by Corollary~\ref{cor:sum}. For $n$ large enough, $c=1$ is ruled out by Lemma~\ref{lem:xi-ub-lb}, hence
$c=0$, and
\[\xi_n(\vec z'_t)\le\frac32\xi_n(\vec1)+2^{-n}k.\]
Repeating the same argument, either $\vec z'_t=\vec0$ and $\vec z_t=\vec1$, or $\vec z''_t=\vec z'_t-\vec1$ satisfies
\[\xi_n(\vec z''_t)\le\frac12\xi_n(\vec1)+2^{1-n}k,\]
in which case we must have $\vec z''_t=\vec0$ by Lemma~\ref{lem:xi-ub-lb}, hence $\vec z_t=\vec2$.

If $\xi_n(\vec z_t)\ge1-\frac32\xi_n(\vec1)$, a similar argument yields $\vec z_t\equiv-\vec1\pmod{\vec m}$.

Now, \eqref{eq:44} immediately gives $b_t\in\{-1,0,1,2\}$ and $\vec z_t\equiv b_t\vec1\pmod{\vec m}$. Moreover,
Lemma~\ref{lem:sum} gives
\begin{align*}
\xi_n(\vec2)&=2\xi_n(\vec1)\PM0{2^{-n}k},\\
\xi_n(-\vec1)&=1-\xi_n(\vec1)\PM{2^{-n}k}0,
\end{align*}
and then
\begin{align*}
\xi_n(\vec y_t)&=\xi_n(2\vec y_{t+1})+\xi_n(b_t\vec1)-c_t\PM0{2^{-n}k}\\
&=2\xi_n(\vec y_{t+1})+\xi_n(b_t\vec1)-c_t\PM0{2^{-n}(2k)}\\
&=2\xi_n(\vec y_{t+1})+b_t\xi_n(\vec1)\PM{2^{-n}k}{2^{-n}(3k)}
\end{align*}
follows.%
\footnote{A subtle point here is that we rely on $-\vec1\nequiv\vec2\pmod{\vec m}$: otherwise, if $c_t=0$ and $\vec
z_t=\vec2$, then Definition~\ref{def:crr-rec} makes $b_t=-1$ rather than $b_t=2$, in which case $b_t\xi_n(\vec1)$ is off
by~$1$ from $\xi_n(b_t\vec1)-c_t$ in the argument above. That is, the given proof only works unless $k=1$ and $m_0=3$.
However, in the latter case, all the numbers involved are standard, and one can check that in actual reality,
always $b_t\in\{0,1\}$, hence the bad case does not arise.}
We did not pay attention to how large $n$ need to be, but we can make sure it holds for $n\ge\lh k+2+\sum_i\lh{m_i}$
using Lemma~\ref{lem:r-indep} as above.
\end{Pf}

We are ready to prove that CRR reconstruction works.
\begin{Thm}\label{thm:crr-rec}
$\vtc(\imulm)$ proves: if $\vec m$ is a nonempty sequence of distinct odd primes, and $\vec x<\vec m$, then
$X=\rec(\vec m;\vec x)$ satisfies $0\le X<2^{\sum_i\lh{m_i}}$ and $\vec x=X\rem\vec m$.
\end{Thm}
\begin{Pf}
Using the notation from Definition~\ref{def:crr-rec}, we define
\[Y_t=\sum_{u<s-t}2^ub_{t+u}\]
for all $t\le s$, where $b_t\in\{-1,0,1,2\}$ by Lemma~\ref{lem:rec-bits}. Clearly, $Y_s=0$, and we see that
\begin{equation}\label{eq:46}
Y_t=2Y_{t+1}+b_t
\end{equation}
for $t<s$. By the definition of~$\vec z_t$ and Lemma~\ref{lem:rec-bits}, we have $\vec y_s=\vec0$ and
\[\vec y_t\equiv2\vec y_{t+1}+b_t\vec1\pmod{\vec m}\]
for $t<s$, hence by reverse induction on~$t$, we obtain
\[\vec y_t=Y_t\rem\vec m.\]
In particular, $Y_0=X$ satisfies $\vec x=X\rem\vec m$.

At this point, $X$ may be negative; we only know $-2^s<X<2^{s+1}$. However, combining \eqref{eq:46}
with~\eqref{eq:45}, we obtain for large enough~$n$
\[\xi_n(\vec y_t)=Y_t\xi_n(\vec1)\pm2^{s-t-n}(3k)\]
by reverse induction on~$t$, hence in particular
\[\xi_n(\vec x)=X\xi_n(\vec1)\pm2^{s-n}(3k).\]
This ensures $X\ge0$, and in view of Lemma~\ref{lem:xi1-ub}, also $X<2^{\sum_i\lh{m_i}}$.
\end{Pf}

\begin{Cor}\label{cor:crr-rec}
$\vtc(\imulm)$ proves: if $\vec m$ is a nonempty sequence of distinct odd primes, and $\vec x=X\rem\vec m$, where $\lh
X<\sum_{i<k}\bigl(\lh{m_i}-1\bigr)$, then $\rec(\vec m;\vec x)=X$.
\end{Cor}
\begin{Pf}
Let $h=\lh X$. For large enough $n$, we have
\[\xi_n(\vec y_h)\le(1-2^{-h})\xi_n(\vec1)+2^{-2s}+2^{-n}k<\xi_n(\vec1)-2^{-n}(3k)\]
by Lemmas \ref{lem:xi-yt} and~\ref{lem:Xcrr}, thus $\vec y_h=\vec0$ by Lemma~\ref{lem:xi-ub-lb}. Likewise, $\vec y_t=\vec0$ for all
$t\ge h$, thus $b_t=0$ for $t\ge h$. It follows that $\rec(\vec m,\vec x)<2^{h+1}$, hence
$X\equiv\rec(\vec m;\vec x)\pmod{\vec m}$ implies $X=\rec(\vec m;\vec x)$ by Corollary~\ref{cor:crr-inj}.
\end{Pf}

It is now straightforward to infer $\imul$: we can compute $\prod_{i<n}X_i$ by performing the iterated product in CRR
and applying~$\rec$; the soundness of the reconstruction procedure easily implies that the result satisfies the
required recurrence.
\begin{Thm}\label{thm:imulm-imul}
$\vtc(\imulm)$ proves $\imul$.
\end{Thm}
\begin{Pf}
Given a sequence $\p{X_i:i<n}$, let us fix a sequence of distinct odd primes $\vec m$ such that
\begin{equation}\label{eq:59}
\sum_{i<k}\bigl(\lh{m_i}-1\bigr)>\sum_{i<n}\lh{X_i}
\end{equation}
using Theorem~\ref{thm:primes}. For each $i<n$, let $\vec x_i=X_i\rem\vec m$, and for each $u\le v\le n$, we define
\begin{align*}
\vec y_{u,v}&=\prod_{i=u}^{v-1}\vec x_i\rem\vec m,\\
Y_{u,v}&=\rec(\vec m;\vec y_{u,v})
\end{align*}
(this is elementwise modular product). Clearly, $\vec y_{u,u}=\vec1$, hence $Y_{u,u}=1$ by Corollary~\ref{cor:crr-rec}. For any
fixed $u\le n$, we prove
\begin{equation}\label{eq:48}
\lh{Y_{u,v+1}}\le\sum_{i=u}^v\lh{X_i}\qquad\text{and}\qquad Y_{u,v+1}=Y_{u,v}\cdot X_v
\end{equation}
by induction on $v=u,\dots,n-1$: for $v=u$, we have $\vec y_{u,u+1}=\vec x_u$, hence $Y_{u,u+1}=X_u$ by
Corollary~\ref{cor:crr-rec}. Assuming \eqref{eq:48} holds for $v-1$, we have
\[\lh{Y_{u,v}X_v}\le\lh{Y_{u,v}}+\lh{X_v}\le\sum_{i=u}^v\lh{X_i}<\sum_{i<k}\bigl(\lh{m_i}-1\bigr),\]
and
\[\vec y_{u,v+1}=\vec y_{u,v}\times\vec x_v\equiv Y_{u,v}X_v\pmod{\vec m}\]
by Theorem~\ref{thm:crr-rec}, hence
\[Y_{u,v}X_v=\rec(\vec m;\vec y_{u,v+1})=Y_{u,v+1}\]
by Corollary~\ref{cor:crr-rec}, which gives \eqref{eq:48} for~$v$.

Thus, $\p{Y_{u,v}:u\le v\le n}$ witnesses that $\imul$ holds.
\end{Pf}

For purposes of the next section, it will be convenient to observe that Theorem~\ref{thm:imulm-imul} also gives a proof of
$\imul$ in the basic theory corresponding to logspace:
\begin{Cor}\label{cor:vl-imul}
$\vl$ proves $\imul$.
\end{Cor}
\begin{Pf}
Since $\vl$ is a CN theory and includes~$\vtc$, it suffices to show $\vl\vdash\tot_\imulm$. Now, $\tot_{\MF{Iter}}$
clearly implies its variant where we start the iteration at a different element than~$0$, and then we can construct the
sequence witnessing the computation of $\prod_{i<n}a_i\rem m$ by iterating the function $F(\p{i,x})=\p{i+1,xa_i\rem
m}$ starting from~$\p{0,1}$.
\end{Pf}

\section{The polylogarithmic cut}\label{sec:polylogarithmic-cut}

After putting iterated multiplication in $\tc(\powm)$, Hesse, Allender, and Barrington~\cite{hab} go on to
show that iterated multiplication restricted to \emph{polylogarithmically small inputs} is in~$\Ac$, essentially by
proving that $\Ac$ includes the polylogarithmically scaled-down version of~$\tc(\powm)$. In fact, although they do not
state it that way, this is a consequence of Nepomnja\v s\v cij's theorem~\cite{nepom}, which implies more generally
that $\Ac$ includes the polylogarithmically scaled-down version of~$\cxt L$, and even~$\cxt{NL}$ (which is
essentially $\cxt{NSPACE}(\log\log n)$, as $\log\bigl((\log n)^{O(1)}\bigr)=O(\log\log n)$).

The counterpart of such scaling-down arguments in arithmetic is the following model-theoretic construction:
\begin{Def}\label{def:polylog}
If $\cM=\p{M_1,M_2,\in,\lh\cdot,0,1,+,\cdot,<}$ is a model of~$V^0$, the \emph{polylogarithmic cut $\cM_\plog$
of~$\cM$} is the substructure of~$\cM$ with first-order and second-order domains
\begin{align*}
M_{\plog,1}&=\{x\in M_1:\exists c\in\omega\:\cM\model\exists z\:x\le\lh z^c\},\\
M_{\plog,2}&=\{X\in M_2:\lh X\in M_{\plog,1}\}=\{X\in M_2:X\sset M_{\plog,1}\}.
\end{align*}
\end{Def}

By formalizing Nepomnja\v s\v cij's construction, M\"uller~\cite{muller:polylog} proved that polylogarithmic cuts of
models of~$V^0$ are models of~$\vnc$ (see \cite{cook-ngu} for a definition):
\begin{Thm}[M\"uller~\cite{muller:polylog}]\label{thm:plog-vnc1}
If $\cM\model V^0$, then $\cM_\plog\models\vnc$.
\noproof\end{Thm}
In fact, earlier Zambella~\cite{zamb:endext} effectively proved that polylogarithmic cuts are even models of the
stronger theory~$\vl$, though the result was presented in a different way. For definiteness, we include a
self-contained proof while strengthening the theory further to $\vnl$, again following the idea of Nepomnja\v s\v
cij~\cite{nepom}. A similar formalization of Nepomnja\v s\v cij's theorem in $\idz(\alpha)$ was given by
Atserias~\cite{ats:wphp-conf,ats:wphp}.
\begin{Thm}\label{thm:plog-vl}
If $\cM\model V^0$, then $\cM_\plog\models\vnl$.
\end{Thm}
\begin{Pf}
Work in~$V^0$. Let $0<a\le\lh z^c$ and $E\sset[0,a]\times[0,a]$. For $l=0,\dots,2c$, We define $\Sig0$~formulas
$\fii_l(d,s,t)$ with parameter~$E$ that express $E$-reachability in $\le d\le w^l$ steps, where $w=\cl{\lh
z^{1/2}}$:
\begin{align*}
\fii_0(d,s,t)&\EQ d\le1\land s\le a\land t\le a\land\bigl(s=t\lor(d=1\land E(s,t))\bigr),\\
\fii_{l+1}(d,s,t)&\EQ\exists\p{x_i:i\le k}\:\bigl(k<w\land kw^l\le d\land\forall i\le k\:x_i\le a\\
&\qquad\qquad\land x_0=s\land\forall i<k\:\fii_l(w^l,x_i,x_{i+1})\land\fii_l(d-kw^l,x_k,t)\bigr).
\end{align*}
Notice that (using our efficient sequence encoding) the sequence quantified in the definition of~$\fii_{l+1}$ has
bit-length $O\bigl(k+\sum_{i\le k}\lh{x_i}\bigr)=O(w\lh a)=O(\lh z^{1/2}\lh{\lh z})=O(\lh z)$, hence it can be encoded
by a small number bounded by a polynomial in~$z$, thus the formulas $\fii_l$ are indeed~$\Sig0$.

By (meta)induction on~$l$, we claim that $V^0$ proves
\begin{gather}
\label{eq:49}
\fii_l(d,s,t)\to d\le w^l\land s\le a\land t\le a,\\
\label{eq:50}
\forall s,t\le a\:\bigl(\fii_l(0,s,t)\eq s=t\bigr),\\
\label{eq:51}
\forall d<w^l\:\forall s,t\le a\:\bigl(\fii_l(d+1,s,t)
    \eq\exists u\le a\:\bigl[\fii_l(d,s,u)\land\bigl(u=t\lor E(u,t)\bigr)\bigr]\bigr).
\end{gather}
The properties \eqref{eq:49} and~\eqref{eq:50} are straightforward. We have \eqref{eq:51} for $l=0$ from the
definition of~$\fii_0$. Assuming \eqref{eq:51} holds for~$l$, we prove it for $l+1$.

Left to right: if $\fii_{l+1}(d+1,s,t)$, let $\vec x=\p{x_i:i\le k}$ be the sequence that witnesses the definition.
By~\eqref{eq:49}, we have $kw^l\le d+1\le(k+1)w^l$; if $d+1=kw^l$, we may drop the last element $x_k=t$ from~$\vec x$
and the definition will still be satisfied, hence we may assume $kw^l\le d<(k+1)w^l$. By \eqref{eq:51} for~$l$,
$\fii_l(d+1-kw^l,x_k,t)$ implies $\fii_l(d-kw^l,x_k,u)$ for some~$u\le a$ such that $u=t$ or $E(u,t)$. Then $\vec x$
witnesses that $\fii_{l+1}(d,x_k,u)$ holds.

For the right-to-left implication, we reverse the process: if $\vec x$ witnesses $\fii_{l+1}(d,s,u)$, where $u=t$ or
$E(u,t)$, we can ensure $kw^l\le d<(k+1)w^l$ by extending $\vec x$ with $u$ if necessary; then $\fii_l(d-kw^l,x_k,u)$
implies $\fii_l(d+1-kw^l,x_k,t)$ by \eqref{eq:51} for~$l$, whence $\vec x$ witnesses $\fii_{l+1}(d+1,s,t)$.

It follows that
\[Y=\bigl\{\p{d,u}:d,u\le a\land\fii_{2c}(d,0,u)\bigr\},\]
which exists by $\Sig0\text-\comp$, witnesses the truth of~$\tot_{\MF{Reach}}$ (the defining axiom of~$\vnl$)  in the
polylogarithmic cut.
\end{Pf}

\begin{Cor}\label{cor:plog-vl}
If $\vnl$ proves $\forall X\,\fii(X)$, where $\fii\in\Sigma^1_1$, then
\[V^0\vdash\forall z\:\forall X\:\bigl(\lh X\le\lh z^c\to\fii(X)\bigr)\]
for every constant~$c$.
\end{Cor}
\begin{Pf}
$\Sigma^1_1$ formulas are preserved upwards from cuts.
\end{Pf}

\begin{Cor}\label{cor:vl-imul-plog}
$V^0$ proves $\forall w\,\imul\bigl[\lh w^c\bigr]$, $\forall w\,\tot^*_\Divf\bigl[\lh w^c\bigr]$, and
$\forall w\,\tot^*_\imulm\bigl[\lh w^c,-\bigr]$ (even modulo arbitrary $m>0$, not just primes) for every constant~$c$.
\end{Cor}
\begin{Pf}
$\vl\sset\vnl$ proves $\imul$, hence $\Div$, by Corollary~\ref{cor:vl-imul}, hence $V^0$ proves $\imul\bigl[\lh w^c\bigr]$ and
$\tot^*_\Divf\bigl[\lh w^c\bigr]$ by Corollary~\ref{cor:plog-vl}. Then $\tot^*_\imulm\bigl[\lh w^c,-\bigr]$ also follows:
given $m$ and $\p{x_i:i<n}$ where $n\le\lh w^c$ and $w\ge\max_ix_i$, we can compute $Y=\prod_{i<n}x_i$ using
$\imul\bigl[\lh w^{c+1}\bigr]$, and $Y\rem m$ using $\tot^*_\Divf\bigl[\lh w^{c+1}\bigr]$.
\end{Pf}
\begin{Rem}\label{rem:imul-plog}
Using the arguments in Corollary~\ref{cor:vl-imul} and Theorem~\ref{thm:plog-vl}, it is easy to prove in~$V^0$ directly
$\tot_\imulm^*$ restricted to products $\prod_{i<n}a_i\rem m$ where $n\le\lh w^c$ and $\lh m\le\lh w^{1-\ep}$ for some
constant $\ep>0$. However, a nontrivial result like Theorem~\ref{thm:imulm-imul} seems to be required to get to
larger~$m$.

As a consequence of Corollary~\ref{cor:vl-imul-plog}, $\prod_{i<\min\{n,\lh w^c\}}a_i\rem m$ is in $\ob{V^0}$ definable by an
$L_{\ob{V^0}}$~function $f_c(A,n,m,w)$ (where $A$ encodes $\p{a_i:i<n}$), and consequently, $\Sig0(f_c)=\Sig0$
over~$\ob{V^0}$. In other words, we may, and will, use modular products of polylogarithmic length freely in
$\Sig0$~formulas.
\end{Rem}

\section{Modular exponentiation}\label{sec:modular-powering}

While \cite{hab} show modular powering $a^r\rem m$ of small integers to be in~$\Ac$, we do not know how to prove the
corresponding result in~$V^0$; instead, we will work in the theory $V^0+\wphp\sset\vtc$.

The argument in~\cite{hab} involves computation with $a^{\fl{n/d}}$, where $n=m-1$ is the size of the group, and $d$ a
logarithmically small prime. This means it suffers from chicken-vs-egg problems as the analysis of the modular powering
algorithm needs powering with non-polylogarithmic exponents, which is only defined after the modular powering algorithm
is proved to work. Moreover, the expression of $a^{\fl{n/d}}$ in terms of $(a^{-n\rem d})^{1/d}$ relies on Fermat's
little theorem, which again cannot be stated, let alone proved, without having a means to express $a^n$ in the group.
(Actually, Fermat's little theorem is not even known to be provable in the theory $V_0+\Omega_1\Sset V_0+\wphp$, which
\emph{can} define modular exponentiation with no difficulty; it appears that the strong pigeonhole principle is
required to prove it. See \cite[\S4]{ej:flt}.) 

It turns out we can avoid both problems by using a modified (and arguably simpler) algorithm that exploits the basic
idea of~\cite{hab}, viz.\ Chinese remaindering of exponents, more directly. We formulate the results for prime moduli
here, but this is only to simplify the bounds; the construction as such works for any finite abelian group.

First, we need to make sure there are enough polylogarithmically small primes $d$ such that $x\mapsto x^d$ is a
bijection on $\zg m$. (In the real world, these are exactly the primes not dividing $m-1$.) We obtain this with two
applications of $\wphp$: one ensures that $x\mapsto x^d$ is surjective whenever it is injective, and the other shows
that the number of primes~$d$ for which it is not injective (i.e., such that $\zg m$ contains an element of order~$d$)
is quite limited, essentially because $\zg m$ contains a subgroup whose order is the product of all such ``bad'' primes.
\begin{Lem}\label{lem:inj-surj}
For any constant~$c$, $V^0+\wphp$ proves: if $m$ and $d\le\lh w^c$ are primes such that $x^d\nequiv1\pmod m$ for all
$1<x<m$, then for all $y$ coprime to~$m$, there exists a unique $x<m$ such that $x^d\equiv y\pmod m$. We will write
$x=y^{1/d}$.
\end{Lem}
\begin{Pf}
Since $x\mapsto x^d$ is a group homomorphism, the fact that it has trivial kernel implies it is injective. Assume for
contradiction that it is not surjective, and fix $y$ outside its image. Since $m$ is prime, the residues coprime to~$m$
comprise the interval $[1,m-1]$. Thus, we can define an injective function $F\colon\{0,1\}\times[1,m-1]\to[1,m-1]$ by
$F(u,x)=y^ux^d\rem m$, contradicting $\php^{2(m-1)}_{m-1}$.
\end{Pf}
\begin{Lem}\label{lem:bad-primes}
For any constant~$c$, $V^0+\wphp$ proves: if $m$ is a prime, and $\p{d_i:i<k}$ a sequence of distinct primes $d_i\le\lh
w^c$ such that for each~$i$, $x\mapsto x^{d_i}\rem m$ is not a bijection on~$\zg m$, then
$\sum_{i<k}\bigl(\lh{d_i}-1\bigr)\le\lh m$.
\end{Lem}
\begin{Pf}
Using Lemma~\ref{lem:inj-surj}, for each~$i$, let $x_i$ be the least number in $[2,m-1]$ such that $x_i^{d_i}\equiv1\pmod
m$. (This is $\Sig0$~definable, hence $\p{x_i:i<k}$ exists.)

Notice that using $x_i^{d_i}\equiv1$ and $\tot^*_\imulm\bigl[\lh w^c,-\bigr]$, we can define $x_i^u\rem m$ for arbitrary $u$ as
$x_i^{u\rem d_i}\rem m$; this will satisfy $x_i^{u+v}\equiv x_i^ux_i^v\pmod m$ by induction on~$v$. Since $d_i$ is
prime and $x_i\nequiv1$, we have $x_i^u\equiv1$ only if $d_i\mid u$.

Assume first that $\sum_{i<k}\lh{d_i}\le2\lh m+c\lh{\lh w}$, thus $d=\prod_{i<k}d_i$ exists, and $d\le2^{c+2}m^2\lh w^c$
is a small number. Using $\tot^*_\imulm\bigl[k\lh w^c,-\bigr]$, we can define a function $F\colon[0,d)\to[1,m-1]$ by
$F(u)=\prod_ix_i^{u_i}\rem m$, where $u_i=\fl{u/\prod_{j<i}d_j}\rem d_i$ (that is, we use $[0,d)$ to encode
$\prod_i[0,d_i)$). We claim that $F$ is injective, hence $d<2m$ by~$\php^{2m}_m$, which implies
\begin{equation}\label{eq:47}
\sum_{i<k}\bigl(\lh{d_i}-1\bigr)\le\lh{2m}-1=\lh m
\end{equation}
by~\eqref{eq:1}. Since $F$ is a group homomorphism w.r.t.\ the elementwise sum of sequences modulo~$\vec d$, it suffices
to show that it has trivial kernel. Thus, let $\vec u<\vec d$ be such that $\prod_ix_i^{u_i}\equiv1$. By induction
on~$v$, we can prove
\[\prod_{i<k}x_i^{u_iv}\equiv1\]
for all~$v$. In particular, for any $j<k$, taking $v_j=\prod_{i\ne j}d_i$ gives
\[1\equiv\prod_{i<k}x_i^{u_iv_j}\equiv x_j^{u_jv_j},\]
thus $d_j\mid u_jv_j$. Since $v_j$ is coprime to $d_j$, this shows $d_j\mid u_j$, i.e., $u_j=0$; thus, $\vec u=\vec0$,
as $j$ was arbitrary.

If $\sum_i\lh{d_i}>2\lh m+c\lh{\lh w}$, let $k'<k$ be maximal such that $\sum_{i<k'}\lh{d_i}\le2\lh m+c\lh{\lh w}$. By
the proof above, we have $\sum_{i<k'}\bigl(\lh{d_i}-1\bigr)\le\lh m$, thus
\[\sum_{i<k'+1}\lh{d_i}\le2\sum_{i<k}\bigl(\lh{d_i}-1\bigr)+\lh{d_{k'}}\le2\lh m+c\lh{\lh w},\]
contradicting the choice of~$k'$.
\end{Pf}

We now get to the construction of modular exponentiation $a^r\rem m$. As we already mentioned, the basic idea
(following~\cite{hab}) is to express exponents in CRR modulo a list $\vec d$ of polylogarithmic primes such that
$x\mapsto x^{1/d_i}$ is well-defined. Unlike~\cite{hab}, the way we employ this idea here is to define $a^{x/d}$
for $x=O(d)$, where $d=\prod_id_i$, using a form of~\eqref{eq:5}. We then extend it to all $x$ by periodicity,
allowing us to define $a^r$ as $a^{(rd)/d}$.
\begin{Thm}\label{thm:pow}
$V^0+\wphp$ proves $\tot^*_\powm$.
\end{Thm}
\begin{Pf}
Since $V^0+\wphp$ is a CN theory, it suffices to prove~$\tot_\powm$. Given a prime~$m$, let $\p{d_i:i<k'}$ be the list%
\footnote{It may not be immediately apparent why we can construct a \emph{sequence} consisting of all these primes.
Note that the $i$th element of the sequence is $\delz$-definable using Theorem~\ref{thm:count-polylog} as the unique prime
$d$ satisfying~\eqref{eq:75} and $\forall x<m\,(x>1\to x^d\nequiv1\pmod m)$ such that there are exactly $i$~smaller
primes with this property.}
of all primes
\begin{equation}\label{eq:75}
d_i\le2\lh m\bigl(\lh{\lh m}+1\bigr)^{17}
\end{equation}
such that $x^{d_i}\nequiv1\pmod m$ for all $x\nequiv1\pmod m$. We have
\[\sum_{d\le2\lh m(\lh{\lh m}+1)^{17}}\bigl(\lh d-1\bigr)\ge2\lh m\]
by Theorems \ref{thm:primes} and~\ref{thm:plog-vl}, hence
\[\sum_{i<k'}\bigl(\lh{d_i}-1\bigr)\ge2\lh m-\lh m=\lh m\]
by Lemma~\ref{lem:bad-primes}. Let $k\le k'$ be smallest such that
\[\sum_{i<k}\bigl(\lh{d_i}-1\bigr)\ge\lh m.\]
Then
\[\sum_{i<k}\bigl(\lh{d_i}-1\bigr)\le\lh m-1+\lh{d_{k-1}}\le\lh m+\lh{\lh m}+17\bigl|\lh{\lh m}+1\bigr|
  =O(\lh m),\]
hence $d=\prod_{i<k}d_i$ exists as a small number, while
\[d\ge2^{\sum_i(\lh{d_i}-1)}\ge2^{\lh m}>m.\]
By Lemma~\ref{lem:inj-surj}, $x\mapsto x^{d_i}\rem m$ is a
bijection on $\zg m$ for each $i<k$. Put $\tilde d_i=\prod_{j\ne i}d_j=d/d_i$.

Let $0<a<m$ be given. For every $r\le2d$, we define
\begin{equation}\label{eq:55}
a^{r/d}=a^{u(r)}\prod_{i<k}(a^{1/d_i})^{u_i(r)}\rem m
\end{equation}
using the notation of Lemma~\ref{lem:inj-surj}, where
\begin{align*}
u_i(r)&=r\tilde d_i^{-1}\rem d_i,\\
u(r)&=\frac1d\Bigl(r-\sum_{i<k}u_i(r)\tilde d_i\Bigr).
\end{align*}
Here,
\[\sum_{i<k}u_i(r)\tilde d_i\equiv u_j(r)\tilde d_j\equiv r\pmod{d_j}\]
for each $j<k$, hence $\sum_{i<k}u_i(r)\tilde d_i\equiv r\pmod d$, i.e., $u(r)$ is an integer, and $-k\le u(r)\le2$,
where $k\le\lh m$. Thus, $a^{r/d}$ can be evaluated using $\tot^*_\imulm\bigl[\lh m^{O(1)},-\bigr]$.

We claim that
\begin{equation}\label{eq:54}
a^{(r+s)/d}\equiv a^{r/d}a^{s/d}\pmod m
\end{equation}
for all $r,s$ such that $r+s\le2d$. Indeed, we have $u_i(r+s)=u_i(r)+u_i(s)-c_id_i$ with $c_i\in\{0,1\}$, hence
$u(r+s)=u(r)+u(s)+\sum_{i<k}c_i$, and
\begin{align*}
a^{(r+s)/d}&\equiv a^{u(r)+u(s)+\sum_ic_i}\prod_{i<k}(a^{1/d_i})^{u_i(r)+u_i(s)-c_id_i}\\
&\equiv a^{u(r)}a^{u(s)}a^{\sum_ic_i}\prod_{i<k}(a^{1/d_i})^{u_i(r)}(a^{1/d_i})^{u_i(s)}a^{-c_i}\\
&\equiv a^{r/d}a^{s/d}.
\end{align*}

Using $\wphp$, there exist $r<s\le2m\le2d$ such that $a^{r/d}=a^{s/d}$. Putting $t=s-r$, we have $0<t\le2m$ and
$a^{t/d}=1$ by~\eqref{eq:54} (which implies $a^{(rt)/d}=1$ for all $r$ such that $rt\le2d$ by induction on~$r$). We then
extend the definition of $a^{r/d}$ to arbitrary small~$r$ by putting
\[a^{r/d}=a^{(r\rem t)/d}.\]
This agrees with the original definition for $r\le2d$ using~\eqref{eq:54}, and the new definition also
satisfies~\eqref{eq:54}. Finally, we define
\[a^r=a^{(rd)/d}.\]
Direct computation shows that $u_i(0)=u_i(d)=0$, $u(0)=0$, and $u(d)=1$, hence $a^{0/d}=1$ and $a^{d/d}=a$. Thus, we
obtain the defining recurrence for $\powm$:
\begin{align*}
a^0&=1,\\
a^{r+1}&=a^ra\rem m.
\end{align*}
We only defined it for $0<a<m$, but we can simply put
\[0^r=\begin{cases}1,&r=0,\\0,&r>0\end{cases}\]
for $a=0$.
\end{Pf}

As in Remark~\ref{rem:imul-plog}, it follows that we can use $\powm$ freely in $\Sig0$~formulas
(as long as we stick to extensions of $V^0+\wphp$):
\begin{Cor}\label{cor:powm-sig0}
$\Sig0(\powm)=\Sig0$ over $\ob{V^0}+\wphp$.
\noproof\end{Cor}

Once we have exponentiation, let us show for further reference that any element of $\zg m$ has a well-defined order,
and that orders have the expected basic properties.
\begin{Lem}\label{lem:order}
$V^0+\wphp$ proves: if $m$ is a prime, then every $0<a<m$ has a unique \emph{order} $0<o_m(a)<2m$ which satisfies
\[a^r\equiv1\pmod m\iff o_m(a)\mid r\]
for all~$r$.
\end{Lem}
\begin{Pf}
Using~$\wphp$, there are $r<r'<2m$ such that $a^r\equiv a^{r'}\equiv a^ra^{r'-r}\pmod m$, thus $r'-r>0$ and
$a^{r'-r}\equiv1\pmod m$ as $a^r$ is invertible. Let $o_m(a)=t$ be the least $t>0$ such that $a^t\equiv1\pmod m$. On
the one hand, this implies $a^{tr}\equiv1\pmod m$ for all~$r$. On the other hand, if $a^r\equiv1\pmod m$, we have
\[1\equiv a^r\equiv a^{(r\rem t)+t\fl{r/t}}\equiv a^{r\rem t}(a^t)^{\fl{r/t}}\equiv a^{r\rem t}\pmod m,\]
hence $r\equiv0\pmod t$ by the minimality of~$t$.
\end{Pf}

We note that $o_m(a)$ is $\Sig0$-definable (using Corollary~\ref{cor:powm-sig0}) as the least $t>0$ such that
$a^t\equiv1\pmod m$.
\begin{Lem}\label{lem:orders}
$V^0+\wphp$ proves that for any prime~$m$ and $0<a,a'<m$:
\begin{enumerate}
\item\label{item:4}
For any~$r$, $o_m(a^r)=o_m(a)/\gcd\{o_m(a),r\}$. Thus, if $r\mid o_m(a)$, then $o_m(a^r)=o_m(a)/r$.
\item\label{item:5}
There exists $0<b<m$ such that $o_m(b)=\lcm\{o_m(a),o_m(a')\}$.
\end{enumerate}
\end{Lem}
\begin{Pf}
\ref{item:4}: Let $t=o_m(a)$ and $d=\gcd\{t,r\}$. Then for any~$s$, $a^{rs}\equiv1$ iff $t\mid rs$ iff
$\frac td\mid\frac rds$ iff $\frac td\mid s$ as $\frac td$ and~$\frac rd$ are coprime.

\ref{item:5}: Put $t=o_m(a)$ and $t'=o_m(a')$. First, we claim that if $\gcd\{t,t'\}=1$, then $o_m(aa')=tt'$: on the
one hand, $(aa')^{tt'}\equiv(a^t)^{t'}(a'^{t'})^t\equiv1$. On the other hand, if $(aa')^r\equiv1$, then
$1\equiv(aa')^{rt'}\equiv a^{rt'}$, hence $t\mid rt'$, which implies $t\mid r$ as $t$ and~$t'$ are coprime. A symmetric
argument gives $t'\mid r$, hence $tt'=\lcm\{t,t'\}\mid r$.

We prove the general case by induction on~$t$. If $t=1$, we may take $b=a'$. Otherwise, $t$ is divisible by a
prime~$p$; write $t=sp^e$ and $t'=s'p^{e'}$, where $p\nmid s,s'$. Since $o_m(a^{p^e})=s<t$ and
$o_m(a'^{p^{e'}})=s'$ by~\ref{item:4}, there exists $b$ such that $o_m(b)=\lcm\{s,s'\}$ by the induction hypothesis.
Moreover, one of $a^s$ and $a'^{s'}$ has order $p^{\max\{e,e'\}}$, hence $ba^s$ or $ba'^{s'}$ has order
$\lcm\{s,s'\}p^{\max\{e,e'\}}=\lcm\{t,t'\}$ by the coprime case.
\end{Pf}

\section{Generators of multiplicative groups}\label{sec:gener-mult-groups}

We could finish the proof of the main result at this point if we could show that $\vtc$ (possibly using, say,
$\tot^*_\powm$ and $\tot^*_\imulm\bigl[\lh w^c,-\bigr]$) proves $\tot_\imulm$. In the real world, iterated
multiplication modulo a prime~$m$ reduces easily to powering modulo~$m$ as $\zg m$ is cyclic, and we can do iterated
sums of the corresponding discrete logarithms in~$\tc$. Thus, it would suffice to prove in $\vtc$ that multiplicative
groups of prime fields are cyclic.

Unfortunately, we do not know how to do that directly. However, as a starting point to further investigation, let us at
least establish that $\imul$ is \emph{equivalent} to the cyclicity of multiplicative groups of prime fields
over~$\vtc$.
\begin{Prop}\label{prop:imul-cyc}
The following are equivalent over~$\vtc$.
\begin{enumerate}
\item\label{item:1} $\imul$.
\item\label{item:2} For all primes~$m$, the groups $\zg m$ are cyclic:
\[\exists g<m\:\forall a<m\:\bigl(a\ne0\to\exists r<m\:g^r\equiv a\pmod m\bigr).\]
\item\label{item:3} For all primes $m$ and~$p$, if $a,b<m$ are such that $a\ne1$ and $a^p\equiv b^p\equiv1\pmod m$,
then $b\equiv a^r\pmod m$ for some~$r<m$.
\end{enumerate}
\end{Prop}
\begin{Pf}

\ref{item:2}\txto\ref{item:1}: In view of Theorem~\ref{thm:imulm-imul}, it suffices to prove $\tot^*_\imulm$. Given a prime
$m$ and $\p{a_i:i<n}$, where w.l.o.g.\ $0<a_i<m$ for each $i<n$, we define $g$ to be the least generator of~$\zg m$, a
sequence $\p{r_i:i<n}$ such that $r_i<m$ is least such that $g^{r_i}\equiv a_i\pmod m$, and a sequence $\p{b_i:i\le n}$
by
\[b_i=g^{\sum_{j<i}r_j}\rem m.\]
Then $b_0=1$ and $b_{i+1}=b_ia_i\rem m$, hence $\vec b$ witnesses that $\tot_\imulm$ holds. Now, we need to
apply this argument in parallel several times to get the aggregate function, but this is not a problem.

\ref{item:3}\txto\ref{item:2}: Let $g$ be an element of $\zg m$ of maximal order, and put $t=o_m(g)$. (While
Lemma~\ref{lem:order} only claims $t<2m$, we have in fact $t<m$, as $\vtc$ implies~$\php^m_{m-1}$.)
Assume for contradiction that $g$ is not a generator of~$\zg m$, and fix $0<a<m$ such that $a\nequiv g^r$ for all
$r<t$. Let $s\le o_m(a)$ be minimal such that $a^s\equiv g^r$ for some $r<t$. Since $s>1$, there is a prime $p\mid s$;
replacing $a$ with $a^{s/p}$ if necessary, we may assume $s=p$. This implies $p\mid o_m(a)$: otherwise
$a\equiv(a^p)^{p^{-1}\rem o_m(a)}\equiv g^r$ for some~$r$, a contradiction.

By Lemma~\ref{lem:orders}~\ref{item:5}, the maximality of~$t$ implies $p\mid o_m(a)\mid t$, thus $b=g^{t/p}$ has
order~$p$. But then $a\equiv b^r\equiv g^{rt/p}$ for some~$r$ by~\ref{item:3}, a contradiction.

\ref{item:1}\txto\ref{item:3}: Let $o_m(a)=p$ and $b^p\equiv1\pmod m$. The basic idea is to use $\imul$ to construct
the polynomials $f_i(x)=\prod_{j<i}(x-a^j)\pmod m$ for $i\le p$, aiming to show $f_p(x)\equiv x^p-1$, which yields
$\prod_{j<p}(b-a^j)\equiv0$.

Fix $n\ge p\lh m$, put $\alpha_j=(-a^j)\rem m$ for $j<p$, and write
\[\prod_{j<i}(2^n+\alpha_j)=\sum_{j\le i}C_{i,j}2^{jn},\qquad0\le C_{i,j}<2^n.\]
By induction on~$i\le p$, we claim that
\begin{align}
\label{eq:52}
\sum_{j\le i}C_{i,j}&\le m^i,\\
\label{eq:53}
C_{i,j}&=\begin{cases}1,&j=i,\\
          C_{i-1,j-1}+\alpha_{i-1}C_{i-1,j},&0<j<i,\\
          \alpha_{i-1}C_{i-1,0},&0=j<i.
         \end{cases}
\end{align}
For $i=0$, \eqref{eq:52} and~\eqref{eq:53} are obvious. Assuming the statements hold
for~$i$, we have
\begin{align*}
\sum_{j\le i+1}C_{i+1,j}2^{nj}&=(2^n+\alpha_i)\sum_{j\le i}C_{i,j}2^{jn}\\
&=C_{i,i}2^{(i+1)n}+\sum_{j=1}^i(C_{i,j-1}+\alpha_iC_{i,j})2^{jn}+\alpha_iC_{i,0}.\numberthis\label{eq:56}
\end{align*}
Here,
\[C_{i,i}+\sum_{j=1}^i(C_{i,j-1}+\alpha_iC_{i,j})+\alpha_iC_{i,0}=(1+\alpha_i)\sum_{j\le i}C_{i,j}\le m\cdot m^i=m^{i+1}\]
by the induction hypothesis, hence also the individual terms in this sum are bounded by $m^{i+1}\le m^p<2^n$. Thus,
matching up the terms in~\eqref{eq:56} gives \eqref{eq:53} for $i+1$, hence also~\eqref{eq:52}.

If we also define $C_{i,j}=0$ for $j<0$ or $j>i$ for notational convenience, \eqref{eq:53} gives the recurrence
\begin{align*}
C_{0,0}&=1,\\
C_{i+1,j}&\equiv C_{i,j-1}-a^iC_{i,j}\pmod m
\end{align*}
for all $j$ and $0\le i<p$, which amounts to saying that $\sum_jC_{i,j}x^j$ is the polynomial
$\prod_{j<i}(x-a^j)$; formally, for any $u<m$, we can prove
\begin{equation}\label{eq:57}
\prod_{j<i}(u-a^j)\equiv\sum_{j\le i}C_{i,j}u^j\pmod m
\end{equation}
by induction on~$i\le p$.

We now wish to formalize the symmetry property $f_p(x)\equiv f_p(ax)$ (or equivalently, $f_p(x)\equiv f_p(a^{-1}x)$),
which will imply that most coefficients of~$f_p$ vanish. To this end, we claim that $C_{i,j}$ satisfies the recurrence
\begin{equation}\label{eq:58}
C_{i+1,j}\equiv a^{i-j+1}C_{i,j-1}-a^{i-j}C_{i,j}
\end{equation}
for all $j$ and $0\le i<p$, which expresses the identity of polynomials
\[\prod_{j\le i}(x-a^j)\equiv a^i(x-1)\prod_{j<i}(a^{-1}x-a^j)\pmod m.\]
Since \eqref{eq:58} holds trivially for $j<0$ or $j>i+1$, and the cases $j=0$ and $j=i+1$ amount to the identities
$C_{i+1,0}\equiv-a^iC_{i,0}$ and $C_{i+1,i+1}\equiv1\equiv C_{i,i}$, it suffices to prove by induction on $i$ that
\eqref{eq:58} holds for all $0<j\le i$. For $i=0$, this statement is vacuous. Assuming it holds for $i$, we prove it for
$i+1$ as follows:
\begin{align*}
C_{i+2,j}&\equiv C_{i+1,j-1}-a^{i+1}C_{i+1,j}\\
&\equiv(a^{i-j+2}C_{i,j-2}-a^{i-j+1}C_{i,j-1})-a^{i+1}(a^{i-j+1}C_{i,j-1}-a^{i-j}C_{i,j})\\
&\equiv a^{i-j+2}C_{i,j-2}-(a^{i-j+1}+a^{2i-j+2})C_{i,j-1}+a^{2i-j+1}C_{i,j}\\
&\equiv a^{i-j+2}(C_{i,j-2}-a^iC_{i,j-1})-a^{i-j+1}(C_{i,j-1}-a^iC_{i,j})\\
&\equiv a^{i-j+2}C_{i+1,j-1}-a^{i-j+1}C_{i+1,j}.
\end{align*}

Applying \eqref{eq:58} with $i=p-1$, we obtain
\begin{align*}
C_{p,j}&\equiv a^{p-j}C_{p-1,j-1}-a^{p-j-1}C_{p-1,j}\\
&\equiv a^{-j}(C_{p-1,j-1}-a^{p-1}C_{p-1,j})\\
&\equiv a^{-j}C_{p,j},\\
\intertext{which implies}
C_{p,j}&\equiv0
\end{align*}
for all $0<j<p$. We also have $C_{p,p}=1$, and then \eqref{eq:57} for $i=p$ and $u=1$ gives
\[0\equiv\sum_{j\le p}C_{p,j}\equiv1+C_{p,0},\]
thus $C_{p,0}\equiv-1$; that is, $f_p(x)\equiv x^p-1$. Then, assuming $b^p\equiv1$, \eqref{eq:57} for $u=b$ gives
\[\prod_{j<p}(b-a^j)\equiv b^p-1\equiv0,\]
whence $b\equiv a^j$ for some $j<p$.
\end{Pf}

The proof of \ref{item:2}\txto\ref{item:1} in Proposition~\ref{prop:imul-cyc} does not quite require the cyclicity of~$\zg m$.
Recalling that (apart from $\powm$) we have $\tot^*_\imulm\bigl[O\bigl(\lh m\bigr),-\bigr]$, it would be enough to find
a ($\Sig0(\card)$-definable) set $X\sset[1,m-1]$ of cardinality $O\bigl(\lh m\bigr)$ such that every element of~$\zg m$
can be written as $\prod_{y\in Y}y\bmod m$ for some $Y\sset X$; in particular, such an~$X$ can be constructed if we can
find a set~$G$ of generators of~$\zg m$ such that $\sum_{a\in G}\lh{o_m(a)}=O\bigl(\lh m\bigr)$.

Ignoring issues of definability, the \emph{structure theorem for finite abelian groups} (stating that any such group
is the product of cyclic groups of prime power orders) ensures that such a generating
set exists in the real world for \emph{every} finite abelian group, obviating the need for a condition
like~\ref{item:3}. The structure theorem for finite abelian groups was proved in~\cite{ej:flt} in the theory
$\sss+\wphp(\sig1)$, which, in our present setup, is a fragment of $V^0+\Omega_1$; unfortunately, the $\Omega_1$ is
needed in the argument not just to prove $\wphp$ (which we have in~$\vtc$ anyway), but also to quantify over subsets
of~$\zg m$ of cardinality $O\bigl(\lh m\bigr)$, and thus of bit-size $O\bigl(\lh m^2\bigr)$. As such, we do not know
how to make the proof work in~$\vtc$.

However, a key insight is that we can smoothly combine this approach with a \ref{item:3}-like condition. Namely, assume
that for a given~$m$, we know \ref{item:3} to hold for $p<x$. Then the argument in \ref{item:3}\txto\ref{item:2}
ensures that $\zg m$ has a cyclic subgroup that includes the $p$-torsion components of~$\zg m$ for all $p<x$, thus, when
looking for other generators as in the structure theorem, we may assume their orders are powers of primes $p\ge x$. In
particular, this restricts the number of generators to about $\lh m/\lh x$, reducing the bit-size of the generating set
to $O\bigl(\lh m^2/\lh x\bigr)$.

We will show below (Lemma~\ref{lem:cyc-imulm}) how to make this idea formal, and use it to break the circular argument in
Proposition~\ref{prop:imul-cyc}: by paying attention to how large numbers are needed in each step, we will see that if we assume
\ref{item:3} to hold up to~$x$, and go around the circle, we end up with \ref{item:3} holding up to something larger
than~$x$, setting the stage for a \emph{coup de gr\^ace} by induction.

\begin{Def}\label{def:bounded-axi}
Let $\cyc[z,x]$ denote condition \ref{item:3} in Proposition~\ref{prop:imul-cyc} restricted to $m\le z$ and $p<x$. Notice that
$\cyc$ is a $\Sig0$~formula.
\end{Def}
\pagebreak[2]
\begin{Lem}\label{lem:imul-cyc}
$\vtc$ proves $\imul\bigl[x^2\lh z\bigr]\to\cyc[z,x]$.
\end{Lem}
\begin{Pf}
The main instance of $\imul$ used in the proof of \ref{item:1}\txto\ref{item:3} in Proposition~\ref{prop:imul-cyc} was
$\prod_{j<p}(2^n+\alpha_j)$, where $n=p\lh m$, thus $\sum_{j<p}\lh{2^n+\alpha_j}=p(n+1)\le(p+1)^2\lh m\le x^2\lh z$.
Moreover, we need products of length~$p$ modulo~$m$ in~\eqref{eq:57}, simulated with $\imul$ followed by division
by~$m$ (using~$\powm$); these instances have size $p\lh m\le x\lh z$.
\end{Pf}
\begin{Lem}\label{lem:imulm-imul}
$\vtc$ proves $\tot^*_\imulm[-,x^3]\to\imul[x]$.
\end{Lem}
\begin{Pf}
We need to examine the usage of $\imulm$ in Section~\ref{sec:chin-rema-repr}. For Subsection~\ref{sec:auxil-prop-crr},
the reader can easily verify that as we already announced at the beginning of~\ref{sec:auxil-prop-crr}, the proof of
each result in Section~\ref{sec:auxil-prop-crr} uses only instances of $\imulm$ modulo primes that actually appear in
the statement of the result (generally $\vec m$, as well as the various $\vec a$ and~$\vec b$); the only place where we
introduce a new auxiliary prime~$p$ to work modulo~$p$ is in Lemma~\ref{lem:half-odd}, where $p=2$, and we can do products
modulo~$2$ already in~$V^0$.

As for Subsection~\ref{sec:chin-rema-reconstr}, all the results up to Corollary~\ref{cor:crr-rec} need only instances of
$\imulm$ modulo $\vec m$ as given in the statements, and modulo the primes $\vec a$ introduced in Definition~\ref{def:crr-rec}.
Finally, the proof of Theorem~\ref{thm:imulm-imul} that we are actually interested in uses $\imulm$ modulo~$\vec m$ as
introduced in the proof, and modulo the corresponding primes~$\vec a$ from Definition~\ref{def:crr-rec} in order to apply the
preceding results.

In order to estimate $\vec m$ and~$\vec a$, let $\sum_{i<n}\lh{X_i}\le x$. The only requirement on~$\vec m$ was that
$\vec m\perp2$ and~\eqref{eq:59}. Now, in view of $\lh2=2$, Theorem~\ref{thm:primes} ensures that it suffices to take
for~$\vec m$ all odd primes up to $(x+2)\lh{x+2}^{17}=O\bigl(x\,\lh x^{17}\bigr)$ as long as $x$ is larger than a
suitable standard constant (which
we may assume w.l.o.g.\ as $\imul[x]$ is trivially provable for each standard~$x$). Going back to
Definition~\ref{def:crr-rec}, we have $s=O(x)$; we claim that in order to find $\vec a$ satisfying the requirements, it
suffices to take the list of all primes below $t=O\bigl(s^2\lh s^{17}\bigr)$, omit $2$ and~$\vec m$, and split it into
sublists $\vec a_u$, $u<s$, of minimal length that satisfy~\eqref{eq:29}. Since the individual primes on the list have
length $O\bigl(\lh s\bigr)$, this will make
\[\sum_{j<l}\bigl(\lh{a_{u,j}}-1\bigr)\le2s+O\bigl(\lh s\bigr)\]
for each $u<s$, while
\[\lh2-1+\sum_{i<k}\bigl(\lh{m_i}-1\bigr)\le s,\]
thus there will be enough primes available as long as
\[\sum_{p\le t}\bigl(\lh p-1\bigr)\ge2s^2+O\bigl(s\,\lh s\bigr),\]
and Theorem~\ref{thm:primes} guarantees that a suitable $t=O\bigl(s^2\lh s^{17}\bigr)=O\bigl(x^2\lh x^{17}\bigr)$ will
satisfy this. This makes $t<x^3$ for $x$ larger than a suitable standard constant.
\end{Pf}
\begin{Lem}\label{lem:cyc-imulm}
For any polynomial~$p$, $\vtc$ proves $\cyc[z,x]\to\tot^*_\imulm\bigl[-,\min\bigl\{z,p(x,\lh z)\bigr\}\bigr]$.
\end{Lem}
\begin{Pf}
Consider a prime $m\le z$ such that $\lh m=O\bigl(\lh x+\lh{\lh z}\bigr)$. As in the proof of
\ref{item:3}\txto\ref{item:2} in Proposition~\ref{prop:imul-cyc}, let $g$ be an element of $\zg m$ of maximal order
$t=o_m(g)<m$. By Lemma~\ref{lem:orders}, $o_m(a)\mid t$ for all $a\in\zg m$. We will expand $\{g\}$ to a
not-too-large generating set by mimicking the proof of \cite[Thm.~3.12]{ej:flt}.

Let us say that $\p{g_i:i<k}$ is a \emph{good independent sequence with exponents} $\p{t_i:i<k}$ if
$\sum_{i<k}\lh{t_i}\le2\lh m$, each $t_i$ is a prime power $p_i^{e_i}$ where $p_i\ge x$, $g_i^{t_i}\equiv1\pmod m$, and
\begin{equation}\label{eq:61}
\forall r<t\:\forall\vec r<\vec t\:\Bigl(g^r\prod_{i<k}g_i^{r_i}\equiv1\pmod m\implies\p{r,\vec r}=\vec0\Bigr).
\end{equation}
Here, the product modulo~$m$ can be evaluated using $\tot^*_\powm$ and~$\tot^*_\imulm\bigl[\lh m,-\bigr]$, and the
conditions on~$\vec t$ ensure that $\vec r$ can be encoded by a bounded first-order quantifier (using the efficient
sequence encoding scheme), hence the definition of good independent sequences is $\Sig0$.

If $\vec g$ is a good independent sequence with exponents $\vec t$, then $t_i=o_m(g_i)<m$ for each $i<k$, and the
mapping
\[\fii_{\vec g}(r,\vec r)=g^r\prod_{i<k}g_i^{t_i}\rem m\qquad(r<t,\vec r<\vec t)\]
is a group homomorphism $C_t\times\prod_{i<k}C_{t_i}\to\zg m$ with a trivial kernel; as such, it is injective.
Moreover, $\fii_{\vec g}$ is $\Sig0$-definable, hence it exists as a set. Since $t_i\ge x$, we have $k\le2\lh m/\lh x$;
it follows that the sequence $\vec g$ can be encoded using
$O\bigl(k\,\lh m\bigr)=O\bigl(\lh m^2/\lh x\bigr)=O\bigl(\lh z\bigr)$ bits, that is, by a bounded first-order
variable. Consequently, we can use bounded $\Sig0$-maximization to find a good independent sequence $\vec g$ such that
$\sum_{i<k}\lh{t_i}$ is maximal possible.

We claim that $\fii_{\vec g}$ is surjective. Assume for contradiction that $b\notin\im(\fii_{\vec g})$. Since
$\im(\fii_{\vec g})$ is $\Sig0$-definable, there exists a least $r>0$ such that $b^r\in\im(\fii_{\vec g})$. We have
$r>1$, thus $r$ has a prime divisor~$p$. By replacing $b$ with $b^{r/p}$ if necessary, we may assume $r=p$. Thus, we
can write
\[b^p=g^s\prod_{i<k}g_i^{s_i}\]
for some $s<t$ and $\vec s<\vec t$. We define $s'<t$, $\vec s'<\vec t$, and $b'=g^{s'}\prod_ig_i^{s'_i}$ as follows:
\begin{itemize}
\item Since $p\mid o_m(b)\mid t$, we have $g^{\frac tps}\prod_ig_i^{\frac tps_i}\equiv1$, thus $t\mid\frac tps$ by
independence, that is, $p\mid s$. We put $s'=s/p$ so that $g^{s'p}=g^s$.
\item For any $i<k$ such that $p_i\ne p$, let $s'_i=s_ip^{-1}\rem t_i$, so that $g_i^{s'_ip}\equiv g_i^{s_i}$.
\item For any $i<k$ such that $p_i=p$ and $p\mid s_i$, we put $s'_i=s_i/p$ so that $g_i^{s'_ip}=g_i^{s_i}$.
\item Otherwise, $s'_i=0$.
\end{itemize}
Since $b'=\fii_{\vec g}(s',\vec s')$, $bb'^{-1}\rem m$ is still outside $\im(\fii_{\vec g})$, while
$(bb'^{-1})^p$ is inside. Thus, we may replace $b$ with $bb'^{-1}$; this ensures $s=0$, and
\begin{equation}\label{eq:63}
s_i\ne0\implies p=p_i\land p\nmid s_i
\end{equation}
for each $i<k$. We distinguish two cases.

If $\vec s=\vec0$, then $b^p\equiv1$. We claim that $\p{\vec g,b}$ is a good independent sequence with exponents
$\p{\vec t,p}$, contradicting the maximality of $\sum_i\lh{t_i}$. Since $p\mid t$, the elements $b$ and $a=g^{t/p}$
have both order~$p$, while $b$ cannot be a power of~$a$ as it is outside $\im(\fii_{\vec g})$; thus, $\cyc[z,x]$
implies $p\ge x$. The independence of~$\vec g$ together with $b^i\notin\im(\fii_{\vec g})$ for $0<i<p$ implies that
$\p{\vec g,b}$ satisfies~\eqref{eq:61}. This means that $\fii_{\vec g,b}$ is injective, hence $pt\prod_it_i$ (which
exists by Theorem~\ref{thm:imul-log}) is less than~$m$ by~$\php$; in particular,
\[\lh p+\sum_{i<k}\lh{t_i}\le2\Bigl(\lh p-1+\sum_{i<k}\bigl(\lh{t_i}-1\bigr)\Bigr)<2\lh m,\]
as required by the definition of a good independent sequence.

If $\vec s\ne\vec0$, let $i_0<k$ be such that $s_{i_0}\ne0$ (thus $p_{i_0}=p$ and $p\nmid s_{i_0}$ by~\eqref{eq:63}),
and such that $e_{i_0}$ is maximal possible among these. Without loss of generality, assume $i_0=0$. We claim that
$\p{b,g_1,\dots,g_{k-1}}$ is a good independent sequence with exponents $\p{pt_0,t_1,\dots,t_{k-1}}$, again
contradicting the maximality of~$\vec g$. The maximality of~$e_0$ along with~\eqref{eq:63} implies $b^{pt_0}\equiv1$.
What remains to show is that the sequence satisfies~\eqref{eq:61}; the bound $\lh{pt_0}+\sum_{i\ge1}\lh{t_i}\le2\lh m$
then follows from~$\php$ as above. So, assume that
\begin{equation}\label{eq:62}
g^rb^{r'_0}\prod_{i=1}^{k-1}g_i^{r_i}\equiv1,
\end{equation}
where $r<t$, $r'_0<pt_0$, and $r_i<t_i$ for $0<i<k$. By taking the $p$th power, this implies
\[g^{pr}g_0^{r'_0s_0}\prod_{i=1}^{k-1}g_i^{pr_i+r'_0s_i}\equiv1,\]
hence in particular $p\mid t_0\mid r'_0$ by the independence of~$\vec g$, as $p\nmid s_0$. Thus, writing $r_0=r'_0/p$,
\eqref{eq:62} can be written as
\[g^rg_0^{r_0s_0}\prod_{i=1}^{k-1}g_i^{r_i+r_0s_i}\equiv1.\]
Then the independence of~$\vec g$ gives $r=0$, $r_0=0$ (using $p\nmid s_0$ and $r_0<t_0$), and then $r_i=0$ for all
$0<i<k$, as required.

This finishes the proof that $\fii_{\vec g}$ is a bijection, thus $\{g\}\cup\{g_i:i<k\}$ generates~$\zg m$. In order to
save us from the trouble of dealing with exponents, let
\[X=\bigl\{g^{2^j}:j<\lh t\bigr\}\cup\bigl\{g_i^{2^j}:i<k,j<\lh{t_i}\bigr\};\] then $X\sset\zg m$ has size
$\card(X)=\lh t+\sum_i\lh{t_i}=O\bigl(\lh m\bigr)$, and every $a\in\zg m$ can be written as $a=\prod Y\rem m$ for some
$Y\sset X$. Notice that having fixed~$X$, we can represent $Y$ by $\card(X)$ bits, and therefore by a single small
number; in particular, we can $\Sig0$-define the $Y_a$ with the least code such that $a\equiv\prod Y_a$. Then we can
compute iterated products modulo~$m$ using $\tot^*_\imulm\bigl[O\bigl(\lh m\bigr),-\bigr]$ and~$\tot^*_\powm$ by
\[\prod_{i<n}a_i\equiv\begin{cases}
  0,&\text{if $a_i\equiv0$ for some $i<n$,}\\
  \displaystyle\prod_{x\in X}x^{\card\{i<n:x\in Y_{a_i}\}},&\text{otherwise.}
\end{cases}\]
This definition provably satisfies the recurrence
\begin{align*}
\prod_{i<0}a_i&\equiv1,\\
\prod_{i<n+1}a_i&\equiv a_n\prod_{i<n}a_i.
\end{align*}

We have proved $\tot_\imulm[-,w]$ for $w=\min\bigl\{z,p(x,\lh z)\bigr\}$. In order to show $\tot^*_\imulm[-,w]$, we
have to deal with a sequence of iterated products modulo different $m\le w$ in parallel. As usual, it suffices to show
that given~$m$, we can $\Sig0$-define a suitable set~$X$ as above. Now, we have already seen that a good
independent sequence $\vec g$ for~$m$ can be encoded using $O\bigl(\lh z\bigr)$ bits; the corresponding exponents
$\vec t$ are $\Sig0$-definable from $\vec g$ as $t_i=o_m(g_i)$, thus we can $\Sig0$-define the maximum of
$\sum_i\lh{t_i}$ among such sequences, and then $\Sig0$-define a good independent sequence with least code that
achieves the maximum. Then we can define~$X$ from~$\vec g$.
\end{Pf}

We note that the argument in Lemma~\ref{lem:cyc-imulm} actually shows $\cyc[z,x]\to\tot^*_\imulm[-,w]$ whenever $w\le z$
and $\lh w^2\le\lh x\,\lh y$ for some~$y$. However, we will only need the formulation given in Lemma~\ref{lem:cyc-imulm}
to proceed, while in the end, we will obtain full $\tot^*_\imulm$ anyway.

We are now ready to finish the proof of the main result of this paper.
\begin{Thm}\label{thm:vtc-imul}
$\vtc$ proves $\imul$.
\end{Thm}
\begin{Pf}
For any fixed~$z$, we can prove
\begin{equation}\label{eq:60}
x^6\lh z^3\le z\to\cyc[z,x]
\end{equation}
by induction on~$x$: $\cyc[z,0]$ holds vacuously, and $\vtc$ proves
\begin{align*}
\cyc[z,x]\land(x+1)^6\lh z^3\le z
&\to\tot^*_\imulm\bigl[-,(x+1)^6\lh z^3\bigr]\\
&\to\imul\bigl[(x+1)^2\lh z\bigr]\\
&\to\cyc[z,x+1]
\end{align*}
by Lemmas \ref{lem:imul-cyc}, \ref{lem:imulm-imul}, and~\ref{lem:cyc-imulm}.

This implies $\imul[x]$ for all~$x$: taking $z$ such that $z\ge x^6\lh z^3$, we have $\cyc[z,x]$ by~\eqref{eq:60}, thus
$\tot^*_\imulm[x^3]$ by Lemma~\ref{lem:cyc-imulm}, and $\imul[x]$ by Lemma~\ref{lem:imulm-imul}.
\end{Pf}

\begin{Cor}\label{cor:vtc-cyc}
$\vtc$ proves that $\zg m$ is cyclic for all primes~$m$.
\noproof\end{Cor}

\begin{Cor}\label{cor:vtc-div}
$\vtc$ proves $\Div$: for every $X>0$ and~$Y$, there are $Q$ and $R<X$ such that $Y=QX+R$.
\noproof\end{Cor}

By results of Je\v r\'abek~\cite{ej:vtc0iopen}, we obtain the following consequence of Theorem~\ref{thm:vtc-imul} relating
$\vtc$ to Buss's single-sorted theories of arithmetic (see \cite{ej:vtc0iopen} for background):
\begin{Cor}\label{cor:vtc-sig0-min}
$\vtc$ proves the $\rsuv$ translations of $\sig0\text-\ind$ and $\sig0\text-\Min$.
\noproof\end{Cor}

Using the $\rsuv$-isomorphism of $\vtc$ to $\delt1\text-\M{CR}$, we can formulate the results in terms of the theories
of Johannsen and Pollett:
\begin{Cor}\label{cor:c02}
$\delt1\text-\M{CR}$ and $C^0_2$ prove $\sig0\text-\ind$, $\sig0\text-\Min$, and (a suitable
single-sorted formulation of) $\imul$. Moreover, $C^0_2[\M{div}]$ is an extension of~$C^0_2$ by a definition, and
therefore a conservative extension.
\noproof\end{Cor}

We stress that in Corollary~\ref{cor:c02}, $\sig0$ refers to sharply bounded formulas in Buss's original language, not in the
expanded language employed in \cite{joh-pol:c02,joh-pol:d1cr}. (In the latter language, $\sig0\text-\ind$ is equivalent
to $\tpv$, and $\sig0\text-\Min$ to $T^1_2$, which is strictly stronger than $C^0_2$ unless the polynomial hierarchy
collapses to~$\tc$, provably in the theory.)

\section{Tying up loose ends}\label{sec:tying-loose-ends}

Our arguments leading to the proof of Theorem~\ref{thm:vtc-imul} involved a few side results that might be
interesting in their own right, but we only proved them in a minimal form sufficient to carry out the
main argument. In this section, we polish them to more useful general results.

\subsection{Chinese remainder reconstruction}

The first side-result concerns the CRR reconstruction procedure. The statement of
Theorem~\ref{thm:crr-rec} gives only a loose bound on~$\rec(\vec m;\vec x)$, and involves unnecessary constraints on~$\vec
m$. These restrictions carry over to Corollary~\ref{cor:crr-rec}, whose statement also imposes an unnecessary bound on~$X$.

Once we prove $\imul$ and~$\Div$ in~$\vtc$, it is not particularly difficult to improve the bounds in
Theorem~\ref{thm:crr-rec} and Corollary~\ref{cor:crr-rec} to $X<\prod_{i<k}m_i$, and to generalize $\rec(\vec m;\vec x)$ so that it also applies
to $m_i=2$. Alternatively, we may abandon Definition~\ref{def:crr-rec} altogether in favour of a more obvious algorithm (note
that we do not require $\vec m$ to consist of primes):
\begin{Def}\label{def:rec-gen}
(In $\vtc$.) Given a sequence $\vec m$ of pairwise coprime nonzero numbers, and $\vec x<\vec m$, let
\[\rec^+(\vec m;\vec x)=\Bigl(\sum_{i<k}x_ih_i\prod_{j\ne i}m_j\Bigr)\rem\prod_{i<k}m_i,\]
where
\[h_i=\prod_{j\ne i}m_j^{-1}\rem m_i.\]
\end{Def}
\begin{Thm}\label{thm:crr-gen}
$\vtc$ proves the following for any pairwise coprime sequence~$\vec m$.
\begin{enumerate}
\item\label{item:6}
For every $\vec x<\vec m$, $\rec^+(\vec m;\vec x)$ is the unique $X<\prod_im_i$ such that $\vec x=X\rem\vec m$.
\item\label{item:7}
For every $X$, $\rec^+(\vec m;X\rem\vec m)=X\rem\prod_im_i$.
\end{enumerate}
\end{Thm}
\begin{Pf}

\ref{item:6}:
Put $M=\prod_{i<k}m_i$. It is easy to show by induction on~$k$ that if $X$ is divisible by $m_i$ for each $i<k$, then
it is divisible by~$M$. Thus, also $X\equiv X'\pmod{\vec m}$ implies $X\equiv X'\pmod M$. This shows uniqueness. We have
$\rec^+(\vec m;\vec x)<M$ by definition, and
\[h_i\prod_{j\ne i}m_j\equiv\left\{\begin{matrix}1,&i'=i\\[\smallskipamount]0,&i'\ne i\end{matrix}\right\}\pmod{m_{i'}}\]
implies $\rec^+(\vec m;\vec x)\equiv x_i\pmod{m_i}$.

\ref{item:7}: By definition, $X'=X\rem M$ satisfies $X'<M$ and $X\equiv X'\pmod{\vec m}$, thus $X'=\rec^+(\vec
m;X\rem\vec m)$ by~\ref{item:6}.
\end{Pf}

\begin{Rem}\label{rem:crr-not-coprime}
It is possible to generalize CRR reconstruction further to \emph{arbitrary} sequences~$\vec m$. First, $\vtc$ can
define $M=\lcm(\vec m)$ as $\prod_{j<l}p_j^{e_j}$, where $\vec p$ is a list collecting all prime factors of~$\vec m$,
and $e_j=\max_iv_{p_j}(m_i)$. Then, $\vtc$ can prove that for any $\vec x<\vec m$ which satisfies
$x_i\equiv x_{i'}\pmod{\gcd(m_i,m_{i'})}$ for all $i<i'<k$, there exists a unique $X<M$ such that $\vec x=X\rem\vec m$
by applying Theorem~\ref{thm:crr-gen} modulo $\p{p_j^{e_j}:j<l}$. We leave the details to the reader.
\end{Rem}

\subsection{Modular powering}

In Theorem~\ref{thm:pow}, we proved that $V^0+\wphp$ can do powering modulo (small) primes. We will generalize it in two
ways: first, we can formalize powering modulo arbitrary small nonzero numbers, and second, we will indicate how to
formulate the result purely in the single-sorted theory $\idz+\wphp(\delz)$.
\begin{Thm}\label{thm:powm-compos}
$V^0+\wphp$ proves that for every $m$, $a<m$, and~$r$, there exists an elementwise unique sequence $\p{a_i:i\le r}$ such
that $a_i<m$, $a_0\equiv1\pmod m$, and $a_{i+1}\equiv aa_i\pmod m$ for each~$i$.
\end{Thm}
\begin{Pf}
Uniqueness follows by induction on~$i$.

For existence, assume first that $m=p^e$ is a prime power. Then we can define powering in~$\zg m$ in the same way as in
Section~\ref{sec:modular-powering}: as already noted there, the basic method applies to arbitrary abelian groups
(provided we can do products of logarithmic length, which we can here as the proof of
$\tot^*_\imulm\bigl[\lh w,-\bigr]$ in Corollary~\ref{cor:vl-imul-plog} works modulo arbitrary~$m$); we only need to be a bit
more careful with applications of~$\wphp$, as $\zg m$ no longer consists of the entire interval $[1,m-1]$. However, we
may construct (as a set) a bijection between $\zg m$ and $\bigl[0,\fii(m)\bigr)$, where $\fii(m)=(p-1)p^{e-1}$: e.g.,
we can map $x<\fii(m)$ to $p\fl{x/(p-1)}+\bigl(x\rem(p-1)\bigr)+1\in\zg m$. With this in mind, we can prove
Lemma~\ref{lem:inj-surj} (for $x$ coprime to~$m$) using an instance of $\php^{2\fii(m)}_{\fii(m)}$. The proof of
Lemma~\ref{lem:bad-primes} then works unchanged (making sure the $x_i$ are coprime to~$m$), and so does the proof of
Theorem~\ref{thm:pow} as long as $a$ is coprime to~$m$. For general~$a$, we write $a\equiv p^u\tilde a$ with $u\le e$ and
$\tilde a\in\zg m$, and we define
\[a^i\equiv\begin{cases}0,&ui\ge e,\\p^{ui}\tilde a^i,&\text{otherwise.}\end{cases}\]

If $m$ is not a prime power, we find its prime factorization $m=\prod_{j<k}p_j^{e_j}$. We apply the construction above
in parallel to define $\p{a_{i,j}:i\le r,j<k}$ where $a_{i,j}=a^i\rem p_j^{e_j}$, and then we define $a^i\rem m$
as the unique $a_i<m$ such that $a_i\equiv a_{i,j}\pmod{p_j^{e_j}}$ for each $j<k$. (This form of the Chinese remainder
theorem is provable already in~$V^0$, cf.\ D'Aquino~\cite{daqui:cheb}.)
\end{Pf}

In order to get the result already in $\idz+\wphp(\delz)$, one way would be to chase the proofs in Sections
\ref{sec:polylogarithmic-cut} and~\ref{sec:modular-powering} as well as of Theorem~\ref{thm:powm-compos}, and make sure that
we can formulate everything without explicit usage of second-order objects, using only $\delz$-definable ``classes''.
However, it is perhaps less work to infer it directly from Theorem~\ref{thm:powm-compos} using the witnessing theorem
for~$V^0$ and the conservativity of $V^0$ over~$\idz$:
\begin{Prop}\label{thm:wit}
If $V^0\vdash\forall x\,\exists X\,\fii(x,X)$, where $\fii\in\Sig0$, there exists a polynomial $p$ and a
$\delz$~formula $\theta(x,u)$ such that
\begin{equation}
\label{eq:71}\idz\vdash\forall x\,\fii\bigl(x,\{u<p(x):\theta(x,u)\}\bigr).
\end{equation}
Here, $\fii\bigl(x,\{u<p(x):\theta(x,u)\}\bigr)$ denotes the $\delz$ formula obtained from $\fii(x,X)$ by replacing all
atomic subformulas $t\in X$ with $t<p(x)\land\theta(x,t)$, and atomic subformulas $\alpha\bigl(\lh X,\dots\bigr)$ with
$\exists z\le p(x)\,\bigl(\alpha(z,\dots)\land\forall w\le p(x)\,\bigl(z\le w\eq\forall u<p(x)\,(\theta(x,u)\to
u<w)\bigr)\bigr)$.

The same holds for $V^0+\wphp$ and $\idz+\wphp(\delz)$ in place of $V^0$ and~$\idz$, respectively.
\end{Prop}
\begin{Pf}
By \cite[Thm.~V.5.1]{cook-ngu} (which is basically Herbrand's theorem for~$\ob{V^0}$), there is an $L_{\ob{V^0}}$
function symbol~$F$ such that $\ob{V^0}\vdash\forall x\,\fii(F(x))$, and $F$ is $\Sig0$ bit-definable by
the Claim in the proof of \cite[V.6.5]{cook-ngu}, i.e., $\ob{V^0}\vdash F(x)=\{u<p(x):\theta(x,u)\}$ for some term~$p$
and $\theta\in\Sig0$. Thus, \eqref{eq:71} by the conservativity of $\ob{V^0}$ over~$V^0$ and over~$\idz$.

In the presence of~$\wphp$, we have
$V^0\vdash\forall x\,\exists X\,\exists n\,\bigl(\fii(x,X)\lor\neg\php^{2n}_n(X)\bigr)$, thus there is an
$L_{\ob{V^0}}$ function $F(x)=\{u<p(x):\theta(x,u)\}$ such that
$\ob{V^0}\vdash\forall x\,\exists n\,\bigl(\fii(x,F(x))\lor\neg\php^{2n}_n(F(x))\bigr)$ as above. Then
$\idz\vdash\forall x\,\exists n\,\bigl(\fii(x,\{u<p(x):\theta(x,u)\})\lor\neg\php^{2n}_n(\{u<p(x):\theta(x,u)\})\bigr)$
by conservativity, hence $\idz+\wphp(\delz)$ proves $\forall x\,\fii\bigl(x,\{u<p(x):\theta(x,u)\}\bigr)$.
\end{Pf}

Alternatively, Proposition~\ref{thm:wit} has an easy direct model-theoretic proof as in \cite[L.~V.1.10]{cook-ngu}.

\begin{Cor}\label{cor:powm-compos-delz}
There exists a $\delz$~formula $\pi(a,r,m,b)$ such that $I\delz+\wphp(\delz)$ proves
\begin{align*}
\pi(a,r,m,b)&\to b<m,\\
m\ne0&\to\exists!b\:\pi(a,r,m,b),\\
m\ne0&\to\pi(a,0,m,1\rem m),\\
\pi(a,r,m,b)&\to\pi(a,r+1,m,ab\rem m).
\end{align*}
\end{Cor}
\begin{Pf}
By applying Proposition~\ref{thm:wit} to Theorem~\ref{thm:powm-compos}, we obtain a $\delz$ formula $\pi'(a,r,m,i)$ that, provably in
$\idz+\wphp(\delz)$, defines the bit-graph of a function $\p{a,r,m}\mapsto a_r+2^{\lh m}A$, where $A$ is a code of
a sequence $\p{a_i:i\le r}$ satisfying $a_0=1\rem m$ and $a_{i+1}=aa_i\rem m$. We can then define
$\pi(a,r,m,b)$ as $b<m\land\forall i<\lh m\,\bigl(\bit(b,i)=1\eq\pi'(a,r,m,i)\bigr)$.
\end{Pf}
\begin{Rem}\label{rem:powm-unique}
Using $\delz$-induction, it is easy to show in $\idz+\wphp(\delz)$ that the formula $\pi$ in
Corollary~\ref{cor:powm-compos-delz} is unique up to provable equivalence, and that it satisfies the Tarski high-school
identities $a^{r+s}\equiv a^ra^s$, $(ab)^r\equiv a^rb^r$, and $a^{rs}\equiv(a^r)^s$ modulo~$m$.
\end{Rem}

Since the statements in Corollary~\ref{cor:vl-imul-plog} are $\forall\Sigma^1_1$, they can be translated to~$\idz$ in a
similar way. Not all of these translations are genuinely interesting, though. In particular, functions with non-small
integers as inputs or outputs are rather awkward to formulate, using $\delz$~formulas describing individual bits of the
numbers, etc. On the other hand, when restricted to \emph{small} numbers, the translation of $\imul\bigl[\lh w^c\bigr]$
(actually, the result is small only if $c=1$, barring uninteresting products with lots of~$1$s) to $\idz$ is already
known from~\cite{ber-daqui}. Likewise, division of small numbers is trivial. Concerning $\imulm$, if $\p{a_i:i<n}$ is
given in $\idz$ explicitly by a sequence, we can again do $\prod_{i<n}a_i\rem m$ by the results of~\cite{ber-daqui} as
we can just compute $\prod_{i<n}a_i$ and reduce it modulo~$m$, but the result is new in the more general case that
$\p{a_i:i<n}$ is only given by a $\delz$-definable function:
\begin{Cor}\label{cor:imulm-delz}
For every $\delz$~formula $\fii(\vec z,n,a)$ and every constant~$c$, there is a $\delz$~formula $\pi(\vec z,n,m,w,y)$
such that $\idz$ proves: for all $m>0$, $w$, and~$\vec z$, if $\forall n<\lh w^c\,\exists!a\,\fii(\vec z,n,a)$, then
$\forall n\le\lh w^c\,\exists!y\,\pi(\vec z,n,m,w,y)$, and for all $n<\lh w^c$ and all $y,a$,
\begin{align*}
&\phantom{{}\to{}}\pi(\vec z,0,m,w,1\rem m),\\
\pi(\vec z,n,m,w,y)\land\fii(\vec z,n,a)&\to\pi(\vec z,n+1,m,w,ya\rem m).
\end{align*}
(That is, if $\fii$ with parameters~$\vec z$ defines a function $f(n)$, then $\pi$ defines a function $g(n,m)$
satisfying $g(0,m)\equiv1\pmod m$ and $g(n+1,m)\equiv g(n,m)f(n,m)\pmod m$ for all $n<\lh w^c$.)
\noproof\end{Cor}

\section{Conclusion}\label{sec:conclusion}

We proved that $\vtc$ can formalize the Hesse, Allender, and Barrington $\tc$~algorithms for integer division and
iterated multiplication. While this result is hopefully interesting in its own right, on a broader note it contributes
to our understanding of $\vtc$ as a robust and surprisingly powerful theory, capable of adequate formalization of
common $\tc$-computable predicates and functions and their fundamental properties. In particular, it makes a strong
case that $\vtc$ is indeed the right theory corresponding to~$\tc$; previous results of~\cite{ej:vtc0iopen} suggested
that $\vtcim$ might be another viable choice, perhaps more suitable than $\vtc$ itself, but results of the present
paper render this distinction moot.

A possible area for further development of $\vtc$ is to try and see what it can prove about approximations of analytic
functions such as $\exp$, $\log$, trigonometric and inverse trigonometric functions. In view of bounds on primes in
Section~\ref{sec:primes} and in Nguyen~\cite{ngu:thesis}, another intriguing question is if $\vtc$ can prove the
prime number theorem.

On a different note, our result on formalization of a $\delz$-definition of modular exponentiation essentially relied
on several instances of the weak pigeonhole principle, but it is not clear to what extent is this really necessary. We
leave it as an open problem if we can we construct a well-behaved modular exponentiation function in a substantially
weaker theory than $\idz+\wphp(\delz)$, or even in $\idz$ itself. The latter problem was first posed by
Atserias~\cite{ats:wphp-conf,ats:wphp}.

\section*{Acknowledgements}
I am grateful to the anonymous referee for helpful comments, in particular drawing my attention to
\cite{ats:wphp-conf,ats:wphp}. The research was supported by grant 19-05497S of GA \v CR. The Institute of Mathematics
of the Czech Academy of Sciences is supported by RVO: 67985840.

\bibliographystyle{mybib}
\bibliography{mybib}
\end{document}